\documentclass{iopart}
\usepackage{amssymb,mathptmx,times,rangcite,graphicx}
\usepackage[varg]{txfonts}
\makeatletter
0
\long\def\@makefntext#1{\parindent 1em\noindent
 \makebox[1em][l]{\footnotesize\rm$^\arabic{footnote}$}%
 \footnotesize\rm #1}
\def\@makefnmark{\hbox{$^\arabic{footnote}$}}
\def\@thefnmark{\arabic{footnote}}

\def\be{\begin{equation}}
\def\ee{\end{equation}}
\def\bea{\begin{eqnarray}}
\def\eea{\end{eqnarray}}
\def\bse{\numparts}
\def\ese{\endnumparts}
\eqnobysec
\def\numparts{\refstepcounter{equation}%
     \setcounter{eqnval}{\value{equation}}%
     \setcounter{equation}{0}%
     \def\theequation{\arabic{section}.\arabic{eqnval}{\it\alph{equation}}}}
\def\endnumparts{\def\theequation{\arabic{section}.\arabic{equation}}%
     \setcounter{equation}{\value{eqnval}}}

\let\truesum=\sum
\let\trueint=\int
\let\trueoint=\oint
\let\trueprod=\prod
\let\truecirc=\circ
\def\sum{\mathop{\textstyle\truesum}\limits}
\def\int{\mathop{\textstyle\trueint}\limits}
\def\oint{\mathop{\textstyle\trueoint}\limits}
\def\prod{\mathop{\textstyle\trueprod}\limits}
\def\circ{{\ifmmode\truecirc\else$\truecirc$\fi}}
\def\txtfrac#1#2{{\textstyle\frac{#1}{#2}}}
\newcommand\partialderiv[3][]{\frac{\partial^{#1}#2}{\partial {#3}^{#1}}}
\def\overl@ss#1#2{\vcenter{\offinterlineskip
        \ialign{$\m@th#1\hfil##\hfil$\crcr#2\crcr<\crcr } }}
\def\overgr@at#1#2{\vcenter{\offinterlineskip
        \ialign{$\m@th#1\hfil##\hfil$\crcr#2\crcr>\crcr } }}
\def\gl{\mathrel{\mathpalette\overl@ss>}}
\def\lg{\mathrel{\mathpalette\overgr@at<}}
\def\fbf#1{\setbox0=\hbox{$#1$}\kern-0.10\wd0
  \lower0.04em\copy0\kern-\wd0 \lower0.04em\hbox{\kern+0.05em\copy0}\kern-\wd0
  \raise0.00em\copy0\kern-\wd0 \raise0.00em\hbox{\kern-0.05em\box0}}
\renewenvironment{pmatrix}{\left(\!\!\begin{array}{cc}}{\end{array}\!\!\right)}
\def\_#1{{\mathsf{#1}}}
\def\@#1{{\mathbf{#1}}}
\makeatother
\def\Real{{\mathbb{R}}}
\def\Complex{{\mathbb{C}}}
\def\Natural{{\mathbb{N}}}
\def\Integer{{\mathbb{Z}}}
\def\Re{\mathop{\rm Re}\nolimits}
\def\Im{\mathop{\rm Im}\nolimits}
\def\Wr{\mathop{\rm Wr}\nolimits}
\def\diag{\mathop{\rm diag}\nolimits}
\def\Res{\mathop{\rm Res}\limits}
\def\sech{\mathop{\rm sech}\nolimits}
\let\.=\dot
\let\^=\hat
\let\~=\tilde
\let\==\bar
\let\#=\rm
\def\e{{\rm e}}
\def\d{{\rm d}}
\def\o#1{^{(#1)}}
\let\epsilon=\varepsilon
\def\half{{\textstyle\frac12}}
\def\hsigma{\^\sigma_3}
\def\esh#1{\e^{#1\^\sigma_3}}
\def\Zhat{\^{\_Z}}
\def\I{\mathrm{I}}
\def\II{\mathrm{II}}
\def\III{\mathrm{III}}
\def\IV{\mathrm{IV}}
\def\punct{^{\,[\raise0.08ex\hbox{\scriptsize$\slash$}\kern-0.34em0]}}

\begin{document}
\title[IBVPs for discrete evolution equations: DLS and IDNLS]%
{Initial-boundary value problems for discrete evolution equations:
discrete linear Schr\"odinger and integrable discrete nonlinear
Schr\"odinger equations}
\author{Gino Biondini and Guenbo Hwang}
\address{State University of New York at Buffalo, Department of Mathematics, Buffalo, NY 14260}
\date{\small\today}

\begin{abstract}
We present a method to solve initial-boundary value problems for
linear and integrable nonlinear differential-difference evolution equations.
The method is the discrete version of the one developed by
A. S. Fokas to solve initial-boundary value problems for linear and
integrable nonlinear partial differential equations
via an extension of the inverse scattering transform.
The method takes advantage of the Lax pair formulation for both linear and
nonlinear equations,
and is based on the simultaneous spectral analysis of both parts
of the Lax pair.
A key role is also played by the global algebraic relation that
couples all known and unknown boundary values.
Even though additional technical complications arise in discrete problems
compared to continuum ones,
we show that a similar approach can also solve
initial-boundary value problems for linear and integrable nonlinear
differential-difference equations.
We demonstrate the method by solving initial-boundary value problems
for the discrete analogue of
both the linear and the nonlinear Schr\"odinger equations,
comparing the solution to those of the corresponding continuum problems.
In the linear case we also explicitly discuss Robin-type boundary conditions
not solvable by Fourier series.
In the nonlinear case we also identify the linearizable boundary conditions,
we discuss the elimination of the unknown boundary datum,
we obtain explicitly the linear and continuum limit of the solution,
and we write down the soliton solutions.
\par\medskip\noindent\today
\par\kern-\bigskipamount
\end{abstract}

%\special{!userdict begin /bop-hook{gsave 40 255 translate
%90 rotate /Helvetica findfont 20 scalefont setfont
%0 0 moveto 0.5 setgray (Draft -- Please Do Not Distribute) show grestore}def end}

%%%%%%%%%%%%%%%%%%%%%%%%%%%%%%%%%%%%%%%%%%%%%%%%%%%%%%%%%%%%%%%%%%%%%%%%%%%%%%
%%%%%%%%%%%%%%%%%%%%%%%%%%%%%%%%%%%%%%%%%%%%%%%%%%%%%%%%%%%%%%%%%%%%%%%%%%%%%%
\section{Introduction and outline}

The development of the theory of infinite-dimensional integrable systems
was a remarkable advance of mathematical physics over the last forty years.
One of the key properties of such systems
is that they can be written as the compatibility condition of an
overdetermined linear system, called the Lax pair.
In turn, the existence of Lax pair is deeply related to many other
features of these systems.
Among them is the inverse scattering transform (IST),
a nonlinear analogue of the Fourier transform which can be used
to solve the initial value problem (IVP).
The IST was successfully used in the late 1960's and early 1970's
to solve IVPs on infinite domains
or with periodic or quasi-periodic boundary conditions (BCs)
for a variety of
nonlinear partial differential equations (PDEs), differential-difference
fully discrete, integro-differential equations, etc.\
(e.g., see Refs.~\cite{AblowitzClarkson,AS1981,BBEIM1994,FT1987} and
references therein).

Following the solution of IVPs, a natural issue was
the solution of initial-boundary value problems (IBVPs).
After some early results \cite{JMP16p1054,NLTY2p37,JMP32p99,PhysD35p167}, however,
the issue remained essentially open for over twenty years.
Recently, renewed interest in the problem
has lead to a number of developments
(e.g., see Refs.\ \cite{JPA30p3505,JPA23p2507,IP16p1813,JETPL74p481,%
 PRSLA453p1411,JMP41p4188,IMA67p559,CMP230p1,JNLMP10p47,CPAM58p639,%
 NLTY18p1771,PRSLA456p805,JMPv41p414,IP22p209,FAA21p86}
and references therein).
Particularly important among these is the method developed by A.~S.\ Fokas
\unskip~\cite{PRSLA453p1411,JMP41p4188,IMA67p559,CMP230p1,JNLMP10p47,CPAM58p639,NLTY18p1771,PRSLA456p805}.
\unskip\break
Fokas'\ method, which is a significant extension of the IST,
is based on the simultaneous spectral analysis of
both parts of the Lax pair.
A crucial role is also played by a relation called
global algebraic relation %(sometimes referred to as the Fokas relation)
that couples all known and unknown boundary values.
Indeed, it is the
analysis of the global relation that allows one to express the
unknown boundary datum in terms of known ones plus the initial datum.
Importantly, the method also yields a new approach to IBVPs for
linear PDEs,
which allows the solution of new kinds of problems.

At the same time,
the effort to extend the properties of integrable nonlinear PDEs
to discrete integrable systems
has been an ongoing theme in the last thirty years
(e.g., see Refs.~\cite{AblowitzClarkson,NLTY13p889,JMP17p1011,APT2003,%AS1981,Date,
 PRB9p1924,PTP51p703,%PRL67p1825,
 PLA207p263,%PRL81p325,
 JETP40p269,jphysa37p11819,%AAM39p133,PLA126p419,
 PR18p1} %,CSF8p917,CMP145p181
and references therein).
The purpose of this work is to show that, \textit{mutatis mutandis},
an approach similar to that for PDEs can also be used to solve IBVPs
for linear and integrable nonlinear differential-difference equations
(DDEs).\break
We demonstrate this claim by solving IBVPs for
the discrete analogue of the linear and nonlinear Schr\"odinger equations
on the natural numbers.
Note that the integrable discrete nonlinear Schr\"odinger (IDNLS)
equation is an important model
since it arises in a number of physical and mathematical contexts
(e.g., see references in Ref.~\cite{APT2003}).

The outline of this work is the following.
In section~\ref{s:DLS} we solve the IBVP
on the natural numbers for the discrete linear Schr\"odinger (DLS) equation,
namely the linear DDE
\be
i\.q_n + \frac{q_{n+1}-2q_n+q_{n-1}}{h^2}= 0\,
\label{e:DLS}
\ee
where
$q_n=q_n(t)\in\Complex$,
$n\in\Natural$,
$\.f\equiv df/dt$ denotes time derivative
and~$h$ is the lattice spacing.
Then, in sections~\ref{s:IDNLS} and~\ref{s:idnlsdata}
we consider the IBVP for the
integrable nonlinear counterpart of~\eref{e:DLS},
namely the IDNLS equation or
Ablowitz-Ladik (AL) equation\unskip~\cite{JMP16p598,JMP17p1011},
\be
i\.q_n+ \frac{q_{n+1}-2q_n+q_{n-1}}{h^2}-\nu |q_n|^2(q_{n+1}+q_{n-1})= 0\,
\label{e:IDNLS}
\ee
(where as usual the cases $\nu=-1$ and $\nu=1$ will be called respectively
focusing and defocusing).
In particular, in section~\ref{s:idnlsdata}
we discuss the elimination of the unknown boundary datum,
the linearizable boundary conditions,
and we write down the soliton solutions.
Finally, in order to appreciate the similarities and differences between the
method in the discrete versus the continuum case,
in section~\ref{s:continuum} we review the solution of
IBVPs for the continuum limits of both equations,
namely the linear and nonlinear Schr\"odinger equations,
and we discuss explicitly the correspondence between the method
in the discrete case versus the continuum limit.
The proof of various statements in the text is confined to the Appendix,
which also contains
a list of notations and frequently used formulae.

In both the linear and the nonlinear problem
we will require the initial datum to be absolutely summable
and the boundary datum~$q_0(t)$ to be smooth,
even though the method can be formulated under weaker conditions.
The constant~$h$ can be eliminated from~\eref{e:DLS} and~\eref{e:IDNLS}
via the rescalings $t'=t/h^2$ and $q'_n(t)=hq_n(t)$.
Thus, for simplicity we will consider the rescaled problems throughout
(thus effectively setting $h=1$);
however, we will will omit the primes
except when considering the limit $h\to0$
to recover the solution of the continuum cases.
The indended meaning should be clear from the context.
Also, for brevity we will occasionally omit functional dependences
when doing so does not cause ambiguity.

%%%%%%%%%%%%%%%%%%%%%%%%%%%%%%%%%%%%%%%%%%%%%%%%%%%%%%%%%%%%%%%%%%%%%%%%%%%%%%
%%%%%%%%%%%%%%%%%%%%%%%%%%%%%%%%%%%%%%%%%%%%%%%%%%%%%%%%%%%%%%%%%%%%%%%%%%%%%%
\section{Discrete linear Schr\"odinger equation}
\label{s:DLS}

Here we solve the linear problem~\eref{e:DLS},
which serves to introduce some of the tools that will be used in
the nonlinear case.
In section~\ref{s:linearLaxpair} we derive a Lax pair for~\eref{e:DLS}.
Then, in section~\ref{s:1.3} we solve the IVP and in section~\ref{s:1.4}
IBVPs via spectral methods.

%%%%%%%%%%%%%%%%%%%%%%%%%%%%%%%%%%%%%%%%%%%%%%%%%%%%%%%%%%%%%%%%%%%%%%%%%%%%%%
\paragraph{IVP and IBVP for DLS via Fourier methods.}

Let us briefly review the solution of
the IVP and the IBVP via Fourier methods.
Doing so we will serve to introduce quantities that will also be used later.
Consider first the IVP,
namely~\eref{e:DLS} with $n\in\Integer$ and with $q_n(0)$ given.\break
We require that the initial datum $q_n(0)$
decays rapidly enough as $n\to\pm\infty$ to belong to $\ell^1(\Integer)$,
the space of sequences $\{a_n\}_{n\in\Integer}$ such that
$\sum\nolimits_{n=-\infty}^\infty|a_n|<\infty$.
Introduce the transform pair as
\bse
\label{e:Fourierpair}
\bea
\^q(k,t)= \sum_{n=-\infty}^\infty q_n(t)/z^n=
  \sum_{n=-\infty}^\infty \e^{-ink}q_n(t)\,,
\\
q_n(t)= \frac1{2\pi i}\oint_{|z|=1}\!z^{n-1}{\^q(z,t)}\,\d z=
  \frac1{2\pi}\,\, \int_{\!\!-\pi}^{\,\,\pi} \e^{ink} \^q(k,t)\,\d k\,,
\eea
\ese
where $z=\e^{ik}$, and the contour $|z|=1$ is oriented counterclockwise.
The transformation $k\to z$ maps $k\in\Real$ into~$|z|=1$
and $\Im\,k\gl0$ into~$|z|\lg1$
(with $k=\pm i\infty$ corresponding respectively to $z=0$ and $z=\infty$).
Use of~\eref{e:Fourierpair} yields
the solution of the IVP in Ehrenpreis form as
\be
q_n(t)=
  \frac1{2\pi i} \oint_{|z|=1}\!z^{n-1}\e^{-i\omega(z)t}\,{\^q(z,0)}\,\d z=
  \frac1{2\pi}\,\, \int_{\!\!-\pi}^{\,\,\pi} \e^{i(nk-\omega(k)t)}
  \^q(k,0)\,\d k\,,
\label{e:DLSsoln}
\ee
where the linear dispersion relation is
\be
\omega(z)=2-(z+1/z)= 2(1-\cos\,k)\,.
\label{e:DLSdisprel}
\ee
Now consider the IBVP, namely~\eref{e:DLS}
with $n\in\Natural$ and $t\in\Real^+$,
with $q_n(0)$ and $q_0(t)$~given.
We assume $q_n(0)\in\ell^1(\Natural)$ and
$q_0(t)\in{\mathcal C}(\Real^+_0)$.
Introduce the Fourier sine series and its inverse
as
\[
\^q\o{s}(z,t)= \sum_{n=1}^\infty q_n(t)(1/z^n-z^n)\,,\qquad\!\!
q_n(t)= \frac1{4\pi i} \oint_{|z|=1}(z^n-1/z^n)\,\^q\o{s}(z,t)\,\d z/z\,,
\nonumber
\]
Use of this pair yields the solution of the IBVP as
\bea
\fl
q_n(t)= \frac1{4\pi i}
  \oint_{|z|=1}(z^n-1/z^n)/z\,\,\e^{-i\omega(z)t}\,\^q\o{s}(z,0)\,\d z
  - \frac1{4\pi}\oint_{|z|=1}(z^n-1/z^n)/z\,\,\e^{-i\omega(z)t}\^g(z,t)\,\d z\,,
\label{e:dLSIBVPsoln}
\\
\noalign{\noindent where}
\^g(z,t)= (z-1/z)\int_0^t \e^{i\omega(z)t'}\,q_0(t')\,\d t'\,.
\nonumber
\eea

%%%%%%%%%%%%%%%%%%%%%%%%%%%%%%%%%%%%%%%%%%%%%%%%%%%%%%%%%%%%%%%%%%%%%%%%%%%%%%
\subsection{A Lax pair for the discrete linear Schr\"odinger equation}
\label{s:linearLaxpair}

A Lax pair formulation,
first discovered for nonlinear PDEs~\cite{CPAM21p467},
is also possible for linear PDEs,
and in fact it is the key to solving a wide class of IBVPs
\cite{JMP41p4188,PRSLA456p805}.
Here we show how a Lax pair for the DLS equation~\eref{e:DLS}.
can be obtained by taking the linear limit of the the Lax pair of the
IDNLS equation~\eref{e:IDNLS}.
(As in the continuum limit,
an algorithmic way also exists to obtain the Lax pair associated to
any linear discrete evolution equation.
The corresponding formalism will be presented elsewhere.)

It is well-known that the IDNLS~\eref{e:IDNLS} is a reduction of the
Ablowitz-Ladik (AL) system~\eref{e:AL} \cite{JMP17p1011}. A Lax pair
for~\eref{e:AL} is given by the overdetermined linear
system~\eref{e:ALLP}. To obtain the linear limit of~\eref{e:ALLP},
let $\_Q_n=O(\epsilon)$, and take $\Phi_n(z,t)=\@v_n(z,t)=
(v_{1,n},v_{2,n})^t$ to be a two-component vector. The leading order
solution of \eref{e:ALLP} is then $\@v_n(z,t)=
\_Z^n\e^{i(z-1/z)^2\sigma_3t/2}\@v_o$, where
$\@v_o=(v_{1,o},v_{2,o})^t$ is an arbitrary constant vector.
Choosing $v_{2,o}=1$ and keeping terms up to $O(\epsilon)$ then
yields the following \textit{scalar} linear system for $v_{1,n}$:
\bse \bea v_{1,n+1} - z\,v_{1,n} = q_nz^{-n}\e^{-i(z-1/z)^2 t/2}\,,
\label{e:Lp2.1}
\\
\.v_{1,n} - \txtfrac i2(z-1/z)^2 v_{1,n} = i
(zq_n-q_{n-1}/z)z^{-n}\e^{-i(z-1/z)^2 t/2}\,. \label{e:Lp2.2} \eea
\ese Enforcing the compatibility of~\eref{e:Lp2.1}
and~\eref{e:Lp2.2} now yields the discrete linear Schr\"odinger
equation~\eref{e:DLS}. To eliminate the dependence on $z^n$ from the
right-hand side (RHS) of~\eref{e:Lp2.1}, we now perform the
rescaling $z'=z^2$ and $\phi_n= z^{n-1}\e^{i(z-1/z)^2 t/2}v_{1,n}$.
Dropping primes for simplicity, we then obtain the following Lax
pair for~\eref{e:DLS}: \bea \phi_{n+1} - z\,\phi_n = q_n\,, \qquad
\.\phi_n + i\omega(z) \phi_n = i (q_n-q_{n-1}/z)\,,
\label{e:LaxpairL} \eea where $\omega(z)$ is given
by~\eref{e:DLSdisprel} as before. Indeed, although it may not be
obvious at this point, the meaning of the variable $z$
in~\eref{e:LaxpairL} coincides exactly with that of~$z$
in~\eref{e:Fourierpair}.

The rescaling $z'=z^2$ between the linear and the nonlinear problem
is the discrete analogue of the rescaling $k'=2k$ in the continuum limit.
Such rescaling will reflect on the location of the jumps
in the Riemann-Hilbert problem (RHP) for the IBVP in the nonlinear problem,
which will differ from the corresponding locations in the linear problem.

%%%%%%%%%%%%%%%%%%%%%%%%%%%%%%%%%%%%%%%%%%%%%%%%%%%%%%%%%%%%%%%%%%%%%%%%%%%%%%
\subsection{IVP for DLS via spectral analysis of the Lax pair}
\label{s:1.3}

We now solve the IVP for~\eref{e:DLS} using spectral methods.
Doing so will introduce some of the ideas that will be useful
for the IBVP and nonlinear case.
Making use of the integrating factor $z^n\e^{-i\omega(z)t}$
[with $\omega(z)$ as in~\eref{e:DLSdisprel}],
we introduce the modified eigenfunction
%$\phi_n(z,t)= z^n\,e^{-i\omega(z)t}\psi_n(z,t)$.
\be
\psi_n(z,t)= z^{-n}\,e^{i\omega(z)t}\phi_n(z,t)\,,
\label{e:Psidef}
\ee
which
%Then %$\psi_n=\phi_n\,z^{-n}\,e^{i\omega(z)t}$
satisfies the following modified Lax pair: %~\eref{e:LaxpairL}:
\be
\psi_{n+1}-\psi_n= \e^{i\omega(z)t}q_n/z^{n+1}\,,\qquad
\.\psi_n= \e^{i\omega(z)t}i(q_n-q_{n-1}/z)/z^n
\label{e:LaxpairL0}
\ee
Of course the above linear system is also compatible if $q_n(t)$
satisfies~\eref{e:DLS}.
It is then easy to define~$\phi_n\o{1,2}(z,t)$ as
the solutions of~\eref{e:LaxpairL}
which vanish as $n\to\mp\infty$, respectively:
\bea
\phi_n\o1(z,t)= \sum_{m=-\infty}^{n-1}q_m(t)\,z^{n-m-1},
\qquad
\phi_n\o2(z,t)= -\sum_{m=n}^\infty q_m(t)\,z^{n-m-1}.
\label{e:PhiIVPL}
\eea
Note that $\phi_n\o1(z,t)$ is analytic as a function of~$z$ for~$|z|<1$
and continuous on $|z|=1$,
while $\phi_n\o2(z,t)$ is analytic for~$|z|>1$ and bounded for $|z|=1$.
The jump conditions obtained by evaluating $\phi_n\o{1,2}(z,t)$ on $|z|=1$
then yield a scalar RHP:
$\phi_n\o1(z,t)-\phi_n\o2(z,t)= z^{n-1}\^q(z,t)\,$,
where $\^q(z,t)$ is given by~\eref{e:Fourierpair}.
However,
the difference $\phi_n\o1-\phi_n\o2$ solves the \textit{homogeneous} version
of~\eref{e:LaxpairL},
and hence it depends on $n$ and $t$ only through the factor
$z^n\,\e^{-i\omega(z)t}$.
Evaluating~\eref{e:PhiIVPL} at $(n,t)=(0,0)$
we can then rewrite the jump condition as:
\bea
\phi_n\o1(z,t)-\phi_n\o2(z,t)= z^{n-1}\e^{-i\omega(z)t}\^q(z,0)\,,
\qquad |z|=1\,.
\label{e:RHPL}
\eea
Equations~\eref{e:PhiIVPL} imply $\phi_n\o1(0,t)= q_{n-1}(t)\ne0$,
and $\phi_n\o2(z,t)\to0$ as $z\to\infty$.
Thus, the RHP defined by~\eref{e:RHPL} is trivially solved by
applying standard Cauchy projectors, namely:
\be
\phi_n(z,t)= \frac1{2\pi i}\oint_{|\zeta|=1} \zeta^{n-1}\,
  \e^{-i\omega(\zeta)t}\,\frac{\^q(\zeta,0)}{\zeta-z}\,\d\zeta\,,
\label{e:RHPLsoln}
\ee
where the contour is oriented counterclockwise, as usual.
Then, inserting~\eref{e:RHPLsoln} into the LHS of the first
of~\eref{e:LaxpairL},
one obtains the solution of the IVP as~\eref{e:DLSsoln}.

The continuum limit of \eref{e:DLSsoln} yields the solution of
the linear Schr\"odinger equation.
Indeed, reinstating the lattice spacing~$h$,
the solution of the IVP for the DLS~\eref{e:DLSsoln} is
\bse
\label{e:DLSsln2}
\bea
q_n(t)=
  \frac1{2\pi}\,\, \int_{\!\!-\pi/h}^{\,\,\pi/h} \e^{i(nkh-\omega(k)t)}
  \^q(k,0)\,\d k\,,
\\
\noalign{\noindent where now $\omega(k)= 2(1-\cos\,kh)/h^2$ and}
\^q(k,t)= h\sum_{n=-\infty}^\infty \e^{-inkh}q_n(t)\,.
\eea
\ese
Then, taking the limit $h\to0$ of~\eref{e:DLSsln2}
with $x_n=nh$ fixed,
one obtains~\eref{e:LSIVPsoln} and the first of~\eref{e:FTpair}.

%%%%%%%%%%%%%%%%%%%%%%%%%%%%%%%%%%%%%%%%%%%%%%%%%%%%%%%%%%%%%%%%%%%%%%%%%%%%%%
\subsection{IBVP for DLS via spectral analysis of the Lax pair}
\label{s:1.4}

We now use spectral methods to solve the IBVP for the DLS,
namely~\eref{e:DLS} for $n\in\Natural$ and $t\in\Real^+$,
with $q_n(0)$ and $q_0(t)$ given, where
as before we assume $q_n(0)\in\ell^1(\Natural)$
and $q_0(t)\in{\mathcal C}(\Real^+_0)$.
Before we do so, however, we address the issue of the
well-posedness of the linear system~\eref{e:LaxpairL}.

In the continuum limit,
the $t$-part of the Lax pair evaluated at $x=0$
depends on $q(0,t)$ and $q_x(0,t)$, only one of which is given.
Use of the global relation allows one to obtain
the unknown BC in terms of the given one.
In the discrete case,
evaluation of the $t$-part of the Lax pair for $n=0$
requires the knowledge of $q_{-1}(t)$.
Thus, \textit{the role of the unknown boundary datum in the discrete case
is played by the fictitious function $q_{-1}(t)$.}
In analogy with the continuum limit,
the solution method proceeds as though this function is given;
a posteriori we will then show that this unknown boundary datum is
determined in terms of known initial-boundary data via the global relation.

A similar problem arises with Fourier methods,
where one must define an appropriate transform so that the unknown
boundary data do not appear in the expression for the solution.
A similar situation also occurs in IBVPs
for Burgers' equation \cite{NLTY2p37,JMP32p99}, where the solution
depends on an unknown function that must be determined a posteriori.
There, similarly to nonlinear PDEs solvable by the IST,
the IBVP is reduced to a
nonlinear integro-differential equation~\cite{JMP32p99},
which can be linearized for special kinds of BCs~\cite{NLTY2p37}.

\begin{figure}[t!]
\smallskip
\centerline{\includegraphics[width=0.995\textwidth]{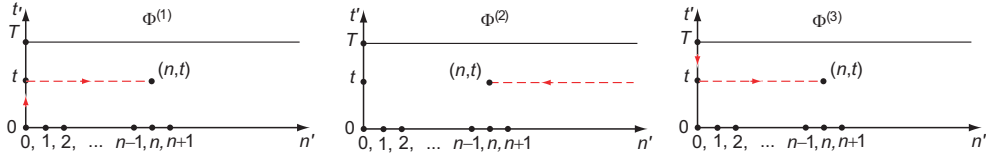}}
\caption{The distinguished points for the eigenfunctions
$\phi_n\o1$, $\phi_n\o2$ and $\phi_n\o3$.} \label{f:IBVPzeropts}
\end{figure}

\paragraph{Eigenfunctions and analyticity.}

\begin{figure}[t!]
\smallskip
\rightline{\includegraphics[width=0.405\textwidth]{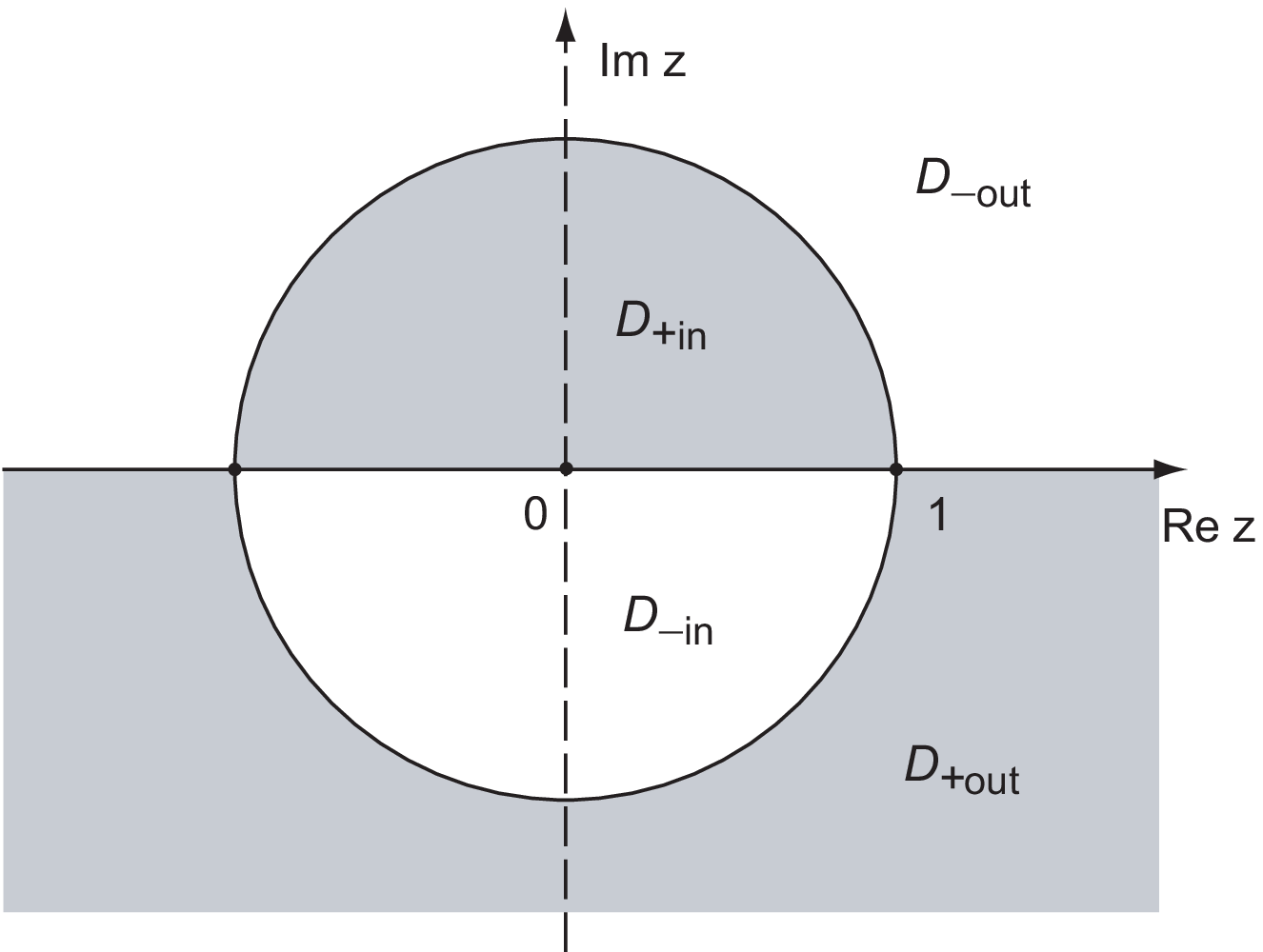}\qquad
\includegraphics[width=0.405\textwidth]{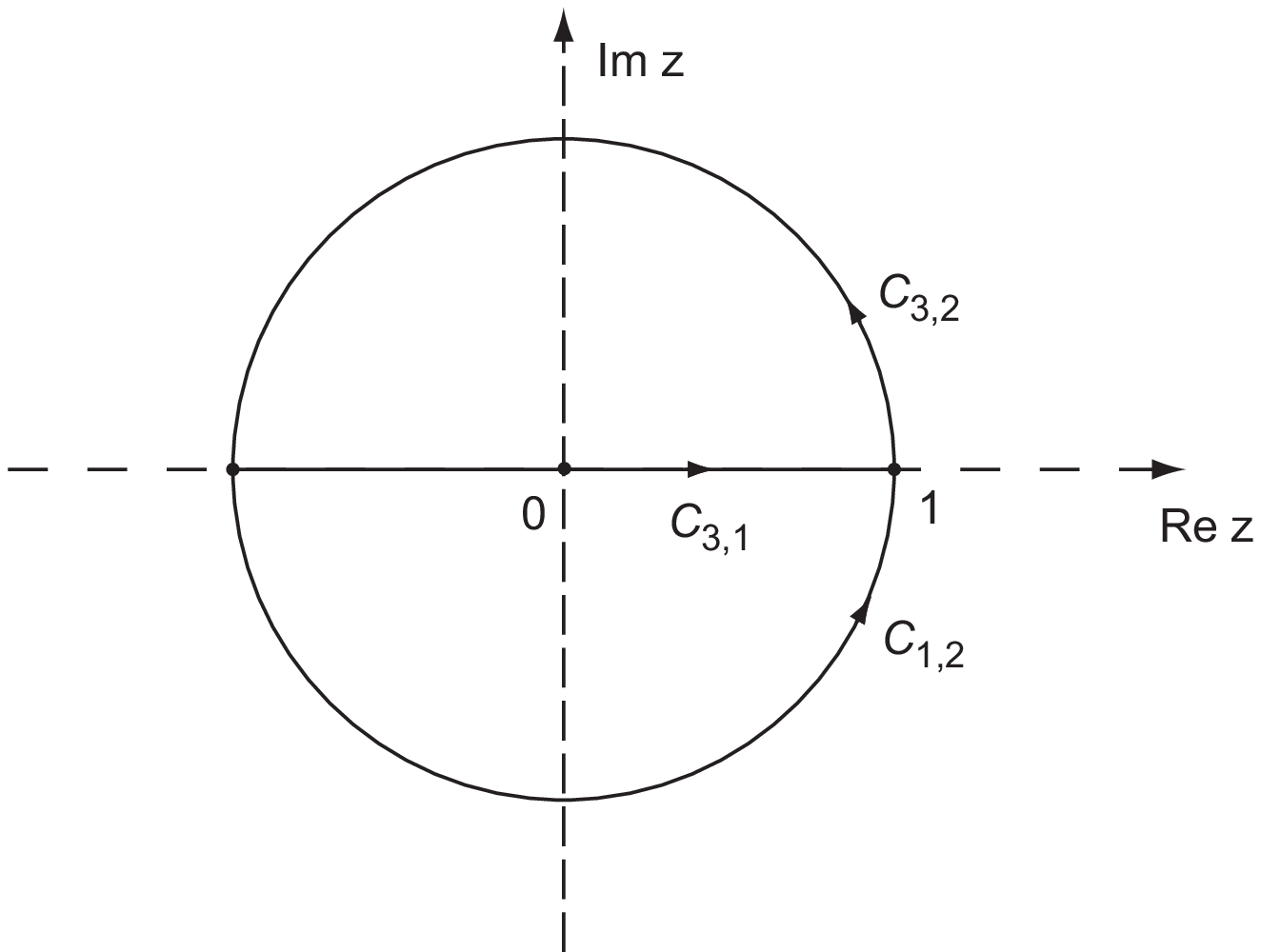}}
\caption{(Left) The regions $D_+$ (shaded) and $D_-$ (white) of the $z$-plane where $\Im[\omega(z)]\gl0$.
(Right) The contours $C_{1,2}$, $C_{2,3}$ and $C_{3,1}$
that define the Riemann-Hilbert problem in the linear case (see text for details).}
\label{f:DpmL}
\end{figure}

As in the continuum case \cite{JMP41p4188,CMP230p1,CPAM58p639},
to solve the IBVP
we consider \textit{simultaneous} solutions of both
the $x$-part and the $t$-part of the Lax pair.
To do this we again use $\psi_n(z,t)$, defined in~\eref{e:Psidef}.
Integrating~\eref{e:LaxpairL0},
we then define three eigenfunctions
uniquely determined in terms of their normalizations:
namely, $\phi_n\o{j}(z,t)$ for $j=1,2,3$,
so that $\phi_n\o{j}(z,t)=0$ respectively at $(n,t)=(0,0)$,
as $(n,t)\to(\infty,t)$ and at $(n,t)=(0,T)$
(cf.\ Fig.~\ref{f:IBVPzeropts}):
\bse
\label{e:PhiIBVP}
\bea
&\phi_n\o1(z,t)= \sum_{m=0}^{n-1}q_m(t)\,z^{n-m-1}
  + iz^n\int_0^t \e^{-i\omega(z)(t-t')}\big(q_0(t')-q_{-1}(t')/z\big)\,\d t',
\\
&\phi_n\o2(z,t)= - \sum_{m=n}^\infty q_m(t)\, z^{n-m-1},
\\
&\phi_n\o3(z,t)= \sum_{m=0}^{n-1}q_m(t)\,z^{n-m-1}
  - iz^n\int_t^T \e^{-i\omega(z)(t-t')}\big(q_0(t')-q_{-1}(t')/z\big)\,\d t'.
\eea
\ese
We introduce the domains $D_\pm=\{z\in\Complex:\Im \omega(z)\gl0\}$,
which will also be convenient to decompose as
$D_\pm=D_{\pm\#in}\cup D_{\pm\#out}$, where
$D_{\pm\#in}$ and $D_{\pm\#out}$ are respectively
the portions of $D_\pm$ inside and outside the unit disk
(cf.~Fig.~\ref{f:DpmL}),
namely
\bea
D_{+\#in}=\{z\in\Complex:|z|<1\,\wedge\,\Im\,z>0\}\,,
\qquad
D_{+\#out}=\{z\in\Complex:|z|>1\,\wedge\,\Im\,z<0\}\,,
\nonumber
\\
D_{-\#in}=\{z\in\Complex:|z|<1\,\wedge\,\Im\,z<0\}\,,
\qquad
D_{-\#out}=\{z\in\Complex:|z|>1\,\wedge\,\Im\,z>0\}\,.
\nonumber
\eea
We then note that:
\begin{itemize}
\item%[(i)]
$\phi_n\o2$ coincides with the eigenfunction in the IVP,
hence it is analytic for $|z|>1$ and continuous and bounded for $|z|\ge1$,
and $\phi_n\o2(z,t)\to0$ as $z\to\infty$;
\item%[(ii)]
$\phi_n\o1$ and $\phi_n\o3$
are analytic in the punctured complex $z$-plane $\Complex\punct$;
\item%[(iii)]
for all $t>0$ it is $\e^{i\omega(z)t}\to0$ as $z\to0,\infty$ in $D_+$
and $\e^{-i\omega(z)t}\to0$ as $z\to0,\infty$ in $D_-$;
as a result,
$\phi_n\o1$ and $\phi_n\o3$ are bounded %inside the unit disk
respectively for $z\in \=D_{-\#in}$ and $z\in \=D_{+\#in}$.
\end{itemize}
Note that~\eref{e:PhiIBVP} do not define
$\phi_0\o1(z,t)$ and $\phi_0\o3(z,t)$ at $z=0$.
In \ref{s:asymptotics}, however, we compute the asymptotics
of these eigenfunctions as $z\to0$,
and we show that
$\phi_0\o1(z,t)=O(1)$ as $z\to0$ with $\Im z\le0$ and
$\phi_0\o3(z,t)=O(1)$ as $z\to0$ with $\Im z\ge0$.

\paragraph{Jump conditions and Riemann-Hilbert problem.}

The difference between eigenfunctions at $|z|=1$
and $z\in[-1,1]$ yields a scalar RHP whose solution
will enable us to reconstruct the potential in terms of the scattering data.
As before, the difference between any eigenfunctions
solves the homogeneous version of~\eref{e:LaxpairL}.
Evaluating these differences at $(n,t)=(0,0)$
we then obtain the jumps as
(of course any two of the jumps uniquely determine
the third one):
\bse
\label{e:Phijumps}
\bea
\fl
\phi_n\o1(z,t) - \phi_n\o2(z,t)= z^{n-1}\e^{-i\omega(z)t}\,\^q(z,0)
&|z|=1~\wedge~\Im\,z\le0\,,
%\nonumber\\[-1ex]
\label{e:Phijumps12}
\\
\fl
\phi_n\o1(z,t) - \phi_n\o3(z,t)= z^{n-1}\e^{-i\omega(z)t}\,\^F(z,T)
&\!\!\Im\,z=0~\wedge~|z|\le1\,,
%\nonumber\\[-1ex]
\label{e:Phijumps13}
\\
\fl
\phi_n\o3(z,t) - \phi_n\o2(z,t)= z^{n-1}\e^{-i\omega(z)t}\,\big(\^q(z,0) - \^F(z,T)\big),\qquad
&|z|=1~\wedge~\Im\,z\ge0\,,
%\nonumber\\[-1ex]
\label{e:Phijumps23}
\eea
\ese
with $\^F(z,t)= i(z\^f_0(z,t)-\^f_{-1}(z,t))$,
and where $\^q(z,t)$ and $\^f_n(z,t)$ are respectively the
$z$-transforms of the initial and boundary data; namely:
\label{e:DLSztransforms}
\bea
\^q(z,t)= \sum_{m=0}^\infty q_m(t)/z^m\,,
\qquad
\^f_n(z,t)= \int_0^t \e^{i\omega(z)t'} q_n(t')\,\d t'\,.
\eea
Note that $\^q(z,t)$ is analytic for $|z|>1$ and continuous and bounded
for $|z|\ge1$,
while the $\^f_n(z,t)$ are analytic $\forall z\ne0$ and
continuous and bounded for $z\in \=D_+$.
Moreover, $\^q(z,t)\to q_0(t)$ as $z\to\infty$,
while $\^f_n(z,t)\to0$ as $z\to0,\infty$ in~$D_+$.
Finally, integration by parts shows that
\be
\^f_n(z,t)= iz\,\big(\,\e^{i\omega(z)t}q_n(t)-q_n(0)\,\big) +O(z^2)
\label{e:fnasymp@z=0}
\ee
as $z\to0$ in~$\partial D_+$ (i.e., along the real $z$-axis).
As shown in \ref{e:ztransfinverse},
\eref{e:DLSztransforms} are inverted by
\bea
q_n(t)= \frac1{2\pi i}\oint_{|z|=1}z^{n-1}\^q(z,t)\,\d z\,,
\qquad
q_n(t)= \frac1{2\pi} \int_{\partial D_{\!+\#out}}\!\!
  \omega'(z)\e^{-i\omega(z)t}\^f_n(z,T)\,\d z\,,
 \nonumber\\[-1ex]
\label{e:LSinvztransf}
\eea
%\ese
for all $0<t<T$, where $\omega'(z)=d\omega/dz$ and
$\partial D_{\!+\#out}$ is oriented so that $\Re z$ is decreasing.

Note that 
$\^F(z,T)/z$ remains bounded as $z\to0$
along the real $z$-axis [cf.\ \ref{s:asymptotics}].
%thanks to the boundedness of$\phi_n\o1(z,t)$ and~$\phi_n\o3(z,t)$ at $z=0$.
%
Thus, the RHS of~\eref{e:Phijumps13} with $n=0$ does not have a pole at $z=0$.
The solution of the RHP defined by~\eref{e:Phijumps} is therefore
simply obtained using standard Cauchy projectors over the
unit circle:
\bea
\phi_n(z,t)= \frac1{2\pi i}\, \oint_{|\zeta|=1}
  \zeta^{n-1}\e^{-i\omega(\zeta)t}\, \frac{\^q(\zeta,0)}{\zeta-z}\,\d\zeta
  - \frac1{2\pi i} \oint_{\partial D_{\!+\#in}}
       \zeta^{n-1}\e^{-i\omega(\zeta)t}\,
       \frac{\^F(\zeta,T)} %{\zeta\^f_0(\zeta,T)-\^f_{-1}(\zeta,T)}
         {\zeta-z}\,\d\zeta\,,
\nonumber\\[-1.4ex]
\label{e:IBVPRHPslnL}
\eea
where $|\zeta|=1$ is taken counterclockwise and
$\partial D_+$ is oriented so as to leave the domain to its left, as usual.
Inserting~\eref{e:IBVPRHPslnL} into the first of~\eref{e:LaxpairL}
then yields the reconstruction formula:
\bea
q_n(t)= \frac1{2\pi i}\!\oint_{|z|=1}
    z^{n-1}\e^{-i\omega(z)t}\,\^q(z,0)\,\d z
  - \frac1{2\pi i}\!\int_{\partial D_{\!+\#in}} z^{n-1}\e^{-i\omega(z)t}\,
     %i\big[z\^f_0(z,T)-\^f_{-1}(z,T)\big]
     \^F(z,T)\,\d z\,.
\label{e:IBVPsoln}
\eea
Of course
the right-hand side of~\eref{e:IBVPsoln} still depends on
the undetermined value $q_{-1}(t)$ via its transform $\^f_{-1}(z,T)$.
We next show how to eliminate this unknown using the global relation.

\paragraph{Global relation and symmetries.}

The global relation, which couples all initial and boundary values,
is obtained in a similar way as in the continuum problem
by integrating~\eref{e:LaxpairL0}
around the edges of the domain $\mathbb{N}_0\times[0,T]$,
namely for $(n,t)$ from $(0,0)$ to $(0,T)$, from there to $(\infty,T)$,
and then to $(\infty,0)$ and back to $(0,0)$:
%\label{e:GR2}
\bea
  i\int_0^t \e^{i\omega(z)t'}\big(q_0(t')-q_{-1}(t')/z\big)\,\d t'
  + \e^{i\omega(z)t}\sum_{m=0}^\infty q_m(t)/z^{m+1}
  = \sum_{m=0}^\infty q_m(0)/z^{m+1}\,.
\nonumber\\[-1.6ex]
\label{e:global}
\eea
Equation~\eref{e:global} holds where all of its terms are defined,
that is, for all $|z|\ge1$.
In terms of the $z$-transforms:
\bea
 i\big[z\^f_0(z,t) - \^f_{-1}(z,t)\big] + \e^{i\omega(z)t}\^q(z,t)
  = \^q(z,0)\,.
\label{e:g2}
\eea
Now note that $\omega(z)$ is invariant under the transformation $z\to1/z$,
and therefore so are the functions $\^f_n(z,t)$.
Moreover, $z\in D_{+\#out}$ implies $1/z\in D_{+\#in}$ and viceversa.
Hence, \eref{e:g2} with $z\to1/z$ gives, for all $0<|z|\le1$:
\bea
  i\big[(1/z)\^f_0(z,t) - \^f_{-1}(z,t)\big] + \e^{i\omega(z)t}\^q(1/z,t)
  = \^q(1/z,0)\,.
\label{e:g3}
\eea
We can then solve for $\^f_{-1}(z,t)$, obtaining, for all $0<|z|\le1$:
\be
\^f_{-1}(z,t) = \^f_0(z,t)/z
  - i\big(\, \e^{i\omega(z)t}\^q(1/z,t)- \^q(1/z,0)\,\big)\,.
\label{e:fm1LS}
\ee

\paragraph{Solution of the IBVP.}

%Equation~\eref{e:fm1LS} can now be inserted in~\eref{e:IBVPsoln}.
%to obtain the solution of the IBVP.
Of course the RHS of~\eref{e:fm1LS} contains
$\e^{i\omega(z)T}\^q(1/z,T)$, which is (apart from the changes
$t\to T$ and $z\to1/z$) the transform of the solution
we are trying to recover.
When this terms is inserted in~\eref{e:IBVPsoln}, however,
the resulting integrand is $z^{n-1}\e^{i\omega(z)(T-t)}\^q(1/z,t)$,
which is analytic and bounded in $D_{+\#in}$,
and whose integral over $\partial D_{+\#in}$ is therefore zero.
[This is analogous to what happens in the continuum limit;
cf.\ section~\ref{s:continuum}.]\,\
Importantly, the result also holds for $n=0$, since $\e^{i\omega(z)(T-t)}$
decays exponentially for all $t<T$ as $z\to0$ in $D_{+\#in}$.
We then have
\bea
\fl
q_n(t)= \frac1{2\pi i}\, \oint_{|z|=1}
    z^{n-1}\e^{-i\omega(z)t}\,\^q(z,0)\,\d z
  + \frac1{2\pi} \int_{\partial D_{+\#in}} z^{n-1}\e^{-i\omega(z)t}\,
     \big[i\^q(1/z,0)-(z-1/z)\^f_0(z,T)\big]\,\d z\,.
\nonumber\\[-2ex]
\label{e:LSIBVPsoln0}
\eea
Equation~\eref{e:LSIBVPsoln0} provides the solution of the IBVP in
Ehrenpreis form \cite{Ehrenpreis1970,Palamodov1970,Henkin1990},
%(in analogy with Ref.~\cite{JNLMP10p47} in the continuum limit)
since the only dependence of the RHS on~$n$ and~$t$ is
via the terms $z^n\e^{-i\omega(z)t}$, as in the IVP. Performing the
change of variable $z'=1/z$ we can write the second term in the RHS
of~\eref{e:LSIBVPsoln0} as an integral over $\partial D_{+\#out}$.
Then, since the resulting integrand,
$\e^{-i\omega(z)t}\^q(z,0)/z^{n+1}$ is analytic on $D_{-\#out}$,
for that portion we can deform the contour $\partial D_{+\#out}$ onto the
circle $|z|=1$ and combine the result with the first integral
in~\eref{e:LSIBVPsoln0}, obtaining the equivalent representation
\be
\fl
q_n(t)= \frac1{2\pi i}\, \oint_{|z|=1}
    \big(z^n-z^{-n}\big)/z\,\,\e^{-i\omega(z)t}\,\^q(z,0)\,\d z
  - \frac1{2\pi} \int_{\partial D_{+\#out}} (z-1/z)\,z^{-n-1}\e^{-i\omega(z)t}\,
     \^f_0(z,T)\,\d z\,,
\label{e:LSIBVPsoln}
\ee
where, as before, $\partial D_{+\#out}$ is oriented so that $\Re z$ is decreasing.

\paragraph{Continuum limit.}

The representation~\eref{e:LSIBVPsoln} is the discrete analogue of
the solution in the continuum limit.
To see this, one can reinstate the lattice spacing $h$ and follow
the same steps as above.
When expressed in terms of~$k$, the solution of the IBVP then becomes:
%\bse
\bea
q_n(t)=
  \frac2\pi\,\, \int_0^{\,\,\pi/h} \e^{-i\omega(k)t}\sin(nkh)\,\^q\o{s'}(k,0)\,\d k
  + \frac1\pi\,\,
 \int_0^{\,\,\pi/h} \e^{-i\omega(k)t}\sin(nkh) \^g(k,t)\,\d k\,,
\nonumber
\\[-1ex]
\label{e:dLSIBVsolnk}
\\
\noalign{\noindent  where $\omega(k)=2(1-\cos(kh))/h^2$, and with}
\^q\o{s'}(k,t)=h\sum_{n=1}^{\infty} \sin(nkh)q_n(t)\,,
\qquad
%\noalign{\noindent and}
\^g(k,t)= 2i\,{\sin(kh) \over h} \int_0^t
\e^{i\omega(k)t'}q_0(t')\,\d t'\,.
\nonumber
\eea
%\ese
It is then trivial to show that, in the limit $h\to 0$,
\eref{e:dLSIBVsolnk} yield the solution of the continuum problem,
namely~\eref{e:LSIBVPsinetransform}.

\paragraph{Remarks.}

Assuming existence, one can now verify that
the RHS of~\eref{e:LSIBVPsoln0} and~\eref{e:LSIBVPsoln}
indeed satisfies the DDE as well as the initial and BCs.
That the function defined by~\eref{e:LSIBVPsoln0}
solves the DLS equation is a trivial consequence
of the fact that it is in Ehrenpreis form.
When $t=0$ the term proportional to $z^{-n}$ in the first
integral of~\eref{e:LSIBVPsoln}
gives zero contribution, since the corresponding integrand
is analytic, bounded for $|z|>1$, and $O(1/z^{n+1})$ as $z\to\infty$.
Similarly, the second integral vanishes for the same reasons.
The only piece left coincides with the RHS of the first of~\eref{e:LSinvztransf}
at $t=0$, which therefore yields the initial datum~$q_n(0)$.
Finally, for $n=0$ the first integral in~\eref{e:LSIBVPsoln} is obviously zero,
while the second becomes just the inversion integral in~\eref{e:LSinvztransf}.
Hence its result is simply~$q_0(t)$.

Even though $\^f_0(z,T)$ depends on values of the BC $q_0(t)$
at all times~$t$~ from 0 to~$T$,
in practice~\eref{e:LSIBVPsoln} preserves causality, and the solution of
the IBVP at time~$t$ does not depend on future values of the BCs,
%(that is, values of $q_0(t')$ at times $t'>t$),
because one can replace $T$ with $t$ in~\eref{e:LSIBVPsoln}.
The reason is that the difference between the two terms is
\[
\frac1{2\pi}\int_{\partial D_{+\#out}}(z-1/z)\,z^{-n-1}
  \int_t^T \e^{-i\omega(z)(t-t')}q_0(t')\,\d t'\,\d z\,,
\]
%(where the change of variable $z\to1/z$ was performed).
and $\forall n\ne0$
the integrand is analytic and bounded in~$D_{+\#out}$, and vanishes
as $z\to\infty$ in~$D_+$.
Hence, the integral is zero~$\forall n>0$.
% and therefore one can replace $\^f_0(z,T)$ by $\^f_0(z,t)$ in~\eref{e:LSIBVPsoln}.

For all $n\ne0$, the second integrand in~\eref{e:LSIBVPsoln}
%[namely, $(z-1/z)\,z^{-n-1}\e^{-i\omega(z)t}\^f_0(z,t)$]
is analytic and bounded in $D_{-\#out}$.
Hence we can deform the integration contour from
$\partial D_{+\#out}$ to $|z|=1$, and substitute $z\to1/z$
in half of the integral.
The resulting expression for the solution coincides with
the solution of the IBVP via sine series,
namely~\eref{e:dLSIBVPsoln}.
%[Note that $\^q\o{s}(z,t)= \^q(z,t)-\^q(1/z,t)$.]\,\
We reiterate however that~\eref{e:LSIBVPsoln} also holds for $n=0$,
unlike~\eref{e:dLSIBVPsoln}.

Unlike sine/cosine transforms, the present method works equally well
for more general BCs, as we show below.
Also, unlike sine/cosine transforms, the present method can solve
IBVPs for arbitrary linear discrete evolution equations.
Finally, the method can be generalized
to solve IBVPs for integrable nonlinear DDEs,
as we show in section~\ref{s:IDNLS}.

\paragraph{Other boundary conditions.}

We now consider a IBVP for the DLS equation~\eref{e:DLS}
in which the BCs are a linear combination
of $q_0(t)$ and $q_{-1}(t)$ with constant coefficients,
namely, when
\be
q_{-1}(t) - \alpha q_0(t) = h(t)\,
\label{e:DLSRobinBC}
\ee
is given,
$\alpha\in\Complex$ is a nonzero but otherwise arbitrary constant,
and where in this case the labeling of the lattice should be such that
$n=-1$, not $n=0$, is the first lattice site.
Such BCs are the discrete analogue of Robin-type BCs in IBVPs for PDEs,
and cannot be solved using sine/cosine series.
The present method however works equally well;
the only difference from the previous case being that one needs to solve
the global relation for a different unknown.
Indeed, in \ref{s:Robin} we show that the solution of this IBVP is given by
\bea
\fl
q_n(t)= \frac1{2\pi i}\!\oint_{|z|=1}
    z^{n-1}\e^{-i\omega(z)t}\,\^q(z,0)\,\d z
  - \frac1{2\pi i}\!\int_{\partial D_{\!+\#in}}\!\!z^{n-1}\e^{-i\omega(z)t}\,
       \frac{\^G(z,T)}{1/z-\alpha}\,\d z
     -\nu_\alpha\alpha^{-n-1}\e^{-i\omega(\alpha)t}\^G(1/\alpha,t)\,,
\nonumber\\[-0.4ex]%\kern16em
\label{e:DLSRobinsoln}
\\
\noalign{\noindent where}
\^G(z,t) =
  i(z-1/z)\^h(z,t) + (z-\alpha)\^q(1/z,0)\,,
\label{e:DLSRobinGdef}
\eea
and where $\nu_\alpha=1$ if $\alpha\in
D_{\!+\#out}$, $\nu_\alpha=1/2$ if $\alpha\in\partial D_{\!+\#out}$
and $\nu_\alpha=0$ otherwise, and where the integral along $\partial
D_{\!+\#in}$ is to be taken in the principal value sense when
$\alpha\in\partial D_{\!+\#out}$. As before, one can easily verify
that the expression in~\eref{e:DLSRobinsoln} indeed
solves~\eref{e:DLS} and satisfies the initial condition and the
BC~\eref{e:DLSRobinBC}. Moreover, one can also verify that, in the
limit $\alpha\to\infty$ with $h(t)/\alpha= h'(t)$ finite, the
solution of the IBVP with ``Dirichlet-type'' BCs
[namely~\eref{e:LSIBVPsoln0}], is recovered.

%%%%%%%%%%%%%%%%%%%%%%%%%%%%%%%%%%%%%%%%%%%%%%%%%%%%%%%%%%%%%%%%%%%%%%%%%%%%%%
%%%%%%%%%%%%%%%%%%%%%%%%%%%%%%%%%%%%%%%%%%%%%%%%%%%%%%%%%%%%%%%%%%%%%%%%%%%%%%
\section{Integrable discrete nonlinear Schr\"odinger equation}
\label{s:IDNLS}

We now turn our attention to IVBPs for the IDNLS
equation~\eref{e:IDNLS}. As before, we first review the IVP, which
serves to introduce some of tools that will be used for the IBVP. We
require the same regularity conditions on the initial-boundary data
as in the linear case.

%%%%%%%%%%%%%%%%%%%%%%%%%%%%%%%%%%%%%%%%%%%%%%%%%%%%%%%%%%%%%%%%%%%%%%%%%%%%%%
\subsection{The Ablowitz-Ladik system on the integers}
\label{s:AL}

Consider the AL system~\eref{e:AL} with $n\in\Integer$ and
$t\in\Real^+$, and with $q_n(0)$ given. A Lax pair for~\eref{e:AL}
is given by~\eref{e:ALLP}, where now we take $\Phi_n(z,t)$ to be a
$2\times2$ matrix, $\_Q_n(t)$ and $\_H_n(z,t)$ are defined
in~\eref{e:QH}, and $\omega(z)\equiv\omega_\mathrm{idnls}(z)=
\omega_\mathrm{dls}(z^2)/2$,
where $\omega_\mathrm{dls}(z)$ was defined in~\eref{e:DLSdisprel}.
As in the linear case, we assume
$q_n(0)\in\ell^1(\Integer)$.
(As in the continuum limit, the IST with non-vanishing BCs at infinity
is significantly more involved, see Refs.~\cite{IP23p1711,IP8p889}.)

\paragraph{Jost solutions.}
As customary, we remove the $n$-dependence of the eigenfunctions
as $n\to\pm\infty$
by introducing a modified eigenfunction as
\be
\Phi_n(z,t)= \mu_n(z,t)\,\_Z^n\e^{-i\omega(z)t\sigma_3}\,.
\label{e:ALphimu}
\ee
(This definition differs from the usual one
by the factor $\e^{-i\omega(z)t\sigma_3}$, which has been added
for consistency with the the IBVP, discussed in section~\ref{s:ALIBVP}.
With this choice, the scattering matrix will be independent of time.)
%The ordering of $\_Z$ and $\sigma_3$ is of course inconsequential
%since $[Z,\sigma_3]=0$.)\,\
Then $\mu_n(z,t)$ satisfies
the following modified Lax pair:
%\bse
\bea \mu_{n+1}-\Zhat\mu_n= \_Q_n\mu_n\_Z^{-1}\,,
\qquad \.\mu_n +
i\omega(z)[\sigma_3,\mu_n] = \_H_n\mu_n\,, \label{e:ALLPm} \eea
%\ese
where $\Zhat\_A= \_Z\_A\_Z^{-1}$
%and $[\_A,\_B]=\_A\_B-\_B\_A$ is the matrix commutator.
It is also useful to use the integrating factor
$\esh{i\theta}(\_A)=\e^{i\theta\sigma_3}\_A\,\e^{-i\theta\sigma_3}$
(cf.~\ref{s:notations}). Then, the function \be \Psi_n(z,t)=
\Zhat^{-n}\esh{i\omega(z)t}\mu_n(z,t)\,, \label{e:PsiALdef} \ee
solves \bea \Psi_{n+1}-\Psi_n=
\_Z^{-1}\Zhat^{-n}\esh{i\omega(z)t}(\_Q_n)\Psi_n\,. \qquad \.\Psi_n=
\Zhat^{-n}\esh{i\omega(z)t}(\_H_n)\Psi_n\,.
\label{e:ALmodifiedLPPsi} \eea One can now easily
``integrate''~\eref{e:ALmodifiedLPPsi} and thereby obtain the
solutions of~\eref{e:ALLPm} which reduce to the identity matrix as
$n\to\mp\infty$:  \be  \fl \label{e:ALmusolns} \mu_n\o1(z,t)= \_I +
\_Z^{-1}\sum_{m=-\infty}^{n-1}\Zhat^{n-m}(\_Q_m\mu_m\o1)\,,\quad
\mu_n\o2(z,t)= \_I -
\_Z^{-1}\sum_{m=n}^\infty\Zhat^{n-m}(\_Q_m\mu_m\o2)\,. \ee Of
course, unlike the linear case the eigenfunctions are now defined in
terms of summation equations (the discrete analogue of integral
equations).

As in the linear problem, %if $\_Q_n(0)\in\ell^1(\Integer)$,
\eref{e:ALmusolns} imply certain analyticity properties
for the eigenfunctions.
%for the columns of $\mu_n\o1(z,t)$ and $\mu_n\o2(z,t)$.
More precisely, let $\mu_n\o{j}(z,t)=(\mu_n\o{j,L},\mu_n\o{j,R})$,
$j=1,2$,
where the column vectors $\mu_n\o{j,L}(z,t)$ and $\mu_n\o{j,R}(z,t)$
denote respectively the first and second column of $\mu_n\o{j}(z,t)$.
These columns are analytic in the following regions~\cite{APT2003}:
\[
\mu_n\o{1,L},~\mu_n\o{2,R}:\quad |z|>1\,,\qquad
\mu_n\o{1,R},~\mu_n\o{2,L}:\quad |z|<1\,,
\]
Moreover, these columns are continuous and bounded on the closure of
these domains. These properties immediately yield those of
$\Phi_n\o{j}(z,t)=\mu_n\o{j}(z,t)\,\_Z^n\e^{-i\omega(z)t\sigma_3}$
for $j=1,2$:\break $\Phi_n\o{1,L}(z,t)$ and $\Phi_n\o{2,R}(z,t)$ are
analytic for $|z|>1$, and $\Phi_n\o{1,R}(z,t)$ and
$\Phi_n\o{2,L}(z,t)$ for $|z|<1$.

\paragraph{Scattering matrix.}
Equation~\eref{e:ALLP1} implies\,
$\det\,\Phi_{n+1}=(1-q_np_n)\det\,\Phi_n$.
Therefore
\be
\det\,\Phi_n\o1= \prod_{m=-\infty}^{n-1}(1-q_m p_m)\,,\quad
\det\,\Phi_n\o2= \prod_{m=n}^\infty(1-q_m p_m)^{-1}=:1/C_n\,.
\label{e:ALphidet}
\ee
(Note $\det\Phi_n=\det\mu_n$.)
Equations~\eref{e:ALphidet} mark a significant difference of the discrete case
from the continuum case,
where such determinants are independent of both the potential and the
independent variable (cf.\ section~\ref{s:continuum}).

For the focusing IDNLS [namely, \eref{e:AL} with $p_n=\nu q_n^*$ and $\nu=-1$],
$1-q_np_n=1+|q_n|^2$, and therefore
$\det\mu_n\o{j}\ne0$ $\forall n\in\Integer$ for $j=1,2$.
For the defocusing case ($\nu=1$), however, it is necessary to assume that
$|q_n|\ne1$ $\forall n\in\Integer$ in order that $\det\mu_n\o{j}$ to be
guaranteed to be nonzero.
Hereafter we will assume that $q_np_n\ne1~\forall n\in\Integer$.
Moreover, we will require that the product
\[
C_{-\infty}= \det\,\Phi_\infty\o1= 1/\det\,\Phi_{-\infty}\o2=
\prod_{n=-\infty}^\infty(1-q_np_n)
\]
be finite, which will simplify the study of the scattering
coefficients. Under these hypotheses, the matrices $\Phi_n\o1$ and
$\Phi_n\o2$ are both fundamental solutions of the scattering
problem~\eref{e:ALLP1}. Hence they must be proportional to each
other: $\Phi_n\o1(z,t)= \Phi_n\o2(z,t)\_A(z)$ on $|z|=1$, where
$\_A(z)=\big(a_{jj'}(z)\big)$ is the $2\times2$ scattering matrix.
In terms of the modified eigenfunctions:
\be \mu_n\o1(z,t)=
\mu_n\o2(z,t)\Zhat^n\,\esh{-i\omega(z)t}\_A(z)\,.
\label{e:ALmuscat0} \ee Or, in component form, \bse
\label{e:ALmuscat} \bea \mu_n\o{1,L}(z,t)=
  a_{11}(z)\,\mu_n\o{2,L}(z,t)+z^{-2n}\e^{2i\omega(z)t}a_{21}(z)\,\mu_n\o{2,R}(z,t)\,,
\\
\mu_n\o{1,R}(z,t)=
  z^{2n}\e^{-2i\omega(z)t}a_{12}(z)\,\mu_n\o{2,L}(z,t)+a_{22}(z)\,\mu_n\o{2,R}(z,t)\,.
\eea
\ese
The above relations imply
$\_A(z)=\lim_{n\to\infty}\Zhat^{-n}\esh{i\omega(z)t}\mu_n\o1(z,t)=
\lim_{n\to\infty}\Psi_n\o1(z,t)$, that is,
\be
\_A(z)= \_I + \_Z^{-1}\sum_{n=-\infty}^\infty
  \Zhat^{-n}\esh{-i\omega(z)t}\big(\_Q_n(t)\mu_n\o1(z,t)\big)\,.
\label{e:SmatrixAL}
\ee
%In particular,
%\bea
%a_{11}(z)= 1 + z^{-1}\!\!\sum_{n=-\infty}^\infty
%q_n(t)\mu_{n,21}\o1(z,t)\,,
%\qquad
%a_{22}(z)= 1 +
%z\sum_{n=-\infty}^\infty p_n(t)\mu_{n,12}\o1(z,t)\,.
%\nonumber\\[-1ex]
%\label{e:ALa11a22}
%\eea
The scattering matrix $\_A(z)$ is independent of time,
since $\_A(z)=\lim_{n\to\infty}\Psi_n\o1(z,t)$,
and
$\lim_{n\to\infty}\.\Psi_n(z,t)=0$.
Equation~\eref{e:ALmuscat0} also implies\,
$\det\,\_A(z)= \det\,\Phi_\infty\o1(z,t)=C_{-\infty}$, %= \prod_{n=-\infty}^\infty(1-q_np_n)$,
as well as
\bea
\_A(z)= C_n\begin{pmatrix}
  \Wr\big(\Phi_n\o{1,L},\Phi_n\o{2,R}\big)
  &\Wr\big(\Phi_n\o{1,R},\Phi_n\o{2,R}\big)
\\
  - \Wr\big(\Phi_n\o{1,L},\Phi_n\o{2,L}\big)
  &- \Wr\big(\Phi_n\o{1,R},\Phi_n\o{2,L}\big)
\end{pmatrix}\,.
\label{e:ALWronskian}
\eea
The analyticity of the eigenfunctions then implies that
$a_{11}(z)$ and $a_{22}(z)$ can be analytically continued off the unit circle,
respectively into the domains $|z|>1$ and $|z|<1$,
but $a_{12}(z)$ and $a_{21}(z)$ cannot.
It is also useful to introduce the reflection coefficients
\be
\rho_1(z)= {a_{21}(z)}/{a_{11}(z)}\,,\qquad
\rho_2(z)= {a_{12}(z)}/{a_{22}(z)}\,.
\label{e:ALIVPreflection}
\ee

\paragraph{Symmetries.}
When $p_n(t)=\nu q_n^*(t)$,
the scattering problem~\eref{e:ALLP1} admits an important involution,
which can be conveniently written introducing the matrix~$\sigma_\nu$
defined in~\eref{e:sigmanudef}.
Indeed, when $p_n=\nu q_n^*$, if $\Phi_n(z,t)$ is a solution of~\eref{e:ALLP1},
so is the matrix
\be
\Phi_n'(z,t)= \sigma_\nu \Phi_n^*(1/z^*,t)\,,
\label{e:PhiALsymm}
\ee
Then, comparing the asymptotic behavior of the first and second columns of
the Jost eigenfunctions as $n\to\pm\infty$
one obtains, for $j=1,2$,
\bea
\Phi_n\o{j,L}(z,t)=\sigma_\nu\big(\Phi_n\o{j,R}(1/z^*,t)\big)^*,\quad
\Phi_n\o{j,R}(z,t)=\nu\sigma_\nu\big(\Phi_n\o{j,L}(1/z^*,t)\big)^*.
\label{e:PhiALsymmcol}
\eea
The above relations imply the following symmetries for the elements of the
scattering matrix: %~$\_A(z,t)$:
\be
a_{22}(z)= a_{11}^*(1/z^*)\,,\qquad
a_{21}(z)= \nu\,a_{12}^*(1/z^*)\,.
\label{e:ALscattsymm}
\ee
In turn, these imply $\rho_2(z)=\nu \rho_1^*(1/z^*)$.

\paragraph{Discrete spectrum.}

The proper eigenvalues of the scattering problem~\eref{e:ALLP1}
are the values $z=z_j$ with $|z_j|<1$ and $z=\=z_j$ with $|\=z_j|>1$
for which there exist eigenfunctions bounded $\forall n\in\Integer$.
From the asymptotic behavior of the Jost solutions one can see that
such eigenvalues occur whenever the appropriate left- and right-sided
Jost solutions are proportional, namely
$\Phi_n\o{1,L}(\=z_j,t)= \=b_j\o{o}\Phi_n\o{2,R}(\=z_j,t)$ and
$\Phi_n\o{1,R}(z_j,t)= b_j\o{o}\Phi_n\o{2,L}(z_j,t)$,
or equivalently:
\bea
\fl
\mu_n\o{1,L}(\=z_j,t)= \=b_j\o{o} \=z_j^{-2n}\e^{2i\omega(\=z_j)t}\mu_n\o{2,R}(\=z_j,t)\,,
\quad
\mu_n\o{1,R}(z_j,t)= b_j\o{o} z_j^{2n}\e^{-2i\omega(z_j)t}\mu_n\o{2,L}(z_j,t)\,.
%\nonumber\\[-1ex]
\label{e:ALIVPzeros}
\eea
The Wronskian representations~\eref{e:ALWronskian} then imply
that such eigenvalues are the zeros of the scattering coefficients:
$a_{11}(\=z_j)=0$ and $a_{22}(z_j)=0$, respectively.
(As in Ref.~\cite{APT2003} we assume that $a_{jj}(z)\ne0$ for all $|z|=1$.)
Since no accumulation points of such zeros can exist
(because of the sectional analyticity of the scattering coefficients),
it follows that there is a finite number of them.
As in Ref.~\cite{APT2003} we assume all of these zeros are simple.
(The case of multiple zeros can be studied as the coalescence of simple zeros,
by analogy with the continuum case~\cite{JETP34p62}.)
Since $a_{jj}(z)$ are even functions \cite{APT2003},
$z=z_j$ is a zero of $a_{22}(z)$ iff $z=-z_j$ is,
and similarly for $a_{11}(z)$.
Moreover, the symmetries~\eref{e:ALscattsymm} imply that
$z=z_j$ is a zero of $a_{22}(z)$ iff $\=z_j=1/z_j^*$ is a zero of $a_{11}(z)$.
Thus, discrete eigenvalues appear in quartets.

The inverse problem will involve the modified eigenfunctions
$\mu_n\o{1,L}(z,t)/a_{11}(z)$ and $\mu_n\o{1,R}(z,t)/a_{22}(z)$.
Equations~\eref{e:ALIVPzeros} imply \bea \fl
\Res_{z=\=z_j}\bigg[\frac{\mu_n\o{1,L}(z,t)}{a_{11}(z)}\bigg]
  = \=b_j\=z_j^{-2n}\e^{2i\omega(\=z_j)t}\mu_n\o{2,R}(\=z_j,t)\,,
\quad
\Res_{z=z_j}\bigg[\frac{\mu_n\o{1,R}(z,t)}{a_{22}(z)}\bigg]
  = b_jz_j^{2n}\e^{-2i\omega(z_j)t}\mu_n\o{2,L}(z_j,t)\,,
\nonumber\\[-1ex]
\eea
where $b_j= b_j\o{o}/a'_{22}(z_j)$ and $\=b_j= \=b_j\o{o}/a'_{11}(\=z_j)$
are referred to as the norming constants.
The symmetries of the scattering problem imply
$\=b_j= -\nu(b_j/z_j^2)^*$.

\newpage
\paragraph{Asymptotics.}
The asymptotic behavior of the eigenfunctions as $z\to0$ or $z\to\infty$
can be obtained from~\eref{e:ALmusolns}.
For example, for $\mu_n\o1(z,t)$ it is
\bea
\mu_n\o1(z,t)= \_I + \_Q_{n-1}\,\_Z^{-1}
  + O(\_Z^{-2})\,
\qquad\rm{as}~z\to(\infty,0)\,,
\label{e:ALasymp}
\eea
where $z\to(z_L,z_R)$ indicates $z\to z_L$ in the first
column and $z\to z_R$ in the second one, and the asymptotics
corresponding to $O(\_Z^m)$ is defined in~\ref{s:notations}.
Equation~\eref{e:ALasymp} will allow us
to reconstruct the potentials from the asymptotic behavior of
$\mu_n\o1$:
\[
\_Q_n(t)= \lim_{z\to(\infty,0)} (\,\mu_{n+1}\o1(z,t)-\_I\,)\,\_Z\,.
\]
%\ese
The asymptotic behavior of $\mu_n\o2(z,t)$ is obtained
in a slightly different way as that of $\mu_n\o1(z,t)$,
and the result is also different.
More precisely, in~\ref{s:asymptotics} we show that
%\bse
\bea
C_n\mu_n\o2(z,t)= \_I - \_Q_n\,\_Z +O(\_Z^2)
\qquad\mathrm{as}~z\to(0,\infty)\,.
\label{e:ALasympmu2}
\eea
%\ese
Also, inserting~\eref{e:ALasymp} into the diagonal elements
of~\eref{e:SmatrixAL}
%~\eref{e:ALa11a22},
one obtains the asymptotic behavior of the analytic scattering coefficients:
%\bse
\bea
a_{11}(z)=1+ \frac1{z^2}\sum_{n=-\infty}^\infty q_n(t)p_n(t)+O(1/z^4)\,,
\qquad\mathrm{as}~z\to\infty\,.
\nonumber
%\\
%a_{22}(z)=1+ z^2\sum_{n=-\infty}^\infty q_n(t)p_n(t)+O(z^4)\,,
%\qquad\mathrm{as}~z\to0\,.
%\nonumber
\eea
which by symmetry also determines the behavior of $a_{22}(z)$ as $z\to0$.

\paragraph{Inverse problem.}
The inverse problem is the RHP defined by \eref{e:ALmuscat}
for $|z|=1$:
\bse
\label{e:ALmu2scat}
\bea
\frac{\mu_n\o{1,L}(z,t)}{a_{11}(z)} - \mu_n\o{2,L}(z,t)
  = z^{-2n}\e^{2i\omega(z)t}\rho_1(z,t)\mu_n\o{2,R}(z,t)\,,
\\
\frac{\mu_n\o{1,R}(z,t)}{a_{22}(z)} - \mu_n\o{2,R}(z,t)
  = z^{2n}\e^{-2i\omega(z)t}\rho_2(z,t)\mu_n\o{2,L}(z,t)\,,
\eea
\ese
where $\rho_1(z)$ and $\rho_2(z)$ as in~\eref{e:ALIVPreflection}.
Unlike the continuum case,
the asymptotics of $\mu_n\o{2,L}(z,t)$
as $z\to\infty$ depends on the values of the potentials
$q_m(t)$ and $p_m(t)$ for all $m\ge n$ through~$C_n$ [cf.~\eref{e:ALphidet}]\,.
This problem can be circumvented by introducing
the following renormalizations:
%\bse
\bea
\_M_n^-(z,t)= \begin{pmatrix}1&0\\0&C_n\end{pmatrix}
  \bigg(\frac{\mu_n\o{1,L}(z,t)}{a_{11}(z)}\,,\,\mu_n\o{2,R}(z,t)\bigg)\,,
\nonumber
\\
\_M_n^+(z,t)= \begin{pmatrix}1&0\\0&C_n\end{pmatrix}
  \bigg(\mu_n\o{2,L}(z,t),\,\frac{\mu_n\o{1,R}(z,t)}{a_{22}(z)}\bigg)\,.
\nonumber
\eea
The matrices $\_M_n^\pm(z,t)$ are sectionally meromorphic for
$|z|<1$ and $|z|>1$, respectively.
Moreover, \eref{e:ALmu2scat} yields
the following jump condition for the matrices~$\_M_n^\pm(z,t)$ on $|z|=1$:
\bea
\_M_n^-(z,t)=\_M_n^+(z,t)\big(\_I-\_J_n(z,t)\big)\,,
\label{e:ALRHP}
\\
\noalign{\noindent where the jump matrix $\_J_n(z,t)$ is}
\_J_n(z,t)= \begin{pmatrix}\rho_1(z)\rho_2(z) &z^{2n}\e^{-2i\omega(z)t}\rho_2(z)\\
  -z^{-2n}\e^{2i\omega(z)t}\rho_1(z) &0\end{pmatrix}.
\nonumber
\eea
Moreover, $\_M_n^\pm(z,t)$ have the following asymptotic behavior:
\bse
\label{e:MasympAL}
\bea
\_M_n^-(z,t)= \_I %\begin{pmatrix}1 &0\\ 0 &1\end{pmatrix}
  + \frac1z\begin{pmatrix}0 &-q_n/C_n\\ p_{n-1}C_n &0\end{pmatrix}
  + O(1/z^2)
\qquad\mathrm{as}~z\to\infty\,,
\label{e:M-asympAL}
\\
\_M_n^+(z,t)= \begin{pmatrix}1/C_n &0\\ 0 &C_n\end{pmatrix}
  + z \begin{pmatrix}0 &q_{n-1}\\ -p_n &0\end{pmatrix}
  + O(z^2)
\qquad\mathrm{as}~z\to0\,.
\label{e:M+asympAL}
\eea
\ese
In the absence of a discrete spectrum [that is, if $a_{11}(z,t)\ne 0$
for $|z|>1$ and $a_{22}\ne 0$ for $|z|<1$] the matrix
functions $\_M_n^\pm(x,t,k)$ are analytic in their
respective domains.

In particular, \eref{e:M-asympAL} allows the RHP~\eref{e:ALRHP}
to be solved via the Cauchy projectors $P^\pm$ over the
unit circle, as in the linear case. Of course, unlike the linear case
the solution is now expressed in terms of a matrix integral equation:
\be
\_M_n^+(z,t)= \_I +
  \frac1{2\pi i}\int_{|\zeta|=1}\_M_n^+(\zeta,t)\frac{\_J_n(\zeta,t)}{\zeta-z}\,\d\zeta\,.
\label{e:ALRHPsoln}
\ee
The asymptotic behavior of $\_M_n^+(z,t)$ as $z\to0$ is easily obtained
from~\eref{e:ALRHPsoln}:
\bea
\_M_n^+(z,t)= \_I
  + \frac1{2\pi i}\!\int_{|\zeta|=1}\!\_M_n^+(\zeta,t)\_J_n(\zeta,t)\frac{\d\zeta}\zeta
  + \frac z{2\pi i}\!\int_{|\zeta|=1}\!\_M_n^+(\zeta,t)\_J_n(\zeta,t)\frac{\d\zeta}{\zeta^2}
  + O(z^2).
%\qquad\mathrm{as}~z\to0\,.
\nonumber\\[-2ex]
\label{e:ALasympRHP}
\eea
Comparing the limit as $z\to0$ of~\eref{e:ALasympRHP} with~\eref{e:M+asympAL},
we see that the off-diagonal portion of the first integral in~\eref{e:ALasympRHP}
is zero, a fact which is not entirely obvious otherwise.
(This integral is missing in the corresponding formula in Ref.~\cite{APT2003}.)
Then, comparing the $(1,2)$ components of~\eref{e:ALasympRHP}
and~\eref{e:M+asympAL} we obtain the reconstruction formula
for the solution of the IVP:
\[
q_n(t)= \frac1{2\pi i}\int_{|z|=1}z^{2n}\e^{-2i\omega(z)t}
  \rho_2(z)\big(\mu_{n+1}\o2(z,t)\big)_{11}\,\d z\,.
\]

\paragraph{Linear limit.}
As in the continuum limit, the IST is the nonlinear analogue of
the linear transform pair.
Namely, if $\_Q_n= O(\epsilon)$, then $\mu_n\o1=\_I + O(\epsilon)$ and
\bea
\_A(z)= \_I + \_Z^{-1}\sum_{n=-\infty}^\infty\Zhat^{-n}\esh{-i\omega(z)t}\_Q_n(t)
 + O(\epsilon^2)\,.
\nonumber
\\
\noalign{\noindent Thus}
\rho_2(\zeta)= \sum_{n=-\infty}^\infty \zeta^{-2n-1}\e^{-2i\omega(\zeta)t}q_n(t)+O(\epsilon^2)
  = \frac1\zeta \^q(\zeta^2,0) + O(\epsilon^2)\,,
\nonumber
\eea
where $\^q(z,t)$ is the linear $z$-transform defined in~\eref{e:Fourierpair}.
Similarly,
\bea
\fl
q_n(t)= \frac1{2\pi i}\int_{|\zeta|=1}\zeta^{2n}\e^{-2i\omega(\zeta)t}\rho_2(\zeta)\,\d\zeta
  + O(\epsilon^2)
%\nonumber\\[-2ex]\kern6em{ }
  = \frac1{2\pi i}\int_{|z|=1}z^{n-1}\e^{-i\omega_\mathrm{dls}(z)t}\^q(z,0)\,\d z + O(\epsilon^2)\,,
\label{e:ALlinearlimit}
\eea
where the change of variable $\zeta^2= z$ was performed
in the RHS of~\eref{e:ALlinearlimit},
and where $\omega_\mathrm{idnls}(\zeta)=
\half\omega_\mathrm{dls}(\zeta^2)$,
as discussed in section~\ref{s:linearLaxpair}.

%
%%%%%%%%%%%%%%%%%%%%%%%%%%%%%%%%%%%%%%%%%%%%%%%%%%%%%%%%%%%%%%%%%%%%%%%%%%%%%%
\subsection{The Ablowitz-Ladik system on the naturals}
\label{s:ALIBVP}

We now consider the IBVP for the IDNLS.
That is, we solve~\eref{e:IDNLS} with $n\in\Natural$$, t\in\Real^+$
and with $q_n(0)$ and $q_0(t)$ given.
The approach we will follow is a combination of the method for the IVP
for the IDNLS
on the integers and that for the IBVP for the DLS on the naturals.

\begin{figure}[t!]
\rightline{\includegraphics[width=0.405\textwidth]{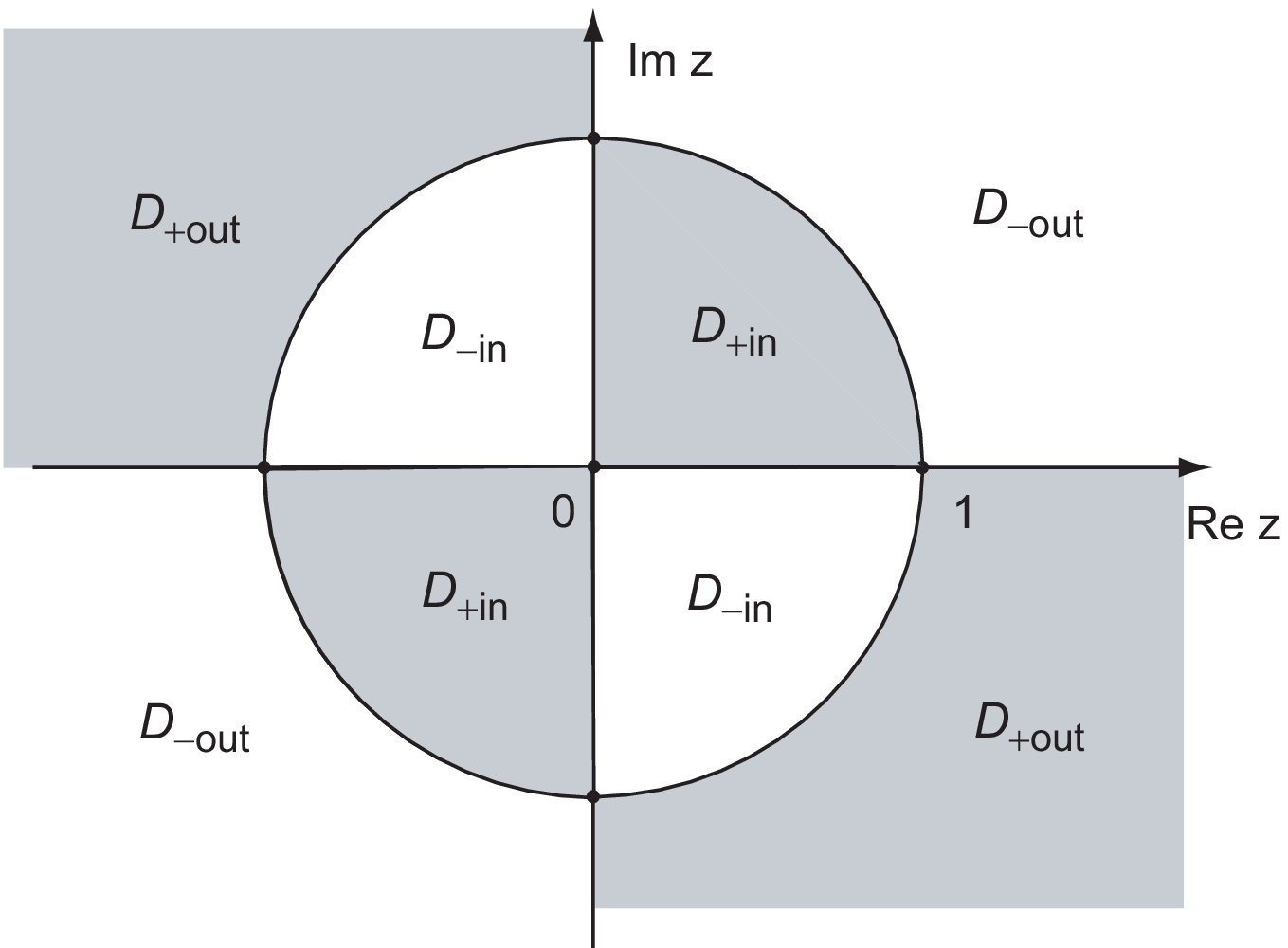}\quad
\includegraphics[width=0.405\textwidth]{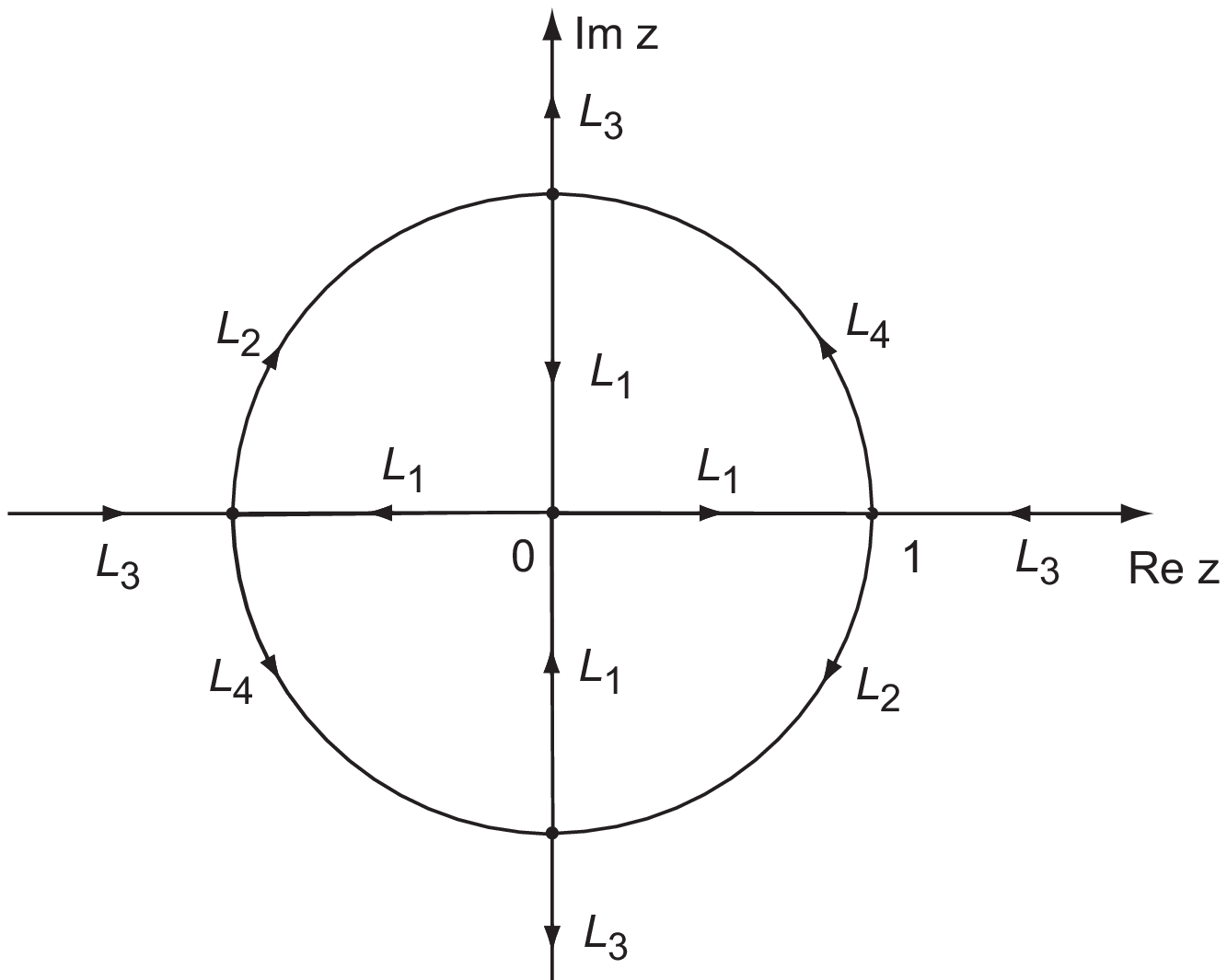}}
\caption{(Left) The regions $D_+$ (shaded) and $D_-$ (white) of the
$z$-plane where $\Im[\omega(z)]\protect\gl 0$ in the nonlinear case,
with $D_\pm= D_{\pm\#in}\cup D_{\pm\#out}$\,.
(Right) The contours $L_1,\dots,L_4$ that define the Riemann-Hilbert problem
(see text for details)}
\label{f:DpmAL}
\end{figure}

\paragraph{Eigenfunctions and analyticity.}
Making use of the modified eigenfunction
$\Psi_n(z,t)$ in~\eref{e:PsiALdef},
we define three eigenfunctions $\mu_n\o{j}(z,t)$ which reduce to the
identity matrix respectively when $(n,t)=(0,0)$,
as $(n,t)\to(\infty,t)$ and at $(n,t)=(0,T)$:
\bse
\label{e:ALmuIBVPsolns}
\bea
\fl
\mu_n\o1(z,t)= \_I
  + \_Z^{-1}\sum_{m=0}^{n-1}\Zhat^{n-m}(\_Q_m(t)\mu_m\o1(z,t))
  + \Zhat^n\int_0^t
      \esh{-i\omega(z)(t-t')}\big(\_H_0(z,t')\mu_0\o1(z,t')\big)\,\d t'\,,
\label{e:ALmu1IBVPsolns}
\\
\fl
\mu_n\o2(z,t)= \_I -
\_Z^{-1}\sum_{m=n}^\infty\Zhat^{n-m}(\_Q_m(t)\mu_m\o2(z,t))\,,
\\
\fl
\mu_n\o3(z,t)= \_I
  + \_Z^{-1}\sum_{m=0}^{n-1}\Zhat^{n-m}(\_Q_m(t)\mu_m\o3(z,t))
  - \Zhat^n\int_t^T
      \esh{-i\omega(z)(t-t')}\big(\_H_0(z,t')\mu_0\o3(z,t')\big)\,\d t'\,.
\eea \ese Note that $\mu_n\o2(z,t)$ coincides with the eigenfunction
in the IVP, defined in~\eref{e:ALmusolns}.
% and hence it enjoys the same domain of analyticity,
As in the linear case,
we partition the complex $z$-plane into the domains $D_\pm$ defined as
$D_\pm=\{z\in\Complex:\Im\omega(z)\gl0\}$.
%which are the image of the four quadrants of the
%complex $k$-plane under the map $z= \e^{ik}$.
We then write $D_\pm= D_{\pm\#in}\cup D_{\pm\#out}$
where the subscripts ``in'' and ``out'' denote the portions of $D_\pm$
inside and outside the unit disk, respectively.
That is (cf.~Fig~\ref{f:DpmAL}),

\vglue-1.4\medskipamount
\bse
\bea
D_{+\#in}= \{z\in\Complex: |z|<1\,\wedge\,\arg z\in(0,\pi/2)\cup(\pi,3\pi/2)\}\,,
\nonumber
\\
D_{-\#in}= \{z\in\Complex: |z|<1\,\wedge\,\arg z\in(\pi/2,\pi)\cup(3\pi/2,2\pi)\}\,,
\nonumber
\\
D_{+\#out}= \{z\in\Complex: |z|>1\,\wedge\,\arg z\in(\pi/2,\pi)\cup(3\pi/2,2\pi)\}\,,
\nonumber
\\
D_{-\#out}= \{z\in\Complex: |z|>1\,\wedge\,\arg z\in(0,\pi/2)\cup(\pi,3\pi/2)\}\,.
\nonumber
\eea
\ese

\smallskip\noindent
Then, in a similar way as in the IVP on the whole line and the
IBVP in the linear problem,
we can obtain the regions of analyticity and boundedness
of the eigenfunctions.
More precisely, writing again $\mu_n\o{j}(z,t)=(\mu_n\o{j,L},\mu_n\o{j,R})$,
%with the superscripts $L$ and $R$ indicating the left and second columns,
we have:
\begin{itemize}
\item
$\mu_n\o1(z,t)$ and $\mu_n\o3(z,t)$ are analytic in the punctured
complex $z$-plane~$\Complex\punct$;
\item
$\mu_n\o{1,L}(z,t)$ is continuous and bounded in $\=D_{+\#out}$;
\item
the restriction of $\mu_n\o{1,R}(z,t)$ to $D_{-\#in}$ is continuous and
bounded in $\=D_{-\#in}$;
\item
$\mu_n\o{3,L}(z,t)$ is continuous and bounded in $\=D_{-\#out}$;
\item
the restriction of $\mu_n\o{3,R}(z,t)$ to $D_{+\#in}$ is continuous and
bounded in $\=D_{+\#in}$;
\item
$\mu_n\o{2,L}(z,t)$ is analytic for $|z|<1$
and continuous and bounded for $|z|\le1$;
\item
$\mu_n\o{2,R}(z,t)$ is analytic for $|z|>1$
and continuous and bounded for $|z|\ge1$.
\end{itemize}
The analyticity of the eigenfunctions is formally proven via
Neumann series as in the IVP~\cite{APT2003} and as in the
IVP for the continuum case~\cite{NLTY18p1771}.
However, showing the continuity of $\mu_n\o{1,R}(z,t)$ and
$\mu_n\o{3,R}(z,t)$ at $z=0$ is nontrivial,
and it requires studying the asymptotic behavior of the eigenfunctions
as $z\to 0$ (see~\ref{s:asymptotics}).

\paragraph{Scattering matrices.}
The relation $\det\,\Phi_{n+1}= (1-q_np_n)\,\det\,\Phi_n$ still holds.
Therefore $\det\,\Phi_n\o1$ and $\det\,\Phi_n\o2$
are still given by~\eref{e:ALphidet}, and
$\det\,\Phi_n\o1= \det\,\Phi_n\o3$.
[Note that $\mu_t=\_L\mu+\mu\_R$ implies
$(\det\mu)_t=\mathop{\rm tr}(\_L+\_R)\,\det\mu$,
and in our case both $\_L$ and $\_R$ are traceless;
cf.~\eref{e:ALLP} and~\ref{s:notations}.]\,\
%\[
%\fl
%\det\,\Phi_n\o1= \det\,\Phi_n\o3= \prod_{m=0}^{n-1}(1-q_mp_m)\,,\qquad
%\det\,\Phi_n\o2= \prod_{m=n}^\infty(1-q_mp_m)^{-1}=:1/C_n\,.
%\label{e:ALIBVPphidet}
%\]
Hence, under the same regularity hypotheses as before,
$\Phi_n\o1$, $\Phi_n\o2$ and $\Phi_n\o3$ are each fundamental
solutions of the Lax pair~\eref{e:ALLP}.
We can therefore write the following relations
among the modified eigenfunctions:
%$\mu_n\o1,\mu_n\o2,\mu_n\o3$,
\bse
\label{e:ALIBVPjump}
\bea
\mu_n\o2(z,t)= \mu_n\o1(z,t)\,\Zhat^n\esh{-i\omega(z)t}\_s(z)\,,
\label{e:ALIBVPjump12}
\\
\mu_n\o3(z,t)= \mu_n\o1(z,t)\,\Zhat^n\esh{-i\omega(z)t}\_S(z,T)\,,
\label{e:ALIBVPjump13}
\eea
\ese
which hold wherever all terms are defined, namely:
the first column of~\eref{e:ALIBVPjump12} holds for
$0<|z|\le1$, the second column for $|z|\ge1$
and~\eref{e:ALIBVPjump13} holds $\forall z\ne0$.
Thus
\be
\_s(z)= \mu_0\o2(z,0)\,,\qquad
\_S(z,T)= \big(\esh{i\omega(z)T}\mu_0\o1(z,T)\big)^{-1}\,.
\label{e:ALscattdef}
\ee
Equation~\eref{e:ALscattdef} allows us to write
integral representations for the scattering matrices:
\bse
\label{e:ALintegralscattering}
\bea
\_s(z)= \_I - \_Z^{-1}\sum_{n=0}^\infty
  \Zhat^{-n}\big(\_Q_n(t)\mu_n\o2(z,0)\big)\,,
\label{e:ALintegralscatterings}
\\
\_S^{-1}(z,T)= \_I + \int_0^T
  \esh{i\omega(z)t}\big(\_H_0(z,t)\mu_0\o1(z,t)\big)\,\d t\,.
\label{e:ALintegralscatteringS}
\eea
\ese
Note that $\_s(z)$ is again independent of time, since $\_s^{-1}(z)=
\lim_{n\to\infty}\Zhat^{-n}\esh{i\omega(z)t}\mu_n\o1(z,t)=
\lim_{n\to\infty}\Psi_n\o1(z,t)$, as in the IVP.
Note also that~\eref{e:ALIBVPjump} implies
\be
\det\,\_s(z)=1/C_0\,,\qquad
\det\,\_S(z,T)=1\,.
\label{e:IDNLSIBVPscattdet}
\ee
The analyticity properties of $\mu_0\o2(z,t)$
are the same as those of $\mu_n\o2(z,t)$.
However, $\mu_0\o1(z,t)$ enjoys
larger domains of analyticity and boundedness than $\mu_n\o1(z,t)$.
The analyticity and boundedness regions of the scattering matrices
are determined correspondingly via~\eref{e:ALscattdef}:
\begin{itemize}
\item
$\_s_L(z)$ is analytic for $|z|<1$ and continuous and bounded for $|z|\le1$;
while $\_s_R(z)$ is analytic in $|z|>1$ and continuous and bounded for $|z|\ge1$;
\item
$\_S(z,T)$ is analytic in $\Complex\punct$;
moreover, $\_S_L(z,T)$ is continuous and bounded in $\=D_-$,
while $\_S_R(z,T)$ is continuous and bounded in $\=D_+$.
\end{itemize}
The above boundedness properties of~$\_S(z,T)$ can be obtained
as follows.
Let us write the matrix $\_S(z,T)$ as
\[
\_S(z,T)= \begin{pmatrix} A(z,T) &\~B(z,T)\\ B(z,T) &
\~A(z,T)\end{pmatrix}.
\]
As we show below, the symmetries of the problem imply that
$\~A(z,T)$ and $\~B(z,T)$ can be obtained respectively in terms of
$A(z,T)$ and $B(z,T)$. Hence, we only need to discuss the properties
of $A(z,T)$ and $B(z,T)$.
Recall that $\_S(z,T)$ is an entire function of~$z$,
and note that~\eref{e:IDNLSIBVPscattdet} implies
\[
\_S^{-1}(z,T)=
  \begin{pmatrix} \~A(z,T) &-\~B(z,T)\\ -B(z,T) & A(z,T)\end{pmatrix}.
\]
Then \eref{e:ALscattdef} and the analyticity properties of
$\mu\o1_0(z,T)$ imply that $A(z,T)$ is bounded in $\=D_-$. Also,
\eref{e:ALintegralscatteringS} and the integral
representation~\eref{e:ALmu1IBVPsolns} with $n=0$ can be used to
write a Neumann series for $\_S^{-1}(z,T)$, which in turn can be
used to prove analyticity and boundedness of $B(z,T)$ in $\=D_-$.

The involution symmetry discussed when dealing with the~IVP is a
local property. Therefore, when $p_n(t)= \nu\,q_n^*(t)$, it also
applies for the~IBVP. That is, \eref{e:PhiALsymmcol} still holds, as
does~\eref{e:NLSsymmetriesPhij} for $j=1,2,3$. This implies \be
\_s(z)= \begin{pmatrix} a(z) &\nu b^*(1/z^*)\\
   b(z) &a^*(1/z^*)\end{pmatrix},
\quad
\_S(z,T)= \begin{pmatrix}A(z,T) & \nu B^*(1/z^*,T)\\
  B(z,T) &A^*(1/z^*,T)\end{pmatrix}.
\label{e:ALIBVPscattdef}
\ee
Note that \eref{e:IDNLSIBVPscattdet} imply
\bea
%\fl
a(z)a^*(1/z^*)-\nu b(z)b^*(1/z^*)=1/C_0 \,,
\nonumber\\
A(z,T)A^*(1/z^*,T)-\nu B(z,T)B^*(1/z^*,T)=1\,.
\nonumber
%\label{e:relationab}
\eea

\paragraph{Asymptotics.}

Since $\mu_n\o2(z,t)$ coincides with~\eref{e:ALmusolns},
%is the same as the eigenfunction in the IVP,
its asymptotics as $z\to (0,\infty)$ is still given
by~\eref{e:ALasympmu2}.
Also, in~\ref{s:asymptotics} we show that, even though the definition
of $\mu_n\o1(z,t)$ and $\mu_n\o3(z,t)$ involves time integrals,
it is still
$\mu_n\o{j}(z,t)=\_I+O(\_Z^{-1})$
as $z\to(\infty,0)$
for $j=1$ and $j=3$ in their respective domains of boundedness.
More precisely, for all $n>0$ it is
\bse
\bea
\mu_n\o{j}(z,t)=\_I+ \_Q_{n-1}(t)\_Z^{-1} + O(\_Z^{-2})\,,
\qquad {\rm as}~ z\to(\infty,0)\,
\label{e:ALIBVPmu13asymp}
\\
\noalign{\noindent for $j=1,3$, and the limits are restricted the
appropriate regions of the complex plane, where the corresponding
columns are bounded. For $n=0$ it is instead}
%\mu_n\o3(z,t)=\_I+ *** + O(\_Z^{-2})\,,
%\qquad {\rm as}~ z\to(0,\infty)\,.
\mu_0\o1(z,t)=
  \_I + \big(\_Q_{-1}(t)-\esh{-i\omega(z)t}\_Q_{-1}(0)\big)\,\_Z^{-1}+O(\_Z^{-2})\,,
\label{e:mu1Oasymp}
\\
\mu_0\o3(z,t)=
  \_I + \big(\_Q_{-1}(t)-\esh{-i\omega(z)(t-T)}\_Q_{-1}(T)\big)\,\_Z^{-1}+O(\_Z^{-2})\,,
\label{e:mu3Oasymp} \eea \ese as $z\to (\infty,0)$. The above yield,
for all $n\ge0$, \be \_Q_{n-1}(t)=
\lim_{z\to(\infty,0)}\big(\mu_n\o{j}(z,t)-\_I\big)\,\_Z\,,\qquad
{\rm for}~j=1,3\,.\label{e:ALQnasymp} \ee

\noindent Also, the asymptotic behavior of the eigenfunctions
determines that of the scattering matrices. In particular, from the
second of~\eref{e:ALscattdef} we have \bse
\label{e:ALIBVPscattcoeffasymp} \bea A^*(1/z^*,T)= 1+O(1/z^2)\,,\,\,
B^*(1/z^*,T)= O(1/z)\,\,\,
\mathrm{as}~z\to\infty~\mathrm{in}~\=D_{+\#out},
%\nonumber\\[-1ex]
\\
\noalign{\noindent while~\eref{e:ALIBVPjump13} implies}
A^*(1/z^*,T)= 1+O(z^2)\,,\quad B^*(1/z^*,T)= O(z)\,\quad
\mathrm{as}~z\to0~\mathrm{in}~\=D_{+\#in}. \eea Similarly,
\eref{e:ALasympmu2} and~\eref{e:ALIBVPjump12} yield \bea a^*(1/z^*)=
1/C_0 + O(1/z^2)\,,\quad b^*(1/z^*)= O(1/z)\quad
\mathrm{as}~z\to\infty~\mathrm{in}~\=D_{+\#out}.
\nonumber\\[-1ex]
\eea
\ese

\paragraph{Riemann-Hilbert problem, solution and reconstruction formula.}

We now formulate the RHP whose solution will enable us to obtain a
representation for the solution of the AL system on the naturals.
For later reference, we introduce the quantities \bea \gamma(z)={\nu
b^*(z) \over a(z)} \,,\qquad R(z,t)= {B^*(1/z^*,t)\over
A^*(1/z^*,t)}\,,\qquad \Gamma(z)={ B(z,T) \over
a^*(1/z^*)d^*(1/z^*)} \,, \nonumber
\\
\noalign{\noindent with}
d(z)= %A^*(1/z^*,T)\,(a(z)-\nu b(z)R(z,T))=
a(z)A^*(1/z^*,T)-\nu b(z)B^*(1/z^*,T) \,. \nonumber \eea Note that
$R(z,T)$ is defined $\forall z\in\Complex$ except where
$A^*(1/z^*,T)=0$, $\Gamma(z)$ is defined for $z\in L_3\cup L_4$,
$d(z)$ for $z\in\=D_{\pm\#in}$, and $\gamma(z)$ for $|z|=1$.
Moreover, $d^*(1/z^*)=1/C_0+O(1/z^2)$ as $z\to\infty$. In the
analysis of linearizable BCs, it will be useful to write
$\Gamma^*(1/z^*)$ in terms of only $a(z)$, $b(z)$ and $R(z,T)$ as
%the ratio of $B(z,T)/A(z,T)$ as
\[
%\Gamma^*(1/z^*)={\nu B(z,T)/A(z,T) \over a^*(1/z^*)\big(a^*(1/z^*)-\nu b^*(1/z^*)B(z,T)/A(z,T)\big)}\,,
\Gamma^*(1/z^*)={R(z,T) \over a(z)\big(a(z)-\nu b(z)R(z,T)\big)}\,.
\]
Finally, %recalling~\eref{e:ALasympmu2},
we introduce the normalization matrix
$\_C_n= \diag(1/C_0,C_n)\,$.
%\begin{pmatrix} 1/C_0 & 0 \\ 0 & C_n \end{pmatrix}.

We are now ready to formulate the RHP, which we do
using~\eref{e:ALIBVPjump}. We introduce the matrix functions
$\_M_n^\pm(z,t)$ defined as: \bse \label{e:IBVPALdefM} \bea
\_M_n^+(z,t)= \left\{\!\begin{array}{l}\displaystyle
\_C_n\,\bigg(\mu_n\o{2,L}(z,t),\displaystyle{\mu_n\o{3,R}(z,t)\over
d(z)}\bigg),\qquad z \in D_{+\#in},
\\[1ex]\displaystyle
%\_M_n^+(z,t)=
\_C_n\,\bigg( {\mu_n\o{1,L}(z,t)\over
a^*(1/z^*)},\mu_n\o{2,R}(z,t)\bigg),\qquad z \in D_{+\#out},
\end{array}\right.
%\nonumber\\[-1ex]
\label{e:IBVPALdefM1}
\\
\_M_n^-(z,t)= \left\{\!\begin{array}{l}\displaystyle
\_C_n\,\bigg(\mu_n\o{2,L} , {\mu_n\o{1,R} \over a(z)} \bigg),\qquad
z \in D_{-\#in}\,\,,
\\[1ex]\displaystyle
%\_M_n^-(z,t)=
\_C_n\,\bigg( {\mu_n\o{3,L} \over d^*(1/z^*)},
\mu_n\o{2,R}\bigg),\qquad z \in D_{-\#out}\,.
\end{array}\right.
%\nonumber\\[-1ex]
\label{e:IBVPALdefM2}
\eea
\ese
Note that $\_M_n^\pm(z,t)$ are sectionally meromorphic respectively
for $z\in D_+$ and $z\in D_-$.
Moreover, after some tedious but straightforward algebra,
equations~\eref{e:ALIBVPjump} yield the jump conditions as
\be
\_M_n^-(z,t)=\_M_n^+(z,t)\,\big(\_I-\_J_n(z,t)\big), \qquad z \in  L\,,
\label{e:IBVPALsystemRHP}
\ee
where the contours $L=L_1\cup L_2\cup L_3 \cup L_4$ are (cf.~Fig.~\ref{f:DpmAL})
\[\fl
L_1=\=D_{+\#in}\cap\=D_{-\#in}\,,\quad
L_2=\=D_{-\#in}\cap\=D_{+\#out}\,,\quad
L_3=\=D_{+\#out}\cap\=D_{-\#out}\,,\quad
L_4=\=D_{+\#in}\cap\=D_{-\#out}\,,
\]
and the jump matrices $\_J_n\o1,\dots ,\_J_n\o4$ are defined by
%\bse
%\label{e:ALjumpmatrix}
\bea
\_J\o1_n(z,t)=\begin{pmatrix}
   0 &\nu z^{2n}\e^{-2i\omega(z)t}\Gamma^* (1/z^*,T)\\
   0 & 0\end{pmatrix} \,, \qquad z \in  L_1\,,
\nonumber
\\
\_J\o2_n(z,t)=\begin{pmatrix} 1-1/C_0 &z^{2n}\e^{-2i\omega(z)t}\gamma (z)\\
  -\nu z^{-2n}\e^{2i\omega(z)t}\gamma^*(z) & 1-C_0\big(1-\nu |\gamma(z)|^2\big)
\end{pmatrix} \,,\qquad z \in L_2\,,
%\nonumber\\[-1ex]
\nonumber
\\
\_J\o3_n(z,t)=\begin{pmatrix}0 & 0\\ -z^{-2n}\e^{2i\omega(z)t}\Gamma (z,T) &0
\end{pmatrix} \,,\qquad z\in L_3\,,
\nonumber
\\
\_J\o4_n(z,t)=\_I- (\_I-\_J\o1_n)(\_I-\_J\o2_n)^{-1}(\_I-\_J\o3_n)\,, \qquad z \in L_4\,.
\nonumber
\eea
%\ese
As in the IVP, we first consider the case in which no discrete spectrum is
present.
For the IBVP, this corresponds to assuming that $a(z)\ne0$ for
$z\in D_{-\#in}$ and $d(z)\ne0$ for $z\in D_{+\#in}$.
In this case,
the matrix functions $\_M_n^\pm(z,t)$ are analytic in their
respective domains. Also, $\_M_n(z,t)\to\_I$ as $z\to\infty$ thanks
to~\eref{e:ALasympmu2}, \eref{e:ALIBVPmu13asymp},
\eref{e:ALIBVPscattcoeffasymp} and \eref{e:IBVPALdefM}. Hence the
matrix RHP \eref{e:IBVPALsystemRHP} is solved by the Cauchy
projectors $P^\pm$ over the contour~$L$, namely $P^\pm=1/(2\pi
i)\,\int\nolimits_L [1/(k'-k)]\,\d k'$. That is, \be
\_M_n^+(z,t)=\_I+ \frac1{2\pi i} \int_L \_M_n^+(\zeta,t)
  {\_J_n(\zeta,t) \over \zeta-z} \,\d\zeta
\,.
\label{e:ALsystemRHPsoln}
\ee
Equation~\eref{e:ALsystemRHPsoln} also yields the asymptotic
expansion of $\_M_n^+(z,t)$ as $z \to 0$, namely,
\bea
\_M_n^+(z,t)= \_I
  + \frac1{2\pi i}\int_L\_M_n^+(\zeta,t)\_J_n(\zeta,t)\,\frac{\d\zeta}\zeta
  + \frac z{2\pi i}\int_L\_M_n^+(\zeta,t)\_J_n(\zeta,t)\,\frac{\d\zeta}{\zeta^2}
  + O(z^2)\,.
\nonumber\\[-2ex]
\label{e:IBVPALasympRHPsoln}
\eea
Note that we can write~\eref{e:IBVPALasympRHPsoln} as
\bea
\_M_n^+(z,t)= %\begin{pmatrix} 1/C_0C_n & 0 \\ 0 & C_0C_n\end{pmatrix}
  \diag[1/(C_0C_n),C_0C_n]
  + \frac z{2\pi i} \int_L \_M_n^+(\zeta,t)
     \big(\_I-\_J_n(\zeta,t)\big) \,{\d\zeta\over \zeta^2} +O(z^2)\,.
\nonumber\\[-2ex]
\label{e:IBVPALasympMn} \eea Now note that the matrix
$\_C_n^{-1}\_M_n(z,t)$ satisfies the $n$-part of the Lax
pair~\eref{e:ALLP1}. Also, thanks to \eref{e:ALasympmu2},
\eref{e:ALIBVPmu13asymp} and~\eref{e:ALIBVPscattcoeffasymp}, it is
\[
\_C_n^{-1}\_M_n^+(z,t)= \diag[1/C_n,C_0] + O(z)\qquad
  \quad{\rm as}~z\to0\,.
\]
Hence, substituting the asymptotic expansion of $\_M_n(z,t)$
into~\eref{e:ALLP1} and comparing the $(1,2)$-components of the
$O(z)$ terms, we can recover the scattering potentials as \be
q_n(t)=\lim_{z\to 0}\big(\_M_{n+1}^+(z,t)-\_I\big)_{12}/z
\label{e:ALreconstruction}\,. \ee Taking the $(1,2)$-component
of~\eref{e:IBVPALasympMn} and comparing
with~\eref{e:ALreconstruction}, we then obtain the reconstruction
formula for the solution of the IDNLS equation on the natural
numbers: \bea \fl q_n(t)=-\frac1{2\pi i}\int_{|z|=1}
    z^{2n}\e^{-2i\omega(z)t}\gamma(z)\big(\_M_{n+1}^+(z,t)\big)_{11}\,\d z
 + \frac\nu{2\pi i} \int_{L_1}z^{2n}
    \e^{-2i\omega(z)t}\Gamma^*(1/z^*)\big(\_M_{n+1}^+(z,t)\big)_{11}\,\d z
\nonumber\\\kern-2em{ }
 + \frac1{2\pi i}\int_{L_2}
   \big(\nu C_0|\gamma(z)|^2-C_0+1\big)
     \big(\_M_{n+1}^+(z,t)\big)_{12}\,{\d z \over z^2}
 + \frac1{2\pi i}\,\bigg(1-\frac1{C_0}\bigg)\int_{L_4}
     \big(\_M_{n+1}^+(z,t)\big)_{12} \,{\d z \over z^2}
\nonumber\\\kern2em{ }
 + \frac1{2\pi i}{\nu\over C_0}\int_{L_4}z^{2n}
     \e^{-2i\omega(z)t}\Gamma^*(1/z^*)\big(\_M_{n+1}^+(z,t)\big)_{11}\,\d z
\,.
\label{e:representationqn}
\eea

\paragraph{Global relation.}

%The scattering matrix $\_S(z,T)$ depends on the unknown boundary datum.
As in the linear problem and the continuum limit,
the unknown boundary datum can be obtained in terms of the known
initial-boundary conditions using the global relation and the
symmetries of the system.

Integrating~\eref{e:ALmodifiedLPPsi} around the boundary of the region
$\Natural_0 \times[0,t]$, one obtains
\bea
\fl
\_Z\int_0^t \esh{i\omega(z)t'}\big(\_H_0(z,t')\mu_0(z,t')\big) \,\d
t'+\esh{i\omega(z)t'}\sum_{n=0}^{\infty} \Zhat^{-n}
\big(\_Q_n(t)\mu_n(z,t)\big)=\sum_{n=0}^{\infty} \Zhat^{-n}
\big(\_Q_n(0)\mu_n(z,0)\big) \,.
\nonumber\\
\label{e:ALIBVPGR0} \eea When~\eref{e:ALIBVPGR0} is evaluated with
$\mu_n(z,t)\equiv\mu_n\o2(z,t)$ and $t=T$, the first and second
columns of the resulting equation are valid respectively for
$z\in\=D_{\pm \#in}$ and $z\in\=D_{\pm \#out}$. Moreover, the RHS
of~\eref{e:ALIBVPGR0} becomes $\_Z(\_I-\_s(z))$ thanks
to~\eref{e:ALintegralscatterings}. Finally,
using~\eref{e:ALIBVPjump}, we can write the first term and the
second term in~\eref{e:ALIBVPGR0} respectively as
$\_Z\,\big(\_S^{-1}(z,T)-\_I\big)\,\_s(z)$ and
$\_Z\,\big(\_I-\e^{i\omega(z)t\hsigma}\mu_0\o2(z,T)\big)$. We
therefore have the following global relation in terms of the
scattering data:
\bea
\_S^{-1}(z,T)\_s(z)=\_I-\esh{i\omega(z)T}\_G(z,T)\,,
\label{e:ALIBVPGR1}
\\
\noalign{\noindent where}
\_G(z,t)=\_Z^{-1}\sum_{n=0}^{\infty} \Zhat^{-n}
\big(\_Q_n\mu_n\o2(z,t)\big)\,.
\nonumber
\eea
Like for~\eref{e:ALIBVPGR0}, the first and second column
of~\eref{e:ALIBVPGR1} are respectively valid for $|z|\le1$ and
$|z|\ge1$. Also, from the analyticity domains of $\mu_n\o2(z,t)$ it
follows that $\_G_L(z,t)$ is analytic in $|z|<1$ and $\_G_R(z,t)$ is
analytic in $|z|>1$. Taking the $(1,2)$ component
of~\eref{e:ALIBVPGR1} we have
\bea
A^*(1/z^*,T)b^*(1/z^*)-B^*(1/z^*,T)a^*(1/z^*)=
-\nu\e^{2i\omega(z)T}G(z,T)
%-\e^{2i\omega(z)T}\sum_{n=0}^{\infty}z^{-2n-1}q_n(T)\big(\mu_n\o2(z,T)\big)_{22}
\,,\,\,  |z|>1,
\nonumber\\[-2ex]
\label{e:ALIBVPGR} \eea where \bea G(z,T)=\sum_{n=0}^{\infty}
z^{-2n-1}q_n(T)\big(\mu_n\o2(z,T)\big)_{22} \,. \nonumber \eea Also
note that the RHS of~\eref{e:ALIBVPGR} is bounded for $z\in
\=D_{+\#out}$. Then, for $z\in \=D_{+\#out}$ the RHS vanishes in the
limit $T\to\infty$, implying
\be
A^*(1/z^*,T)b^*(1/z^*)-B^*(1/z^*,T)a^*(1/z^*)=0 \,,\quad z \in
\=D_{+\#out}\,. \label{e:ALGRTinfty} \ee

\noindent For finite values of $T$, letting $r(z)=b(z)/a(z)$, the
global relation is now
\[
B^*(1/z^*,T)-r(1/z^*)A^*(1/z^*,T)=
\nu\e^{2i\omega(z)T}G(z,T)/a^*(1/z^*)\,.
\]
Since $G(z,t)=O(1/z)$ as $z\to\infty$ for $z\in D_{+\#out}$,
multiplying by $z\e^{-2i\omega(z)t}$ and integrating
over $\partial\~D_{+\#out}$
[where $\~D_{+\#out}=\{D_{+\#out}\wedge\Im z>0\}$]
we obtain the integral relation:
\be
\int_{\partial \~D_{+\#out}}
z\,\e^{-2i\omega(z)t}\,\big(B^*(1/z^*,T)-r(1/z^*)A^*(1/z^*,T)\big)\,\d
z =0\,.
\ee
This is the discrete analogue of the one that
in the continuum case is used to obtain the Dirichlet-to-Neumann map
\cite{CPAM58p639}.
In the discrete case, however, the unknown boundary datum can be obtained
using an alternative, simpler method,
as we will show in section~\ref{s:idnlsdata}.

\paragraph{Linear limit.}

The linear limit of the solution~\eref{e:representationqn} of the
IBVP for the IDNLS equation coincides with the solution of the IBVP
for the DLS equation, as we show next.
Suppose $\_Q_n(t)=O(\epsilon)$. %and $\_Q_{-1}(t)=O(\epsilon)$.
From~\eref{e:ALmuIBVPsolns} it follows that $\mu_n=\_I+O(\epsilon)$.
Recalling~\eref{e:ALintegralscatterings}, we obtain, to $O(\epsilon)$:
\bea
\fl
\gamma(z)= -\frac 1z\,\^q(z^2,0)\,, \quad
d(z)= 1\,, \quad C_0=1\,,\quad
\Gamma^*(1/z^*)= i\nu\,\bigg(\frac 1z\^f_{-1}(z^2,T)-z\^f_0(z^2,T)\bigg)\,.
\nonumber
\eea
Thus~\eref{e:representationqn} yields, to $O(\epsilon)$,
\bea
\fl
q_n(t)= \frac1{2\pi i} \int_{|\zeta|=1}
\zeta^{2n-1}\e^{-i\omega(\zeta^2)t}\^q(\zeta^2,0)\, \d\zeta
%\nonumber\\\kern6em
{ }+\frac1{2\pi}\int_{L_1+L_4}\zeta^{2n} \e^{-i\omega(\zeta^2)t}
  \bigg( \frac1\zeta \^f_{-1}(\zeta^2,T)-\zeta\^f_0(\zeta^2,T) \bigg)\,\d\zeta\,,
\nonumber
\eea
where the integrals are taken in Cauchy's principal value sense.
Now note that the contour $L_1\cup L_4$ can be deformed to
$\partial D_{+\#in}$ by Cauchy's theorem.
Performing the change of variable $\zeta^2=z$,
we then obtain that the linear limit of~\eref{e:representationqn}
coincides with the solution of the IBVP for the DLS on the natural numbers,
namely~\eref{e:LSIBVPsoln0}.

\paragraph{Continuum limit.}

Reinstating the lattice spacing $h$, it is easy to show that the Lax
pair for the NLS is the continuum limit of that for the AL system as
$h\to 0$ \cite{JMP16p598,APT2003}.
The continuum limit is formally obtained by writing the solution of the
discrete case as $Q_n(t)=hq(nh,t)$ and $P_n(t)=hp(nh,t)$. Then for
$z=\e^{ikh}$, the Lax pair~\eref{e:ALLPm} becomes
%\bse
\bea
{\mu_{n+1}-\mu_n \over h}-ik[\sigma_3,\mu_n]=\_Q_n(t)\,\mu_n+O(h^2)\,,
\nonumber
\\
\.\mu_n +i\omega(k)[\sigma_3,\mu_n]=\_H_n(t,k)\,\mu_n+O(h)\,,
\nonumber
\eea
%\ese
where now $\omega(k)=(1-\cos 2kh)/h^2$, with $\mu_n=\mu(nh,t,k)$,
$q_n=q(nh,t)$ and $p_n=p(nh,t)$ for brevity, and where
\[
\fl
\_H_n(t,k)=i\,\begin{pmatrix} -q_n p_{n-1} &
(q_n-q_{n-1})/h+ik(q_n+q_{n-1})\\
 -(p_n-p_{n-1})/h+ik(p_n+p_{n-1}) & q_{n-1}p_n
\end{pmatrix}.
\]
Correspondingly, the Jost solutions are obtained
from~\eref{e:ALmuIBVPsolns}, for example,
\bea \fl
\mu_n\o1(k,t)=\_I+h\sum_{m=0}^{n-1}\esh{ikh(n-m)}(\_Q_m(t)\mu_m\o1(k,t))
+\int_0^t \esh{i[nkh-\omega(k)(t-t')]}(\_H(0,k,t')\mu_0\o1(k,t')) \,
\d t'\,. \nonumber
%\\
\eea
As $h\to 0$ with $x=nh$ fixed, we have $\omega(k)\to2k^2$, together with
$\_H_n(t,k)\to\_H(x,t,k)$ and
$\mu_n\o{j}(k,t)\to\mu\o{j}(x,t,k)$, $j=1,2,3$, where $\mu\o{j}(x,t,k)$
are the Jost solutions for the IBVP of the NLS, namely~\eref{e:NLSLPsolutions}.
Note also that $C_n\to 1$ as $h\to0$.
Hence, in the continuum limit, the solution of the IBVP for the IDNLS
becomes exactly that of the IBVP for NLS.

The result can also be verified directly via the continuum limit of the
solution~\eref{e:representationqn}.
Explicitly, since $C_0=1+O(h^2)$,
as $h\to 0$ we have
\bea
\fl
Q_n(t)=-\frac1{2\pi i}\int_{|\zeta|=1}
\zeta^{2n}\e^{-2i\omega(\zeta)t}\gamma(\zeta)\big(\_M_{n+1}^+(\zeta,t)\big)_{11}
\,\d\zeta
\nonumber\\\fl\kern2em{ }
+\frac \nu{2\pi i} \int_{L_1+L_4} \zeta^{2n}
\e^{-2i\omega(\zeta)t}\Gamma^*(1/\zeta^*)\big(\_M_{n+1}^+(\zeta,t)\big)_{11}
\,\d\zeta
%\nonumber\\{ }
 + \frac \nu{2\pi i}\int_{L_2} |\gamma(\zeta)|^2
\big(\_M_{n+1}^+(\zeta,t)\big)_{12} \,{\d\zeta \over\zeta^2}+O(h^2)\,.
\nonumber
%\\[-2ex]
%\label{e:Continuumrepresentqn}
\eea
The oriented contour $L_1\cup L_4$ can be deformed onto $|\zeta|=1$
since the corresponding integrand is analytic in~$D_{-\#in}$.
In terms of $q(nh,t)=Q_n(t)/h$, and performing the substitution
$\zeta=\e^{ikh}$, we then have
\bea
\fl
q(nh,t)= -\frac h{\pi}\,\, \int_{\!\!-\pi/h}^{\,\,\pi/h}
\e^{2i(nkh-\omega(k)t)}\gamma(k)\big(\_M_{n+1}^+(k,t)\big)_{11}\,\e^{ikh}\,\d k
\nonumber\\\fl\kern3em
+ \frac h{\pi}\,\, \int_{\!\!-\pi/h}^{\,\,\pi/h}
  \nu\e^{2i(nkh-\omega(k)t)}\Gamma^*(k^*)\big(\_M_{n+1}^+(k,t)\big)_{11}
    \,\e^{ikh}\,\d k
-\frac h{\pi}\,\, \int_{\!\!\pi/2h}^{\,\,\pi/h} \nu
  |\gamma(k)|^2 \big(\_M_{n+1}^+(k,t)\big)_{12}\,\e^{-ikh}\, \d k
\nonumber\\\fl\kern16em{ }
  -\frac h{\pi}\,\, \int_0^{\,\,\pi/2h} \nu |\gamma(-k)|^2
  \big(\_M_{n+1}^+(-k,t)\big)_{12}\,\e^{ikh}\, \d k\,.
%\nonumber\\[-1ex]
\label{e:continuumlimit} \eea Now note that, since
$\e^{-4ik^2t}\Gamma^*(k^*)(\_M_{n+1}^+(x,k,t))_{11}$ is analytic and
bounded for $(\Re\,k\in[-\pi/h,0])\wedge(\Im\,k>0)$ [which becomes
$\Complex_\II$ in the limit $h\to0$], the portion of the
corresponding integral on the negative real axis can be deformed
onto the positive imaginary axis. Then, taking the continuum limit
of all the integrals in~\eref{e:continuumlimit} we obtain that
$q(nh,t)$ coincides with the solution of the IBVP for the NLS,
namely~\eref{e:NLSIBVPrepresentationq}, in the limit $h\to 0$.

\paragraph{Remarks.}
A few comments are now in order:
\begin{itemize}
\item
Equation~\eref{e:representationqn} provides
the Ehrenpreis~\cite{Ehrenpreis1970,Palamodov1970,Henkin1990}
representation for the solution of the IBVP for the IDNLS,
in analogy with Ref.~\cite{JNLMP10p47} in the continuum limit.
\item
One can now use~\eref{e:representationqn} as a starting point
to formally prove that the function $q(x,t)$ given by the reconstruction
formula satisfies~\eref{e:IDNLS} as well as the initial-boundary conditions,
using the dressing method, as in Ref.~\cite{NLTY18p1771} in the continuum limit.
\item
In the continuum problem, the location of the jumps is
the union of the jumps for the scattering problem in the
linear case and those of its adjoint. In the discrete problem,
however, this is not the case. Indeed, the extra jump along the
imaginary axis arises as a consequence of the rescaling $z\to z^2$
when going from the linear to the nonlinear case.
\item
The scattering matrix $\_S(z,T)$ involves $T$ explicitly.
In~\ref{s:IndependenceT}, however, we show that the solution of the IBVP
for the AL system on the naturals does not depend on future values of
the boundary datum.
\item
With the due modifications, the method presented here can also be used
to solve the IBVP for all members of the Ablowitz-Ladik hierarchy.
Moreover, the method can be generalized to any integrable
differential-difference evolution equation.
\end{itemize}

%%%%%%%%%%%%%%%%%%%%%%%%%%%%%%%%%%%%%%%%%%%%%%%%%%%%%%%%%%%%%%%%%%%%%%%%%%%%%%%
\section{Elimination of the unknown boundary datum, linearizable BCs and soliton solutions}
\label{s:idnlsdata}

\subsection{Elimination of the unknown boundary datum}

The scattering matrix $\_S(z,T)$ depends on the both the known and
the unknown boundary datum.
In the linear problem, it was possible to
overcome this difficulty by making use of the fact that the
transformation $z\to1/z$ leaves the transforms of the boundary data
unchanged. In the nonlinear problem, however, the matrix $\_S(z,T)$
is \textit{not} invariant under this transformation, because it is
defined in terms of the eigenfunction~$\mu_n\o1(z,t)$, which is not
invariant under $z\to1/z$.
As in the continuum case~\cite{CPAM58p639}, the determination of the
unknown boundary datum in terms of the known initial-boundary conditions
is in general a nontrivial issue.

For linearizable BCs it is possible to
express the RHP only in terms of the initial data,
as we show in section~\ref{s:ALlinearizable}.
This is not possible for generic BCs, however.
In this case one must solve a coupled system of nonlinear
ordinary differential equations (ODEs)
to obtain simultaneously the unknown boundary datum $q_{-1}(t)$
as well as scattering coefficients $A(z,T)$ and $B(z,T)$,
as we show next.

The boundary data enters the RHP only via the ratio
$R(z,T)=B(z,T)/A(z,T)$ appearing in $\Gamma(z)$.
Recalling~\eref{e:ALphimu} and~\eref{e:ALscattdef}, we have
$\_S(z,t)= {\~\Phi}^{-1}(z,t)\,\e^{-i\omega(z)t\sigma_3}$, where
the matrix
\be
\label{e:ALPhitilde}
\~\Phi(z,t)=
\Phi_0\o1(z,t)=\begin{pmatrix}
e^{-i\omega(z)t}A^*(1/z^*) & -\nu e^{-i\omega(z)t}B^*(1/z^*)\\
-e^{i\omega(z)t}B(z) & e^{i\omega(z)t}A(z) \end{pmatrix}\,,
\ee
satisfies the $t$-part of the Lax pair~\eref{e:ALLP} for $n=0$,
namely:
\bea
\.{\~\Phi}= \big(-i\omega(z)\sigma_3 + \_H_0(z,t)\big)\,\~\Phi\,,
\label{e:ALtLP0}
\eea
together with the initial condition $\~\Phi(z,0)=\_I\,$.

The term $\_H_0(z,t)$ in~\eref{e:defHn}
contains $q_{-1}(t)$, of course.
Note however that using~\eref{e:ALQnasymp} with $n=0$, we can express
$q_{-1}(t)$ in terms of $\mu_0\o1(z,t)$:
\be
q_{-1}(t)=
  \lim_{ z\to 0} \big(\mu_0\o1(z,t)\big)_{12}/z\qquad z\in D_{-\#in}\,.
\label{e:ALQnm1}
\ee
The simultaneous solution of~\eref{e:ALPhitilde}
and~\eref{e:ALQnm1} provides the unknown boundary datum as well as
the auxiliary spectral functions $A(z,t)$ and $B(z,t)$, allowing one
to completely define the RHP and therefore we also obtain the
solution of the inverse problem.
Note that this procedure is significantly simpler
than that requried to obtain the generalized Dirichlet-to-Neumann map
in the continuum case~\cite{CPAM58p639}.

%%%%%%%%%%%%%%%%%%%%%%%%%%%%%%%%%%%%%%%%%%%%%%%%%%%%%%%%%%%%%%%%%%%%%%%%%%%%%%%%%%%%%%%%
\subsection{Linearizable boundary conditions}
\label{s:ALlinearizable}

Like in the continuum case,
there is a class of BCs, called \textit{linearizable}, for which
it is possible to obtain the unknown boundary datum via only
algebraic manipulations of the global relation.

Recall that $A(z,t)$ and $B(z,t)$ are given in terms of
$\_\Phi(z,t)=\mu_0\o1\e^{-i\omega(z)t\sigma_3}$ by~\eref{e:ALPhitilde}
which solves the ODE~\eref{e:ALtLP0}
together with the initial condition $\~\_\Phi(z,0)=\_I$.
Since $\omega(1/z)=\omega(z)$, the matrix~$\~\Phi(1/z,t)$
%$\~\Phi'(z,t)=\~\Phi(1/z,t)$
satisfies equations identical to~\eref{e:ALtLP0}
except that $\_H_0(z,t)$ is replaced by~$\_H_0(1/z,t)$.
If there exists a time-independent matrix
$\_N(z)$ such that
\be
\_N(z)\,\big(-i\omega(z)\sigma_3 + \_H_0(z,t)\big)\,
  = \big(-i\omega(z)\sigma_3 + \_H_0(1/z,t)\big)\,\_N(z)\,,
\label{e:ALlinearizcond} \ee it is then easy to show that \be
\~\Phi(1/z,t)= \_N(z)\,\~\Phi(z,t)\,\_N(z)^{-1}\,.
\label{e:ALIBVPphi1zinv} \ee A necessary condition
for~\eref{e:ALlinearizcond} to be satisfied is obviously that
$\det[-i\omega(z)\sigma_3+\_H_0(z,t)]=
  [(z^2-1/z^2)(q_0p_{-1}-q_{-1}p_0)]^2$
be invariant under the transformation~$z\to1/z$.
In turn, for this condition to be satisfied one needs
%\[
%z^2q_0p_{-1}+{q_{-1}p_0\over z^2}={q_0p_{-1}\over z^2}+q_{-1}p_0z^2\,,
%\]
%which implies
\be q_0p_{-1}-q_{-1}p_0= 0\,. \label{e:ALlinearizableBC} \ee In the
reduction $p_n(t)=\nu q_n^*(t)$ to IDNLS, \eref{e:ALlinearizableBC}
is satisfied by the discrete analogue of homogeneous Robin BCs: \be
q_{-1}-\chi q_0=0 \,,\quad \chi\in\Real \,. \label{e:ALIBVPRobinBC}
\ee These BCs had been previously identified via algebraic methods
\cite{PLA207p263}. For the BCs~\eref{e:ALIBVPRobinBC}, we can solve
the system~\eref{e:ALlinearizcond} for $\_N(z)$, obtaining $N_{12}=
N_{21}=0$ and $N_{11}=f(z)N_{22}$, where
\[
f(z)=
{1-\chi z^2 \over z^2-\chi} \,.
\]
Recalling~\eref{e:ALIBVPphi1zinv}, we then find the following
symmetries for the scattering data: \be A^*(z^*,T)=A^*(1/z^*,T)\,,
\quad B^*(z^*,T)=f(z)B^*(1/z^*,T) \label{e:ALsymmetryAB}\,. \ee Note
that $\_N(z)$ is not invertible for
$z=\pm\chi^{1/2},\pm\chi^{-1/2}$. However, \eref{e:ALsymmetryAB} is
still valid at such values of~$z$. Indeed, since $\~\Phi(z,t)$
solves~\eref{e:ALtLP0}, writing a Neumann series for $\~\Phi(z,t)$
one finds $\~\Phi(\pm\chi^{1/2},t)_{12}=0$, which implies that
$B^*(\pm\chi^{-1/2},t)=0$. As a consequence, since $A^*(1/z^*,T)$
and $B^*(1/z^*,T)$ are analytic for $z\in\Complex\punct$, we can
conclude that the limit as $z\to\pm\chi^{1/2}$ of the product
$f(z)\,B^*(1/z^*,T)$ exists and is finite.

The above properties now allow $\Gamma^*(1/z^*)$ to be expressed in
terms of the known functions, $a(z)$ and $b(z)$.
For simplicity, we consider the case in which no discrete spectrum is present.
Consider first the case $T=\infty$. The global relation in this case
is simply given by~\eref{e:ALGRTinfty}. Replacing $1/z$ by $z$ and
using~\eref{e:ALsymmetryAB},
we obtain
\bse
\[
A^*(1/z^*)={a^*(z^*)d(z)\over \Delta (1/z)}\,,\quad
B^*(1/z^*)={f(1/z)b^*(z^*)d(z)\over \Delta (1/z)} \quad z\in
D_{+\#in}\,,
\label{e:ALABLin}
\]
\eject\noindent
where
\[
\Delta(z)=a(1/z)a^*(1/z^*)-\nu f(z)b(1/z)b^*(1/z^*)\,.
\label{e:ALDeltadef}
\]
\ese As a result, we can express the ratio $R(z,T)=
B^*(1/z^*,T)/A^*(1/z^*,T)$ as
\be
R(z,T)=f(1/z)\,{b^*(z^*)\over
a^*(z^*)} \qquad z\in D_{+\#in}\,, \label{e:ALratioAB} \ee and we
therefore obtain $\Gamma^*(1/z^*)$ only in terms of known spectral
functions.
Now consider the case $T<\infty$. The global relation in this case
is~\eref{e:ALIBVPGR}. Replacing $1/z$ by $z$ in~\eref{e:ALIBVPGR}
and using the symmetry~\eref{e:ALsymmetryAB} as before, we obtain
\be
%{B(z,T)\over A(z,T)}
R(z,T)=
  f(1/z){b^*(z^*)\over a^*(z^*)}+\nu\e^{2i\omega(z)T}\,
  {f(1/z)G(1/z,T)\over a^*(z^*)A^*(z^*,T)}\,,\quad z\in D_{+\#in} \,.
\label{e:ALratioAB1} \ee We therefore see that the difference from
the case $T=\infty$ is simply the appearance of an additional term
in the RHS of~\eref{e:ALratioAB}. In~\ref{s:IndependenceT}, however,
we show that the second term in the RHS of~\eref{e:ALratioAB1} does
not affect the solution of the IBVP for the IDNLS. Hence,  even in
the case $T<\infty$, we can use~\eref{e:ALratioAB} in the
RHP~\eref{e:IBVPALsystemRHP}.

%%%%%%%%%%%%%%%%%%%%%%%%%%%%%%%%%%%%%%%%%%%%%%%%%%%%%%%%%%%%%%%%%%%%%%%%%%%%%%%%%%%%%
\subsection{Discrete spectrum and soliton solutions}

Equations~\eref{e:IBVPALdefM} imply that
when the functions $a(z)$ and $d(z)$ possess zeros
the matrices $\_M_n^{\pm}(z,t)$ are only meromorphic
functions in $D_+$ and $D_-$, respectively.
As a consequence, the RHP~\eref{e:IBVPALsystemRHP} formulated
becomes singular.
As in the IVP, however, it can be converted to a regular RHP
by taking into account the appropriate residue relations.
We assume that these discrete eigenvalues are all simple.
More precisely, we assume that:
\begin{itemize}
\item
$a(z)$ has simples zeros in $D_{-\#in}$.
We label such zeros $\pm z_j$ for $j=1,\,\dots\,,J$;
\item
$d(z)$ has simple zeros in $D_{+\#in}$.
We label such zeros $\pm\lambda_j$ for $j=1,\,\dots\,,J'$.
\end{itemize}
We also assume that there are no zeros on the boundaries of these
domains and that there are no common zeros of $a(z)$ and $d(z)$
in $D_{+\#in}$\,.

The fact that the zeros of $a(z)$ and $d(z)$ always appear in opposite pairs
is a trivial consequence of $a(z)$ and $d(z)$ both being
even functions of~$z$ [cf.~\ref{s:asymptotics}].
Also, the symmetry $p_n(t)= \nu q_n(t)$ of the potentials
implies that, corresponding to these zeros,
there is an equal number of zeros of $a^*(1/z^*)$ and $d^*(1/z^*)$
in $D_{+\#out}$ and $D_{-\#out}$, respectively,
which we denote respectively by
$\=z_j=1/z_j^*$ and $\=\lambda_j=1/\lambda_j^*$.
Thus, discrete eigenvalues in the IBVP can appear in two different
kinds of quartets, namely,
\[
\{\pm z_j,\,\pm\=z_j\}_{j=1}^{J}\,,\quad
\{\pm \lambda_j,\,\pm\=\lambda_j\}_{j=1}^{J'}\,.
\]
Similarly to the IVP,
from~\eref{e:ALIBVPjump} and~\eref{e:IBVPALsystemRHP}
we find the following residue relations;
\bse
\label{e:ResRelation}
\bea
\Res_{z=z_j}\big[\_M_n\o{-,R}\big]=a_j\,\_M_n\o{-,L}(z_j)\,,\quad
\Res_{z=\=z_j}\big[\_M_n\o{+,L}\big]=\=a_j\, \_M_n\o{+,R}(\=z_j) \,,\\
\Res_{z=\lambda_j}\big[\_M_n\o{+,R}\big]=d_j\,\_M_n\o{+,L}(\lambda_j)\,,\quad
\Res_{z=\=\lambda_j}\big[\_M_n\o{-,L}\big]=\=d_j\,\_M_n\o{-,R}(\=\lambda_j)
\label{e:ResRelationlambda} \,,
\eea
\ese
where
\bea
\fl
a_j=K_j z_j^{2n}\e^{-2i\omega(z_j)t}\,,\quad
\=a_j=\=K_j\=z_j^{\,-2n}\e^{2i\omega(\=z_j)t}\,,\quad
d_j=\Lambda_j \lambda_j^{2n}\e^{-2i\omega(\lambda_j)t}\,, \quad
\=d_j=\=\Lambda_j \=\lambda_j^{-2n}\e^{2i\omega(\=\lambda_j)t}\,,
\nonumber\\
\fl
K_j=1/(\. a(z_j)\,b(z_j))\,,\quad
\Lambda_j= \nu B^*(\=\lambda_j)/(a(\lambda_j)\,\. d(\lambda_j))\,,\quad
\=K_j=(-z_j^*)^{-2}\nu K_j\,,\quad
\=\Lambda_j=(-\lambda_j^*)^{-2}\nu \Lambda_j\,.
\nonumber
\eea
and as customary $K_j$, $\Lambda_j$, $\=K_j$ and $\=\Lambda_j$
are referred to as norming constants.
Note that since $b(z)$ and $B^*(1/z^*)$ are odd
functions of~$z$ [cf.~\ref{s:asymptotics}],
the norming constants $K_j$ at $z=\pm z_j$ are identical,
and the same follows for $\Lambda_j$ at $z=\pm \lambda_j$.

The RHP is now solved by removing the singularities,
which is done by subtracting the residue contributions at the poles.
As usual, the solution of the RHP then has additional terms compared to
the case of no poles~\eref{e:ALsystemRHPsoln}, and is given by
\bea
\fl
\_M_n(z,t)= \_I
   + \frac1{2\pi i} \int_L \_M_n^+(\zeta,t){\_J_n(\zeta,t) \over \zeta-z} \,\d\zeta\,,
 + \sum_{j=1}^{2J}\bigg( \frac1{z-z_j}\,\Res_{z=z_j}[\_M_n^-(z)]
    + \frac1{z-\=z_j}\,\Res_{z=\=z_j}[\_M_n^+(z)] \bigg)
\nonumber \\
 + \sum_{j=1}^{2J'} \bigg(\frac1{z-\lambda_j}\,\Res_{z=\lambda_j}[\_M_n^+(z)]
    +\frac1{z-\=\lambda_j}\,\Res_{z=\=\lambda_j}[\_M_n^-(z)] \bigg)\,,
\label{e:ALRHPsolnRes}
\eea
where we defined $z_{j+J}= -z_j$ for $j=1,\dots,J$
and $\lambda_{j+J'}= -\lambda_j$ for $j=1,\dots,J'$.
From the asymptotic expansion of~\eref{e:ALRHPsolnRes}
and the symmetries~\eref{e:symmMn}, we then obtain
the reconstruction formula:
\bea
\fl
q_n(t)=-2\sum_{j=1}^J
z_j^{2n}\e^{-2i\omega(z_j)t}K_j\_M_{n+1,11}^-(z_j)-2\sum_{j=1}^{J'}
\lambda_j^{2n}\e^{-2i\omega(\lambda_j)t}\Lambda_j\_M_{n+1,11}^+(\lambda_j)
+\~q_n(t)\,,
\label{e:represnqnRes}
\eea
where $\~q_n(t)$ is given by~\eref{e:representationqn}.

In the reflectionless case with $\nu=-1$, we obtain the soliton
solution solving the following algebraic system of equations for
$\_M_{n+1,11}^-(z_j)$ and $\_M_{n+1,11}^+(\lambda_j)$:
\bea
\fl
\_M_{n,11}^-(z_l)=1+\sum_{j=1}^{J} \=a_j\,\bigg({1\over
z_l-\=z_l}-{1\over
z_l+\=z_j}\bigg)\,\_M_{n,12}^+(\=z_j)+\sum_{j=1}^{J'}
\=d_j\,\bigg({1\over z_l-\=\lambda_j}-{1\over
z_l+\=\lambda_j}\bigg)\,\_M_{n,12}^-(\=\lambda_j)
\nonumber\\
\fl
\_M_{n,12}^+(\=z_l)=\sum_{j=1}^{J} a_j\,\bigg({1\over
\=z_l-z_j}+{1\over
\=z_l+z_j}\bigg)\,\_M_{n,11}^-(z_j)+\sum_{j=1}^{J'}
d_j\,\bigg({1\over \=z_l-\lambda_j}+{1\over
\=z_l+\lambda_j}\bigg)\,\_M_{n,11}^+(\lambda_j)\,,
\nonumber\\
\fl
\_M_{n,11}^+(\lambda_l)=1+\sum_{j=1}^{J} \=a_j\,\bigg({1\over
\lambda_l-\=z_l}-{1\over
\lambda_l+\=z_j}\bigg)\,\_M_{n,12}^+(\=z_j)+\sum_{j=1}^{J'}
\=d_j\,\bigg({1\over \lambda_l-\=\lambda_j}-{1\over
\lambda_l+\=\lambda_j}\bigg)\,\_M_{n,12}^-(\=\lambda_j)\,,
\nonumber\\
\fl
\_M_{n,12}^-(\=\lambda_l)=\sum_{j=1}^{J} a_j\,\bigg({1\over
\=\lambda_l-z_j}+{1\over
\=\lambda_l+z_j}\bigg)\,\_M_{n,11}^-(z_j)+\sum_{j=1}^{J'}
d_j\,\bigg({1\over \=\lambda_l-\lambda_j}+{1\over
\=\lambda_l+\lambda_j}\bigg)\,\_M_{n,11}^+(\lambda_j)\,.
\nonumber
\eea
For a single quartet $\{\pm z_1,\,\pm\=z_1\}$,
the solution of the above system with $J=1$ and $J'=0$
yields the one-soliton solution of the IDNLS as
\be
q_n(t)= \e^{2i[(n+1)\beta+2wt+\phi]}\sinh(2\alpha)
  \sech[2((n+1)\alpha-v t-\delta)]\,,
\ee
where $z_1=\e^{\alpha+i\beta}$ and
\bea
w=\cosh(2\alpha)\cos(2\beta)-1\,,\quad
v=\sinh(2\alpha)\sin(2\beta)\,,
\nonumber\\
\delta=\frac12\log\big(\sinh(2\alpha)\big)-\frac12\log|K_1|+\log|z_1|\,,
\quad \phi=\frac\pi2-\arg z_1 + \frac12\arg K_1\,.
\nonumber
\eea
The soliton solution corresponding to a single quartet
$\{\pm\lambda_1,\,\pm\=\lambda_1\}$ has an identical
functional representation, which also coincides
with the well-known one-soliton solution in the IVP.

Note that the norming constants $\Lambda_j$
contain the unknown scattering datum $q_{-1}(t)$ through
the spectral functions $A(z,t)$ and $B(z,t)$.
In general, this datum must be obtained by solving a nonlinear system
of ODEs, as explained previously.
In the case of linearizable BCs, however,
$\Lambda_j$ can be expressed only in terms of known scattering data.
In particular, with $T=\infty$, the global relation implies
\be
\Lambda_j={f(1/\lambda_j)b^*(\lambda_j^*) \big/
  \big[a(\lambda_j)\.\Delta(1/\lambda_j)}\big]\,,
\label{e:Lambdaj}
\ee
where $\Delta(z)$ was defined in~\eref{e:ALDeltadef}.
This result can then be used in the
residue relations~\eref{e:ResRelationlambda}.
Equation~\eref{e:Lambdaj}
is a consequence of the fact that $d(z)$ and $\Delta(1/z)$
have the same set of zeros in $D_{+\#in}$,
which in turn can be easily proved considering the analyticity of
$A^*(1/z^*)$ and $B^*(1/z^*)$ with~\eref{e:ALABLin}.

%%%%%%%%%%%%%%%%%%%%%%%%%%%%%%%%%%%%%%%%%%%%%%%%%%%%%%%%%%%%%%%%%%%%%%%%%%%%%%
%%%%%%%%%%%%%%%%%%%%%%%%%%%%%%%%%%%%%%%%%%%%%%%%%%%%%%%%%%%%%%%%%%%%%%%%%%%%%%
%%%%%%%%%%%%%%%%%%%%%%%%%%%%%%%%%%%%%%%%%%%%%%%%%%%%%%%%%%%%%%%%%%%%%%%%%%%%%%
\section{Continuum: linear and nonlinear Schr\"odinger equations}
\label{s:continuum}

In order to compare the solution of the IBVP in the discrete case
to its continuum limit, and to appreciate the differences between
the method for discrete problems and its continuum counterpart,
here we briefly review the solution of IBVPs
for the linear Schr\"odinger (LS) equation and the
nonlinear Schr\"odinger (NLS) equation:
\bea
i\partialderiv qt + \partialderiv[2]qx - 2\nu |q|^2q=0\,
\label{e:NLS}
\eea
(with $\nu=0,\pm1$ denoting respectively the linear, defocusing and focusing
cases),
to which~\eref{e:DLS} and~\eref{e:IDNLS} %in sections~\ref{s:DLS} and~\ref{s:IDNLS}
reduce to in the limit $h\to0$.
%
%Our presentation will be concise.
Note that,
even though the IVP for~\eref{e:NLS} was solved in the early days
of integrable systems for both vanishing~\cite{JETP34p62}
and nonzero~\cite{JETP37p823} BCs,
the IBVP on the half line was solved only recently~\cite{NLTY18p1771}.
Also, even though the IVP for the
vector generalization of~\eref{e:NLS} was also solved
early on in the case of vanishing BCs~\cite{JETP38p248},
the analogue problem with nonzero BCs was also
only recently solved~\cite{JMP47p63508}.

%%%%%%%%%%%%%%%%%%%%%%%%%%%%%%%%%%%%%%%%%%%%%%%%%%%%%%%%%%%%%%%%%%%%%%%%%%%%%%
\paragraph{Linear Schr\"odinger equation: IVP and IBVP via Fourier methods.}

%As in the discrete case,
Consider first the initial value problem for the LS equation
with $x\in\Real$, $t>0$ and $q(x,0)$ given.
For simplicity we assume that $q(x,0)$ belongs to the Schwartz class,
which we denote by ${\cal S}(\Real)$.
The IVP is trivially solved using the Fourier transform pair, defined as
\bea
\^q(k,t)= \int_{\!\!-\infty}^{\,\,\infty} \e^{-ikx} q(x,t)\,\d x\,,
\qquad
q(x,t)= \frac1{2\pi}\,\,\int_{\!\!-\infty}^{\,\,\infty} \e^{ikx}\^q(k,t)\,\d k\,.
\label{e:FTpair}
\eea
Use of~\eref{e:FTpair} yields the solution of the IVP as
\be
q(x,t)=
  \frac1{2\pi}\,\, \int_{\!\!-\infty}^{\,\,\infty} \e^{i(kx - k^2t)}
  \^q(k,0)\,\d k\,.
\label{e:LSIVPsoln}
\ee
Now consider the IBVP for the LS equation
on the half line with Dirichlet BCs;
i.e., $x>0$, $t>0$ and %$q(x,0)$ and $q(0,t)$ given,
with $q(x,0)\in{\cal S}(\Real^+)$ and $q(0,t)\in{\cal C}(\Real^+)$ given.
Employing the sine transform pair
%\bse
\bea
\^q\o{s}(k,t)= \int_0^{\,\,\infty} \sin(kx) q(x,t)\,\d x\,,
\qquad
q(x,t)= \frac2\pi\,\,\int_0^{\,\,\infty} \sin(kx) \^q\o{s}(k,t)\,\d k\,,
\label{e:FST}
\eea
%\ese
yields the solution of the IBVP as
\bea
q(x,t)=
  \frac2\pi\,\, \int_0^{\,\,\infty} \e^{-ik^2t}\sin(kx)\,\^q\o{s}(k,0)\,\d k
  + \frac1\pi\,\,
 \int_0^{\,\,\infty} \e^{-ik^2t}\sin(kx) \^g(k,t)\,\d k\,,
\label{e:LSIBVPsinetransform}
\\
\noalign{\noindent  where}
\^g(k,t)=
  2ik \int_0^t \e^{ik^2t'}q(0,t')\,\d t'\,.
\eea

%%%%%%%%%%%%%%%%%%%%%%%%%%%%%%%%%%%%%%%%%%%%%%%%%%%%%%%%%%%%%%%%%%%%%%%%%%%%%%
\subsection{Linear Schr\"odinger equation: IVP and IBVPs via spectral methods}
\label{s:LSIBVP}

An algorithmic method to obtain the Lax pair of
linear PDEs was given in Ref.~\cite{IMA67p559}.
However, one can also obtain the Lax pair for the LS equation
via the linear limit of the Lax pair of the NLS equation,
namely~\eref{e:NLSLP}.
Let $\_Q=O(\epsilon)$
and take $\Phi(x,t,k)=\@v(x,t,k)$ to be a two-component vector.
To leading order it is $\@v(x,t,k)= \esh{i(kx-2k^2t)}\@v_o$,
where $\@v_o=(v_{1,o},v_{2,o})^t$ is an arbitrary constant vector.
Choosing $v_{2,o}=1$ and substituting into the RHS
of~\eref{e:NLSLP1} then yields the following equations
for $\mu(x,t,k)= \e^{i(kx-2k^2t)}v_1(x,t,k)$ up to $O(\epsilon^2)$ terms:
\bea
\mu_x - ik'\mu = q\,,
\qquad
\mu_t + ik'^2\mu = iq_x-k'q\,,
\label{e:LSLP}
\eea
where $k'=2k$.
One can now verify that enforcing the compatibility
of~\eref{e:LSLP} yields the LS equation.
Hereafter, for convenience, we will omit the primes.

%%%%%%%%%%%%%%%%%%%%%%%%%%%%%%%%%%%%%%%%%%%%%%%%%%%%%%%%%%%%%%%%%%%%%%%%%%%%%%
\subsubsection*{Initial value problem.}

Introduce a modified eigenfunction
$\psi(x,t,k)=\e^{-i(kx-k^2t)}\mu(x,t,k)$,
which satisfies the simplified Lax pair
\[
\psi_x= \e^{-i(kx-k^2t)}q\,,\qquad
\psi_t= \e^{-i(kx-k^2t)}\big(iq_x - k q\big)\,.
\]
It is then easy to obtain the solutions of~\eref{e:LSLP}
which decay as $x\to\pm\infty$ respectively as:
%\bse
\bea
\mu\o1(x,t,k)= \int_{\!\!-\infty}^x \e^{ik(x-x')}q(x',t)\,\d x'\,,
\qquad
\mu\o2(x,t,k)= -\int_x^\infty \e^{ik(x-x')}q(x',t)\,\d x'\,.
\nonumber\\[-1ex]
\label{e:LSIVPmusoln2}
\eea
%\ese
Note that $\mu\o{1,2}(x,t,k)$ are analytic for $\Im k\gl0$, respectively
Also, on $\Im k=0$ it is
\bea
\mu\o1(x,t,k) - \mu\o2(x,t,k)=
\e^{ikx}\^q(k,t)= \e^{i(kx-k^2t)}\^q(k,0)\,,
\label{e:RHP0}
\\
\noalign{\noindent where $\^q(k,t)$ is the Fourier transform of
$q(x,t)$:} \^q(k,t)= \int_{\!\!-\infty}^{\infty\!\!}
\e^{-ikx'}q(x',t)\,\d x'\,. \eea Also, $\mu\o{1,2}(x,t,k)=O(1/k)$ as
$k\to\infty$ in their respective half planes. Thus~\eref{e:RHP0}
defines a scalar RHP which is trivially solved via the standard
Cauchy projectors $P^\pm$ over the real line: \be \mu(x,t,k)=
\frac1{2\pi i}\, \int_{\!\!-\infty}^{\infty\!\!}
  \e^{i(k'x-k'^2t)}\,\frac{\^q(k',0)}{k'-k}\,\d k'\,.
\label{e:LSIVPmusoln}
\ee
Inserting~\eref{e:LSIVPmusoln} into~\eref{e:LSLP} then
yields~\eref{e:LSIVPsoln} as the solution of the IVP.

%%%%%%%%%%%%%%%%%%%%%%%%%%%%%%%%%%%%%%%%%%%%%%%%%%%%%%%%%%%%%%%%%%%%%%%%%%%%%%
\subsubsection*{Initial-boundary value problems.}

We now consider the IBVP for the LS equation on the half line.
Define simultaneous solutions of both the $x$-part and the
$t$-part of the Lax pair:
\[
%\fl
\mu\o{j}(x,t,k)= \int_{(x_j,t_j)}^{(x,t)}\e^{ik(x-x')-ik^2(t-t')}
  \big[q(x',t')\,\d x' + \big(iq_{x'}(x',t')-k q(x',t')\big)\,\d t'\big]\,.
\]
In particular,
consider the three eigenfunctions $\mu\o{j}(x,t,k)$, $j=1,2,3$,
defined by the choices $(x_1,t_1)=(0,0)$,\, $(x_2,t_2)=(\infty,t)$ and
$(x_3,t_3)=(0,T)$:
\bse
\label{e:LSLPsolutions}
\bea
%\fl
\mu\o1(x,t,k)= \int_0^x \e^{ik(x-x')}q(x',t)\,\d x'
  + \int_0^t \e^{ikx-ik^2(t-t')} \big(iq_x(0,t')-k q(0,t')\big)\,\d t'\,,
\nonumber\\[-2ex]
\label{e:LSLPsolutions1}
\\
\mu\o2(x,t,k)= - \int_x^\infty\e^{ik(x-x')}\, q(x',t)\,\d x'\,,
\label{e:LSLPsolutions2}
\\
%\fl
\mu\o3(x,t,k)= \int_0^x \e^{ik(x-x')}q(x',t)\,\d x'
  - \int_t^T \e^{ikx-ik^2(t-t')} \big(iq_x(0,t')-k q(0,t')\big)\,\d t'\,.
\nonumber\\[-2ex]
\label{e:LSLPsolutions3}
\eea
\ese
Note that
$\mu\o2$ coincides with the eigenfunction in the IVP.
As for $\mu\o1$ and $\mu\o3$, they are entire functions of~$k$.
These eigenfunctions have the following domains of analyticity and boundedness:
\bea
\mu\o1\!:~  k\in\Complex_\II\,,  %\Re\,k\le0\,,
\qquad
\mu\o2\!:~  k\in\Complex_{\III+\IV}\,,
\qquad
\mu\o3\!:~  k\in\Complex_\I\,, %\Re\,k\ge0\,,
\label{e:muregions}
\eea
%\ese
where $\Complex_{\III+\IV}$ is the lower-half plane.
The two jumps on $\Im\,k=0$ and the jump on
$\Re\,k=0$ (with $\Im\,k\ge0$) then define a scalar RHP:
\bse
\label{e:LSIBVPmujumps}
\bea
%\fl
\mu\o1(x,t,k) - \mu\o3(x,t,k)= \e^{ikx-ik^2t}\,\^F(k,T)
&\Re\,k=0~\wedge~\Im\,k\ge0\,,
\nonumber\\[-1ex]
\label{e:mujumps13}
\\
%\fl
\mu\o1(x,t,k) - \mu\o2(x,t,k) = \e^{ikx-ik^2t}\,\^q(k,0)\,,
&\Im\,k=0~\wedge~\Re\,k\le0\,,
\nonumber\\[-1ex]
\label{e:mujumps12}
\\
%\fl
\mu\o3(x,t,k) - \mu\o2(x,t,k) = \e^{ikx-ik^2t}\,\big(\^q(k,0) - \^F(k,T)\big)\,,~
&\Im\,k=0~\wedge~\Re\,k\ge0\,,
\nonumber\\[-1ex]
\label{e:mujumps23}
\eea
\ese
where $\^F(k,t)= i\^f_1(k,t)-k\^f_0(k,t)$, and with
\be
\^q(k,t)= \int_0^{\infty\!\!} \e^{-ikx}q(x,t)\,\d x\,,
\qquad
\^f_n(k,t)= \int_0^t \e^{ik^2t'} \,\partial^n_x q(x,t')|_{x=0}\,\d t'\,.
\label{e:LSktransforms}
\ee
The one-sided Fourier transform $\^q(k,t)$
is analytic and bounded for $\Im\,k<0$,
while the transforms $\^f_n(k,t)$ of the boundary data are
entire, and are bounded for $\Im\,k^2\ge0$.
Moreover, $\^q(k,t)\to0$ as $k\to\infty$ with $\Im\,k<0$,
and $\^f_n(z,t)\to0$ as $k\to\infty$ with $\Im\,k^2<0$.
The solution of the RHP defined by~\eref{e:LSIBVPmujumps} is thus given by
\bea
\mu(x,t,k)= \frac1{2\pi i}\, \int_{\!\!-\infty}^{\infty\!\!}
    \e^{ik'x-ik'^2t}\,\frac{\^q(k',0)}{k'-k}\,\d k'
  - \frac1{2\pi i} \int_{\partial\Complex_\I} \e^{ik'x-ik'^2t}
     \frac{F(k',T)} %{i\^f_1(k',T)-k'\^f_0(k',T)}
     {k'-k}\,\d k'\,.
\nonumber\\[-1.2ex]
\label{e:IBVPRHPsolution}
\eea
Inserting~\eref{e:IBVPRHPsolution} into the first of~\eref{e:LSLP}
then yields the reconstruction formula:
\be
q(x,t)= \frac1{2\pi}\, \int_{\!\!-\infty}^{\infty\!\!}
    \e^{ikx-ik^2t}\^q(k,0)\,\d k
  - \frac1{2\pi} \int_{\partial\Complex_\I} \e^{ikx-ik^2t}
     %\big[i\^f_1(k,T)-k\^f_0(k,T)\big]
     F(k,T) \,\d k\,.
\label{e:IBVPreconstruction}
\ee
As in the discrete case, \eref{e:IBVPreconstruction} still depends
on the unknown boundary datum $q_x(0,t)$ via its transform in $F(k,t)$.
%Our last task is to obtain the Dirichlet-to-Neumann map,
%namely to determine the unknown boundary datum in terms of the
%given one.
Integrating~\eref{e:LSLP} from $(0,0)$ to $(0,T)$, $(\infty,T)$, $(\infty,0)$
and back yields the global relation as
\[
%\fl
\int_0^T \e^{ik^2t} \big(iq_x(0,t)-k\,q(0,t)\big)\d t
  + \e^{ik^2T}\int_0^{\infty\!\!} \e^{-ikx}q(x,T)\,\d x =
  \int_0^{\infty\!\!} \e^{-ikx}q(x,0)\,\d x\,,
%\label{e:LSGR3}
\]
which holds for $\Im\,k\le0\,\wedge\,\Im\,k^2\le0$,
i.e., $k\in\Complex_\III$.
In terms of the spectral data:
\bse
\be
i\^f_1(k,T)-k\^f_0(k,T) + \e^{ik^2T}\^q(k,T)= \^q(k,0)\,,
\qquad\forall k\in\=\Complex_\III\,.
\label{e:LSGR4}
\ee
Using the the transformation $k\to-k$, which leaves $\^f_n(k,t)$
invariant, from~\eref{e:LSGR4} we obtain
\be
i\^f_1(k,T)+k\^f_0(k,T) + \e^{ik^2T}\^q(-k,T)= \^q(-k,0)\,
\qquad\forall k\in\=\Complex_\I\,.
\label{e:LSGR4a}
\ee
\ese
We then solve for $\^f_1(k,T)$ and insert the result
in~\eref{e:IBVPreconstruction}.
[The first term in the RHS of~\eref{e:LSGR4a}
yields a zero contribution to the solution.]
Thus, the solution of the IBVP is given by
\bea
%\fl
q(x,t)= \frac1{2\pi}\, \int_{\!\!-\infty}^{\infty\!\!}
    \e^{ikx-ik^2t}\^q(k,0)\,\d k
  - \frac1{2\pi} \int_{\partial\Complex_\I} \e^{ikx-ik^2t}
     \big[\^q(-k,0)-2k\^f_0(k,T)\big]\,\d k\,.
\nonumber\\[-1ex]
\label{e:IBVPLSreconst} \eea
Note that one can replace $\^f_0(k,T)$
with $\^f_0(k,t)$. Also, the second integrand
in~\eref{e:IBVPLSreconst} is analytic and bounded for
$\Im\,k\ge0\,\wedge\,\Im\,k^2\le0$. Thus, one can deform the
integration contour on the second integral onto the real $k$-axis
and recover the sine transform
solution~\eref{e:LSIBVPsinetransform}. Unlike sine/cosine transform
approaches, however, the present method can be applied to solve
IBVPs with more complicated BCs, as we show next.

\paragraph{Robin BCs.}

Consider the IBVP for LS equation with Robin BCs:
\be
\alpha q(0,t)+q_x(0,t)= h(t)\,,
\label{e:LSRobinBC}
\ee
with $h(t)$ given and
where $\alpha\in\Complex$ is a nonzero but otherwise arbitrary constant.
In a similar way as shown in~\ref{s:Robin} for the discrete case,
one obtains \cite{PRSLA453p1411,IMA67p559}
\bse
\bea
\^F(k,t)= \frac{\^G(k,t)}{k-i\alpha}
  + \frac{k+i\alpha}{k-i\alpha}\e^{ik^2t}\^q(-k,t)\,,
\label{e:LSRobinF}
\\
\noalign{\noindent where} \^G(k,t)= 2ik\^h(k,t)-(k+i\alpha)\^q(-k,0)
\eea \ese contains the known portion of~\eref{e:LSRobinF} and where
$\^h(k,t)$ is defined according to~\eref{e:LSktransforms}. Then,
again following similar steps as in the discrete case, one obtains
the solution of the IBVP as:
\be \fl
q(x,t)=
  \frac1{2\pi}\, \int_{\!\!-\infty}^{\infty\!\!} \e^{ikx-ik^2t}\^q(k,0)\,\d k
  - \frac1{2\pi} \int_{\partial \Complex_\I}\e^{ikx-ik^2t} {\^G(k,t) \over k-i\alpha} \,\d k
  + i\nu_\alpha \e^{-\alpha x+i\alpha^2t} \^G(i\alpha,t) \,,
\label{e:LSRobinBCsoln}
\ee
where $\nu_\alpha=1$ for
$-\pi/2<\arg\alpha<0$, $\nu_\alpha=1/2$ for $\arg\alpha=0,-\pi/2$
and $\nu_\alpha=0$ for $0<\arg\alpha<3\pi/2$, and where the integral
along $\partial \Complex_\I$ is to be taken in the principal value
sense when $\arg\alpha=0,-\pi/2$. (The last term in the RHS
of~\eref{e:LSRobinBCsoln} is missing in
Refs.~\cite{PRSLA453p1411,IMA67p559}. One can easily show, however,
that without this term $q(x,t)$ does \textit{not} satisfy the BC at
$x=0$.)

%%%%%%%%%%%%%%%%%%%%%%%%%%%%%%%%%%%%%%%%%%%%%%%%%%%%%%%%%%%%%%%%%%%%%%%%%%%%%%
\subsection{Nonlinear Schr\"odinger equation: initial value problem}

%We now move on to the nonlinear Schr\"odinger (NLS) equation~\eref{e:NLS}.
As in the linear case we assume that $q(x,0)\in{\cal S}(\Real)$.
Recall that the Lax pair for the NLS equation~\eref{e:NLS}
is given by~\eref{e:NLSLP}
with $p(x,t)=\nu q^*(x,t)$.
For the present purposes, we consider $\Phi(x,t,k)$ to be a $2\times2$ matrix.

\paragraph{Analyticity.}
Introduce a modified eigenfunction
which has a well-defined limit as $x\to\pm\infty$:
%\unskip\footnote{This definition differs from the usual one~\cite{APT2003}
%by presence of the factor $\e^{-2ik^2t}$, but it is used for consistency
%with the IBVP, discussed in the next section.}
\be
\mu(x,t,k)=\Phi(x,t,k)\,\e^{-i\theta(x,t,k)\sigma_3},
\label{e:NLSmudef}
\ee
with $\theta(x,t,k)=kx-2k^2t$.
Note $\mu(x,t,k)$ satisfies the following modified Lax pair:
%\bse
\bea
\mu_x - ik[\sigma_3,\mu] = \_Q\mu\,,
%\label{e:NLSLP1m}
\qquad
\mu_t + 2ik^2[\sigma_3,\mu]= \_H\mu\,.
%\label{e:NLSLP2m}
\label{e:NLSLPm}
\eea
%\ese
Then, letting $\mu(x,t,k)= \esh{i\theta}\Psi(x,t,k)$,
we obtain the simplified Lax pair:
$\Psi_x= \esh{-i\theta}(\_Q)\,\Psi$
and
$\Psi_t= \esh{-i\theta}(\_H)\,\Psi\,$.
We then define the Jost eigenfunctions as the solutions of~\eref{e:NLSLPm}
that reduce to the identity as $x\to\pm\infty$:
\bse
\label{e:JostNLS}
\bea
\mu\o1(x,t,k)=
  \_I + \int_{-\infty}^x \esh{ik(x-x')}\big(\_Q(x',t)\mu\o1(x',t,k)\big)\,\d x'\,,
\\[-0.4ex]
\mu\o2(x,t,k)=
  \_I - \int_x^\infty \esh{ik(x-x')}\big(\_Q(x',t)\mu\o2(x',t,k)\big)\,\d x'\,.
\label{e:JostNLSb}
\eea
\ese
We have the following regions of analyticity and boundedness~\cite{APT2003}:
%\bse
\bea
\mu\o{1,L},~\mu\o{2,R}\!:\quad &\Im k<0\,,\qquad
\mu\o{1,R},~\mu\o{2,L}\!:\quad &\Im k>0\,,
\nonumber
\eea
where $\mu\o{j}(x,t,k)=\big(\mu\o{j,L}\,,\mu\o{j,R}\big)$,
as before.
The analyticity properties of
$\Phi\o{j}(x,t,k)=\mu\o{j}(x,t,k)\,\e^{i\theta\sigma_3}$, $j=1,2$,
follow trivially. % from those of $\mu\o1$ and $\mu\o2$, namely:
%$\Phi\o{1,L}(z,t,k)$ and $\Phi\o{2,R}(z,t,k)$ are analytic for $\Im k<0$,
%while $\Phi\o{1,R}(z,t,k)$ and $\Phi\o{2,L}(z,t,k)$ are analytic for $\Im k>0$.
%namely $\Phi\o1=(\Phi_1^-,\Phi_1^+)$ and $\Phi\o2=(\Phi_2^+,\Phi_2^-)$.

\paragraph{Scattering matrix.}
Note $\det\Phi\o{j}=\det\mu\o{j}=1$ for $j=1,2$.
Thus $\Phi\o1$ and $\Phi\o2$ are
both fundamental solutions of~\eref{e:NLSLP} $\forall k\in\Real$.
Hence $\Phi\o1(x,t,k)=\Phi\o2(x,t,k)\,\_A(k)$,
where $\_A(k)$ is the scattering matrix.
Equivalently,
\be
\mu\o1(x,t,k)= \mu\o2(x,t,k)\,\esh{i\theta}\_A(k)\,.
\label{e:scatteringNLS}
\ee
Note that $\_A(k)$ is indeed independent of time,
and $\det \_A(k)=1$.
Moreover,
\be
\_A(k)= \_I + \int_{-\infty}^\infty \esh{-i(kx-2k^2t)}\big(\_Q(x,t)\mu\o1(x,t,k)\big)\,\d x\,,
\label{e:NLFT}
\ee
and
\bse
\label{e:NLSwronskians}
\bea
a_{11}(k)=\Wr(\Phi\o{1,L},\Phi\o{2,R})\,,\quad
a_{12}(k)=\Wr(\Phi\o{1,R},\Phi\o{2,R})\,,
\\
a_{21}(k)= -\Wr(\Phi\o{1,L},\Phi\o{2,L})\,,\quad
a_{22}(k)=-\Wr(\Phi\o{1,R},\Phi\o{2,L})\,.
\eea \ese
Thus,
$a_{11}(k)$ and $a_{22}(k)$ can be analytically continued
respectively on $\Im k<0$ and $\Im k>0$, but $a_{12}(k)$ and
$a_{21}(k)$ are nowhere analytic, in general.

\paragraph{Symmetries.}
When $p(x,t)=\nu q^*(x,t)$, with $\nu=\pm1$, the scattering
problem~\eref{e:NLSLP} admits an involution expressed via the matrix
$\sigma_\nu$ in~\eref{e:sigmanudef}: if $\Phi(x,t,k)$ is a solution
of~\eref{e:NLSLP1}, so is
\be \Phi'(x,t,k)= \sigma_\nu
\Phi^*(x,t,k^*)\,. \label{e:NLSsymmetriesPhi} \ee
Comparing the
behavior of the Jost eigenfunctions as $x\to\pm\infty$ we then have
% from the asymptotic behavior as $x\to\pm\infty$ one obtains that
%\bse
\bea
%\fl
\Phi\o{j,L}(x,t,k)=\sigma_\nu\big(\Phi\o{j,R}(x,t,k^*)\big)^*\,,\quad
\Phi\o{j,R}(x,t,k)=\nu\sigma_\nu\big(\Phi\o{j,L}(x,t,k^*)\big)^*\,,%\quad
\nonumber
\\[-1ex]
%\Phi_2^+(x,t,k)=\sigma_\nu\big(\Phi_2^-(x,t,k^*)\big)^*\,,\qquad
%\Phi_2^-(x,t,k)=\nu\sigma_\nu\big(\Phi_2^+(x,t,k^*)\big)^*\,,
\label{e:NLSsymmetriesPhij}
\eea
for $j=1,2$\,.
Hence the following relations hold for the elements of the scattering matrix
$\_A(k)$:
\be
a_{22}(k)= a_{11}^*(k^*)\,,\qquad
a_{21}(k)= \nu\,a_{12}^*(k^*)\,.
\label{e:NLSsymmetries}
\ee
%\ese
Note that, since $\det\_A(k)=1$, \eref{e:NLSsymmetries} imply
$|a_{11}(k)|^2-\nu|a_{12}(k)|^2=1$ $\forall k\in\Real$.

\paragraph{Asymptotics.}
The asymptotics of the Jost solutions as $k\to\infty$
in their half planes is:
\bse
\label{e:asympNLS}
\bea
\mu\o1(x,t,k)= \_I - \frac1{2ik}\sigma_3\_Q
  + \frac1{2ik}\,\sigma_3\int_{-\infty}^x q(x',t)p(x',t)\,\d x'  + O(1/k^2)\,,\\
\mu\o2(x,t,k)= \_I - \frac1{2ik}\sigma_3\_Q
  - \frac1{2ik}\,\sigma_3\int_x^\infty q(x',t)p(x',t)\,\d x' + O(1/k^2)\,.
\label{e:asympNLSb}
\eea
\ese
%These relations imply
%$q(x,t)=-\lim_{k\to\infty}2ik\big(\mu\o{j}(x,t,k)\big)_{12}$\,,
%with $\Im k>0$ for $j=1$ and $\Im k <0$ for $j=2$.
Moreover, from~\eref{e:NLSwronskians} and
%recalling that $a_{22}(k)=\Wr(\Phi\o{1,L},\Phi\o{2,R})$, from
\eref{e:asympNLS} one also obtains
\bea
a_{22}(k)= 1 - \frac1{2ik}\int_{-\infty}^\infty q(x,t)p(x,t)\,\d x + O(1/k^2)\,.
\label{e:asympNLSa}
\eea

\paragraph{Inverse problem.}
The inverse problem is the RHP defined by~\eref{e:scatteringNLS} for $k\in\Real$:%
%\bse
\bea
\_M^-(x,t,k)= \_M^+(x,t,k)(\_I-\_J(k,t))\,,
\label{e:NLSRHP}
\\[0.2ex]
\noalign{\noindent where the matrix-valued sectionally meromorphic functions are}
\nonumber
\\[-2ex]
\fl
\_M^+(x,t,k)= \bigg(\mu\o{2,L}(x,t,k)\,,\frac{\mu\o{1,R}(x,t,k)}{a_{22}(k)}\bigg)\,,
\qquad
\_M^-(x,t,k)= \bigg(\frac{\mu\o{1,L}(x,t,k)}{a_{11}(k)}\,,\,\mu\o{2,R}(x,t,k)\bigg)\,,
\nonumber
\\
\noalign{\noindent the jump matrix is}
%J(k,t)= \begin{pmatrix}0&-\e^{-2ikx}\rho_2(k,t)\\
%  \e^{2ikx}\rho_1(k,t)&\rho_1(k,t)\rho_2(k,t)\end{pmatrix},
\_J(k,t)= \begin{pmatrix}\rho_1(k)\rho_2(k)&\e^{2i\theta}\rho_2(k)\\
  -\e^{-2i\theta}\rho_1(k)&0\end{pmatrix},
\nonumber
\\[0.2ex]
\noalign{\noindent and the reflection coefficients, defined $\forall k\in\Real$, are}
\rho_1(k)= {a_{21}(k)}/{a_{11}(k)} \,, \qquad
\rho_2(k)= {a_{12}(k)}/{a_{22}(k)}\,.
\nonumber
\eea
%\ese
Of course \eref{e:NLSsymmetries} imply $\rho_1(k)=
\nu\rho_2^*(k^*)$ when $p(x,t)= \nu q^*(x,t)$.
In the absence of a discrete spectrum [i.e., if $a_{11}(k)\ne0$
$\forall \Im k<0$ and $a_{22}(k)\ne0$ $\forall\Im k>0$]
%for all $k$ respectively in the lower-half and upper-half plane]
the matrix functions $\_M^\pm(x,t,k)-\_I$ are sectionally analytic
in their respective half planes, and they vanish as $k\to\infty$.
Therefore the RHP~\eref{e:NLSRHP} is solved via the Cauchy
projectors~$P^\pm$, as for the linear case:
\be
\_M^+(x,t,k)= \_I + \frac1{2\pi i}\int_{-\infty}^\infty \_M^+(x,t,k')\frac{\_J(k',t)}{k'-k}\d k'\,.
\label{e:NLSRHPsoln}
\ee
The asymptotic behavior of $\_M(x,t,k)$ as $k\to\infty$ is easily obtained
from~\eref{e:NLSRHPsoln}: for $\Im\,k>0$,
\be
\_M^+(x,t,k)= \_I - \frac1{2i\pi k}\int_{-\infty}^\infty \_M^+(x,t,k')\_J(k',t)\d k' + O(1/k^2)\,.
\label{e:RHPasympNLS}
\ee
Comparing the $(1,2)$-components of~\eref{e:RHPasympNLS}
and~\eref{e:asympNLS} then yields the
reconstruction formula:
\be
q(x,t)= \frac1\pi \int_{-\infty}^\infty \e^{2i(kx-2k^2t)}\rho_2(k)
 \big(\mu\o2(x,t,k)\big)_{11}\,\d k\,.
\label{e:NLSreconstruction}
\ee

\paragraph{Linear limit.}
%Equations~\eref{e:NLFT} and~\eref{e:NLSreconstruction} imply that
%$\_A(k)$ is the nonlinear analogue of the Fourier transform.
%Indeed,
If $\_Q(x,t)=O(\epsilon)$
%from \eref{e:NLSLPm} and the requirement that $Q(x,t)\to0$
%together with all of its derivatives as $x\to\pm\infty$,
one has
$\mu(x,t,k)=\_I+O(\epsilon)$ and, to $O(\epsilon)$,
%$\_A= \_I+\epsilon\,\_A\o1$, where, up to $O(\epsilon^2)$ terms,
\[
\_A(k)= \_I + \int_{-\infty}^\infty \esh{-i(kx-2k^2t)}\_Q(x,t)\,\d
x\,.
\]
From here and~\eref{e:NLSreconstruction} one then obtains, to $O(\epsilon)$,
%(using the fact that $\_A(k)$ is independent on $t$)
%\bse
\bea
q(x,t)= \frac1\pi \int_{-\infty}^\infty
e^{2i(kx-2k^2t)}\rho_2(k)\,\d k\,,
% + O(\epsilon^2)\,,
\qquad
%\\
\rho_2(k)= \int_{-\infty}^\infty \e^{-2ikx}q(x,0)\,\d x\,,
%+ O(\epsilon^2)\,,
\nonumber
\eea
%\ese
which, with the familiar rescaling $k'=2k$, coincide with the Fourier
transform pair~\eref{e:FTpair}.

%%%%%%%%%%%%%%%%%%%%%%%%%%%%%%%%%%%%%%%%%%%%%%%%%%%%%%%%%%%%%%%%%%%%%%%%%%%%%%
\subsection{Nonlinear Schr\"odinger equation: initial-boundary value problem}
\label{s:IBVPNLS}

We now discuss the IBVP for the NLS equation~\eref{e:NLS} on the half line.
%following Ref.~\cite{NLTY18p1771}.
As in the linear case, we assume $q(x,0)\in{\cal S}(\Real^+)$ and
$q(0,t)\in{\cal C}(\Real^+)$.

\paragraph{Eigenfunctions and analyticity.}

Introduce three Jost eigenfunctions as the solutions of
\eref{e:NLSLPm} that reduce to the identity respectively at
$(x,t)=(0,0)$, $(x,t)\to(\infty,t)$ and $(x,t)=(0,T)$:
\par\kern-2\medskipamount
\bse
\label{e:NLSLPsolutions}
\bea
\fl
\mu\o1(x,t,k)= \_I + \int_0^x \esh{ik(x-x')}\big(\_Q(x',t)\mu\o1(x',t,k)\big)\,\d x'
\nonumber\\[-2ex]\kern8em{ }
  + \int_0^t \esh{i[kx-2k^2(t-t')]}\big(\_H(0,t',k)\mu\o1(0,t',k)\big)\,\d t'\,,
\label{e:NLSIBVPmu1}
\\
\fl
\mu\o2(x,t,k)= \_I - \int_x^\infty\esh{ik(x-x')}\big(\_Q(x',t)\mu\o2(x',t,k)\big)\,\d x'\,,
\\
\fl
\mu\o3(x,t,k)= \_I + \int_0^x \esh{ik(x-x')}\big(\_Q(x',t)\mu\o3(x',t,k)\big)\,\d x'
\nonumber\\[-2ex]\kern8em{ }
  - \int_t^T \esh{i[kx-2k^2(t-t')]}\big(\_H(0,t',k)\mu\o3(0,t',k)\big)\,\d t'\,.
\eea
\ese
Note that $\mu\o1(x,t,k)$ and $\mu\o3(x,t,k)$ are entire functions
of~$k$, while $\mu\o2(x,t,k)$ coincides with~\eref{e:JostNLSb}.
Moreover,
\eref{e:NLSLPsolutions} imply the
following domains of analyticity and boundedness:
%\bse
\bea
\mu\o{1,L}:~~ \Complex_\III\,,\qquad
\mu\o{1,R}:~~ \Complex_\II\,,\qquad
\mu\o{3,L}:~~ \Complex_\IV\,,\qquad
\mu\o{3,R}:~~ \Complex_\I\,,
\nonumber
\\
\mu\o{2,L}:~~ \Complex_{\I+\II}\,,\qquad\!\!
\mu\o{2,R}:~~ \Complex_{\III+\IV}\,.
\nonumber
\eea
%\ese
%where, as before, the subscripts $\I,\dots,\IV$ indicate the four
%quadrants of the complex $k$-plane, and the subscripts $L$ and $R$
%indicate the first and second column of the eigenfunctions.

\paragraph{Scattering matrices.}
We still have $\det\Phi\o{j}(x,t,k)=1$ for
all $x,t\in\Real^+$ and for all $j=1,2,3$. Hence the matrices
$\Phi\o{j}(x,t,k)$, $j=1,2,3$ are three fundamental solutions of the
Lax pair~\eref{e:NLSLP}, and they must be proportional to each
other. In terms of the modified eigenfunctions:
\bse
\label{e:NLSIBVPscattering}
\bea
\mu\o2(x,t,k)= \mu\o1(x,t,k)\,\esh{i(kx-2k^2t)}\_s(k)\,,
%&&\kern-6em\!\Re\,k=0~\vee\,\Im\,k=0\,,
\label{e:NLSIBVPscattering1}
\\
\mu\o3(x,t,k)= \mu\o1(x,t,k)\,\esh{i(kx-2k^2t)}\_S(k,T)\,.
%&&\kern-6em k\in\Complex\,.
\label{e:NLSIBVPscattering2}
\eea
\ese
Note that the first column of~\eref{e:NLSIBVPscattering1} is
defined $\forall k\in\=\Complex_{\I+\II}$, the second column
$\forall k\in\=\Complex_{\III+\IV}$ and~\eref{e:NLSIBVPscattering2}
holds $\forall k\in\Complex$.
Also, $\det\,\_s(k)=\det\,\_S(k,T)=1$.
The scattering matrices $\_s(k)$
and~$\_S(k,T)$ are obtained from the boundary values of the
eigenfunctions, namely, $\forall k\in\Complex$,
\be
\_s(k)= \mu\o2(0,0,k)\,,\qquad \_S(k,T)=
\big(\esh{2ik^2T}\mu\o1(0,T,k)\big)^{-1}\,.
\label{e:NLSIBVPscattering3}
\ee
Then, from~\eref{e:NLSLPsolutions} we have the following integral
representations of the scattering matrices:
\bse
\label{e:scattdata}
\bea
\_s(k)= \_I - \int_0^\infty \esh{-ikx}\big(\_Q(x,0)\mu\o2(x,0,k)\big)\,\d x\,,
\\[-1ex]
\_S^{-1}(k,T)= \_I + \int_0^T \esh{2ik^2t}\big(\_H(0,t,k)\mu\o1(0,t,k)\big)\,\d t\,.
\label{e:scattdataS}
\eea
\ese
%The analyticity and boundedness properties of $\mu\o2(0,0,k)$
%are the same as those of $\mu\o2(x,t,k)$.
These imply that:
$\_s_L(k)$ and $\_s_R(k)$ are analytic respectively
for $k\in\Complex_{\I+\II}$ and $k\in\Complex_{\III+\IV}$, and
their restriction to these domains
are continuous and bounded on the boundary;
$\_S(k,T)$ is entire,
and $\_S_L(k,T)$ and $\_S_R(k,T)$ are bounded
respectively for $k\in\=\Complex_{\II+\IV}$ and
$k\in\=\Complex_{\I+\III}$.

\paragraph{Symmetries, discrete spectrum and asymptotics.}

When $p(x,t)=\nu q^*(x,t)$, \eref{e:NLSsymmetriesPhi} still holds,
as does~\eref{e:NLSsymmetriesPhij} for $j=1,2,3$.
This implies that the scattering matrices can be expressed as
\[
\_s(k)= \begin{pmatrix} a(k) &\nu b^*(k^*)\\ b(k)
&a^*(k^*)\end{pmatrix},\qquad \_S(k,T)= \begin{pmatrix} A(k,T) &\nu
B^*(k^*,T)\\ B(k,T) &A^*(k^*,T)\end{pmatrix}.
\]
The properties of $a(k)$, $b(k)$, $A(k,T)$ and $B(k,T)$
follow trivially from those of $\_s(k)$ and $\_S(k,T)$.
Also, one can show that $\mu\o{j}(x,t,k)= \_I+ O(1/k)$ for $j=1,2,3$ as
$k\to\infty$ in the respective domains of boundedness of their
columns. The asymptotics of the eigenfunctions then determines that
of the scattering matrices. In particular, $a(k)=1+O(1/k)$ and
$b(k)=O(1/k)$ as $k\to\infty$ in $\=\Complex_{\I+\II}$, and
$A(k,T)=1+O(1/k)$ and $B(k,T)=O(1/k)$ as $k\to\infty$ in
$\=\Complex_{\III+\IV}$.

\paragraph{Riemann-Hilbert problem, solution and reconstruction formula.}

Equations~\eref{e:NLSIBVPscattering} allow us to formulate the
following RHP:
\be
\_M^-(x,t,k)=\_M^+(x,t,k)\,(\_I-\_J(k,t))\,, \qquad k\in  L\,,
\label{e:NLSIBVPRHP}
\ee
with
$L=\partial\Complex_\I\cup\partial\Complex_\III=L_1\cup L_2\cup L_3\cup L_4$,
where
\[
L_1=\=\Complex_\I\cap\=\Complex_\II\,,\quad
L_2=\=\Complex_\II\cap\=\Complex_\III\,,\quad
L_3=\=\Complex_\III\cap\=\Complex_\IV \,,\quad
L_4=\=\Complex_\I\cap\=\Complex_\IV\,,
\]
and where \bea \fl \_M^+(x,t,k)=  \left\{\!
\begin{array}{l}\displaystyle \left(\mu\o{2,L}, {\mu\o{3,R}\over
d(k)}\right)\,,\quad
 k \in \Complex_\I\,,
\\[0.6ex]\displaystyle
%\_M^+(x,t,k)=
\left({\mu\o{1,L}\over a^*(k^*)}, \mu\o{2,R}\right)\,,\quad
 k \in \Complex_\III\,,
\end{array}\right.
\qquad
%\nonumber\\
%\fl
\_M^-(x,t,k)= \left\{\! \begin{array}{l}\displaystyle
\left(\mu\o{2,L}, {\mu\o{1,R} \over a(k)}\right)\,,\quad
 k \in \Complex_\II\,\,,
\\[0.6ex]\displaystyle
%\_M^-(x,t,k)=
\left({\mu\o{3,L} \over d^*(k^*)}, \mu\o{2,R}\right)\,,\quad
 k \in \Complex_\IV\,.
\end{array}\right.
\nonumber
\eea
The jump matrices $\_J_{\!j}(k,t)$, each defined for $k\in L_j$, are:
%\bse
\bea
\_J_1(k,t)=\begin{pmatrix} 0 &\nu \e^{2i\theta} \Gamma^* (k^*)
  \\ 0 & 0 \end{pmatrix} \,,
\qquad
%\quad k \in  L_1\,,
%\nonumber
%\\
\_J_2(k,t)=\begin{pmatrix} 0 & \e^{2i\theta}\gamma (k)
\\ -\nu \e^{-2i\theta}\gamma^*(k) & \nu |\gamma (k)|^2
\end{pmatrix}\,,
%\quad k \in L_2\,,
\nonumber
\\
\_J_3(k,t)=\begin{pmatrix} 0 & 0
\\ -\e^{-2i\theta}\Gamma (k) & 0 \end{pmatrix} \,,
%\quad k \in L_3\,,
\qquad
%\nonumber
%\\
\_J_4(k,t)= \_I- (\_I-\_J_1)(\_I-\_J_2)^{-1}(\_I-\_J_3)\,,
%\quad k \in L_4\,,
\nonumber
\eea
%\ese
and the reflection coefficients are
\[\fl
\gamma(k)={\nu b^*(k) \over a(k)} \,,\quad
d(k)=a(k)A^*(k^*,T)-\nu b(k)B^*(k^*,T) \,, \quad \Gamma(k)={ B(k,T)
\over a^*(k^*)d^*(k^*)}\,.
\]
Note that $d(k)$ is defined $\forall k\in\=\Complex_{\I+\II}$,
$\Gamma(k)$ for $k\in L_3\cup L_4$ and $\gamma(k)$ $\forall
k\in\Real$. Their asymptotics as $k\to\infty$ follow trivially from
those of $\_s(k)$ and $\_S(k,T)$. As a result, $\_M(x,t,k)\to\_I$ as
$k\to\infty$. Hence, in the absence of a discrete spectrum [that is,
assuming that $a(k)$ and $d(k)$ have no zero respectively for
$k\in\Complex_\II$ and $k\in\Complex_\I$], the
RHP~\eref{e:NLSIBVPRHP} is solved by Cauchy projectors: \be
\_M^+(x,t,k)=
  \_I+ \frac1{2\pi i} \int_L \_M^+(x,t,k'){\_J(k',t) \over k'-k} \,\d k'\,.
\label{e:NLSIBVPRHPsoln}
\ee
Substituting the asymptotic expansion for $\_M(x,t,k)$ into the
$x$-part of the Lax pair and comparing the $(1,2)$ components,
we have
\be
q(x,t)=-2i\lim_{k\to\infty}
 k\big(\_M(x,t,k)-\_I\big)_{12}\,.
\label{e:NLSIBVPreconstructionq}
\ee
Using the asymptotic expansion for $\_M(x,t,k)$ as $k\to\infty$,
from~\eref{e:NLSIBVPRHPsoln} and comparing the $(1,2)$
components, we obtain the solution of the IBVP for the NLS equation as
\bea
\fl
q(x,t)=\frac1{\pi}\int_{\partial \Complex_\I} \nu
\e^{2i\theta(x,t,k')} \Gamma^*(k'^*)\_M^+_{11}(x,t,k') \,\d k'
\nonumber
\\[-1ex]%\kern3em
- \frac1{\pi}\int_{\!\!-\infty}^{\infty\!\!}\e^{2i\theta(x,t,k')}\gamma (k')
  \_M^+_{11}(x,t,k') \,\d k'
- \frac1{\pi}\int_0^{\infty \!\!} \nu
  |\gamma (-k')|^2\_M^+_{12}(x,t,-k') \,\d k'.
%\nonumber\\[-2ex]
\label{e:NLSIBVPrepresentationq}
\eea

\paragraph{Linear limit.}

%We now study the linear limit of~\eref{e:NLSIBVPrepresentationq},
Supppose that $\_Q=O(\epsilon)$.
%and $\_Q_x(0,t)=O(\epsilon)$.
From~\eref{e:NLSLPsolutions} and~\eref{e:NLSIBVPrepresentationq}
we have
$\mu=\_I+O(\epsilon)$ and $\_M=\_I+O(\epsilon)$.
Also, \eref{e:scattdata} imply
%that we evaluate the followings up to $O(\epsilon^2)$:
$\gamma(k)=-\^q(2k,0)+O(\epsilon^2)$ and
$d(k)=1+O(\epsilon^2)$, as well as
\[
\Gamma^*(k^*)=\nu\big(2k\^f_0(2k,T)-i\^f(2k,T)\big)
+O(\epsilon^2) \,.
\]
Thus~\eref{e:NLSIBVPrepresentationq} yields, to $O(\epsilon)$,
\bea
\fl
q(x,t)=\frac1{\pi} \int_{\partial \Complex_\I}
\e^{2i(k'x-2k'^2t)}\big(2k'\^f_0(2k',T)-i\^f_1(2k',T)\big) \,\d k'
%\nonumber\\\kern11em{ }
+\frac1{\pi} \int_{\!\!-\infty}^{\infty\!\!}
\e^{2i(k'x-2k'^2t)}\^q(2k',0) \,\d k'\,.
\nonumber
%\label{e:NLSIBVPlinearlimit}
\eea
Performing the change of variable $2k'=k$, we then see that,
to leading order, this expression yields exactly the
solution of the linear Schr\"odinger equation on the half line,
namely~\eref{e:IBVPreconstruction}.

\paragraph{Global relation and Dirichlet-to-Neumann map.}

Equations~\eref{e:scattdata} involve all initial and boundary data for
$\_Q(x,t)$.
These values are not all independent, however, since they satisfy
the global relation
\bea
\fl
\int_0^T \esh{2ik^2t}\big(\_H(0,t,k)\mu(0,t,k)\big)\,\d t
  + \esh{2ik^2T}\int_0^\infty \esh{-ikx}\big(\_Q(x,T)\mu(x,T,k)\big)\,\d x
\nonumber\\[-2ex]\kern14em
  = \int_0^\infty \esh{-ikx}\big(\_Q(x,0)\mu(x,0,k)\big)\,\d x\,.
\label{e:NLSglobal}
\eea
%obtained as usual by integrating~\eref{e:PsiNLSLP}
%around the domain $[0,\infty)\times[0,T]$ in the $xt$-plane.
When~\eref{e:NLSglobal} is evaluated with $\mu\equiv\mu\o2(x,t,k)$,
its first column is defined $\forall k\in\=\Complex_{\I+\II}$,
its second column $\forall k\in\=\Complex_{\III+\IV}$.
Moreover, when $\mu(x,t,k)=\mu\o2(x,t,k)$,
the RHS of~\eref{e:NLSglobal} equals~$\_I-\_s(k)$.
Using~\eref{e:NLSIBVPscattering2} in the LHS,
%to express the LHS in terms of $\mu\o1(x,t,k)$,
one then obtains a relation between the
scattering matrices:
%spectral functions $\_s(k)$ and $\_S(k,T)$:
%\bse
\bea
\_S^{-1}(k,T)\_s(k)= \_I - \esh{2ik^2T}\_G(k,T)\,,
\label{e:NLSglobal2}
\\
\noalign{\noindent where} \_G(k,t)= \int_0^\infty
\esh{-ikx}\big(\_Q(x,t)\mu\o2(x,t,k)\big)\,\d x\,, \nonumber \eea
%\ese
and $\_G_L(k,t)$ and $\_G_R(k,t)$ are analytic respectively
for $k\in\Complex_{\I+\II}$ and $k\in\Complex_{\III+\IV}$,
and continuous and bounded on the boundary of these domains.
In particular, %taking the $(1,2)$ component of~\eref{e:NLSglobal2}
for $k\in\Complex_{\III+\IV}$ we have
\[
%\fl
A^*(k^*,T)b^*(k^*)-B^*(k^*,T)a^*(k)=
  -\nu\e^{4ik^2T}\int_0^\infty \e^{-2ikx}q(x,T)\,\mu\o2_{22}(x,T,k)\,\d x\,.
\]
Since the integral term in the RHS is of $O(1/k)$,
as $k\to\infty$ in $\Complex_\III$, integrating along $\partial\Complex_\III$
we obtain the following integral relation:
\be
\int_{\partial\Complex_{\III}}
k\,\e^{-4ik^2t}\big( B^*(k^*)-r(k^*)A^*(k^*)\big) \,\d k =0\,,
\label{e:NLSGRintegral}
\ee
where $r(k)=b(k)/a(k)$.
As shown in \cite{CPAM58p639}, this relation can be solved to obtain the
Dirichlet-to-Neumann map, which expresses the unknown boundary datum
$q_x(0,t)$ in terms of the known one, $q(0,t)$.

\paragraph{Linearizable BCs and soliton solutions.}

%As in the discrete case, %the scattering matrix
%$\_S(k,T)$ depends on both known and unknown boundary values.
%The linearizable BCs are those for which it is possible to obtain
%the unknown boundary data via only algebraic manipulations
%of the global relation.
%Using~\eref{e:NLSmudef} and the second of \eref{e:NLSIBVPscattering3}
One can write $\_S(k,t)=
\~\Phi^{-1}(k,t)\esh{-2ik^2t\sigma_3}$,
%Since $\_S(k,T)=(***)^{-1}$
%and $\mu\o3(x,t,k)=\Phi\o3(x,t,k)\,\e^{i\theta\sigma_3}$,
where $\~\Phi(k,t)= \Phi\o1(0,t,k)$ solves the $t$-part of the Lax
pair~\eref{e:NLSLP2} for $x=0$, namely
\be
\~\Phi_t + 2ik^2\sigma_3\~\Phi= \_H(0,t,k)\,\~\Phi\,,
\label{e:Phitilde}
\ee
with $\~\Phi(k,0)=\_I$.
The matrix $\~\Phi(-k,t)$ solves an equation identical to~\eref{e:Phitilde}
except that $\_H(0,t,k)$ is replaced by $\_H(0,t,-k)$.
If there is an invertible time-independent matrix $\_N(k)$ such that
\be
\_N(k)\,\big(2ik^2\sigma_3-\_H(0,t,k)\big)=
  \big(2ik^2\sigma_3-\_H(0,t,-k)\big)\,\_N(k)\,,
\label{e:NLSNdef}
\ee
it then is easy to see that
$\~\Phi(-k,t)= \_N(k)\,\~\Phi(k,t)\,\_N^{-1}(k)\,$.
One can show that
a suitable matrix $\_N(k)$ only exists for homogeneous Robin BCs,
namely,
\[
q_x(0,t)-\chi q(0,t)=0\,,
\]
with $\chi\in\Real$ arbitrary.
In that case, \eref{e:NLSNdef} implies $N_{12}=N_{21}=0$ and
$N_{11}=f(k)\,N_{22}$, where %\quad
$f(k)=-(2ik-\chi)/(2ik+\chi)\,$, which in turn imply
$A^*(k^*,T)=A^*(-k^*,T)$ and $B^*(k^*,T)=f(k)B^*(-k^*,T)$. From
here, similar arguments to those used in the discrete problem can be
applied to the analysis of linearizable BCs.

As in the discrete case, the poles for the IBVP occur at the zeros of
$a(k)$ in $\Complex_\II$ and those of $d(k)$ in $\Complex_\I$, plus
their complex conjugates in $\Complex_\III$ and $\Complex_\IV$
\cite{NLTY18p1771}.
Each of these pairs of zeros, by itself, generates the well-known
one-soliton solution of NLS:
\be
q(x,t)= 2\eta\e^{2i\xi x-4i(\xi^2-\eta^2)t +i(\phi-\pi/2)}
  \sech(2\eta x-8\xi\eta t-2\delta)\,,
\ee
where $k_1=\xi+i\eta$ is the zero of $a(k)$ or of $d(k)$, and
$C_1= 2\eta\,\e^{2\delta+i\phi}$ is the norming constant
(see \cite{NLTY18p1771} for further details).

%%%%%%%%%%%%%%%%%%%%%%%%%%%%%%%%%%%%%%%%%%%%%%%%%%%%%%%%%%%%%%%%%%%%%%%%%%%%%%
%%%%%%%%%%%%%%%%%%%%%%%%%%%%%%%%%%%%%%%%%%%%%%%%%%%%%%%%%%%%%%%%%%%%%%%%%%%%%%
%%%%%%%%%%%%%%%%%%%%%%%%%%%%%%%%%%%%%%%%%%%%%%%%%%%%%%%%%%%%%%%%%%%%%%%%%%%%%%

\section{Conclusion}
\label{s:conclusion}

In conclusion, we have demonstrated a method to solve initial-boundary
value problems for linear and integrable nonlinear discrete evolution
equations.
We have done so by solving the IBVP for the discrete linear Schr\"odinger
(DLS) and integrable discrete nonlinear Schr\"odinger (IDNLS) equations
on the natural numbers.
Moreover, we have illustrated the similarities and differences between
the method for differential-difference equations and PDEs
by showing explicitly the correspondence between the discrete and
its continuum limit.
While the differential form representation of the continuum is lost,
the essential ideas of the method can be carried over to the discrete,
but the actual implementation of the method presents some
additional difficulties.
In particular,
the jump location in the nonlinear case differs because of the rescaling
$z'=z^2$ in the dispersion relation $\omega(z)$ when going from the
linear to the nonlinear case.
This is a significant difference from continuum limit, where the jumps
in the nonlinear case are given by the union of those for the linear
problem and its adjoint
(cf.\ sections~\ref{s:IDNLS} and~\ref{s:continuum}).
Also, the limit $k\to\infty$ in the continumm becomes
$z\to0$ (for $\Im k>0$) and $z\to\infty$ (for $\Im k<0$) in the
discrete.
As a consequence, the behavior of the eigenfuncions and spectral
data as $z\to0$ in the discrete problem must also be studied
in addition to that as $z\to\infty$.
This is the why the point $z=0$ plays such a special role
in the discrete problem, similarly to Ref.~\cite{IP23p1711},
and is one of the reason why discrete problems are more complicated
than their continuum counterparts.

For the DLS, in addition to solving the IBVP with Dirichlet-type
BCs we have shown that, contrary to Fourier series approaches,
the method can deal with more complicated kinds of BCs
just as effectively.
For the IDNLS, in addition to solving the IBVP (showing explicitly
how to eliminate the unknown boundary datum),
we have characterized the linear limit, the linearizable BCs
(showing how they fit within the IST framework),
and we have obtained explicitly the soliton solutions.

It should be clear that, similarly to the continuum,
the method can be generalized to solve IBVPs for both the DLS and
IDNLS equation defined on a finite set of integers.
It would also be straightforward to generalize
this method to any discrete linear evolution equation and to
other integrable discrete nonlinear evolution equations.

Several interesting questions can now be effectively addressed
using the present method.
For example, one can use the expression for the solution to study its
long-time asymptotics,
using the Deift-Zhou method \cite{BAMS26p119},
or to study the ``small dispersion'' or ``anti-continuum'' limit
(i.e., the limit $h\to\infty$),
e.g., using the Deift-Venakides-Zhou method~\cite{IMRN6p286}.
Doing so is a nontrivial task, however,
which is beyond the scope of this work.

%%%%%%%%%%%%%%%%%%%%%%%%%%%%%%%%%%%%%%%%%%%%%%%%%%%%%%%%%%%%%%%%%%%%%%%%%%%%%%
%%%%%%%%%%%%%%%%%%%%%%%%%%%%%%%%%%%%%%%%%%%%%%%%%%%%%%%%%%%%%%%%%%%%%%%%%%%%%%
%%%%%%%%%%%%%%%%%%%%%%%%%%%%%%%%%%%%%%%%%%%%%%%%%%%%%%%%%%%%%%%%%%%%%%%%%%%%%%

\section*{Acknowledgements}

It is a pleasure to thank Mark Ablowitz, Athanassios Fokas,
Beatrice Pelloni and Barbara Prinari for many insightful discussions.
This work was partially supported by the National Science Foundation
under grant number DMS-0506101.

%%%%%%%%%%%%%%%%%%%%%%%%%%%%%%%%%%%%%%%%%%%%%%%%%%%%%%%%%%%%%%%%%%%%%%%%%%%%%%
%%%%%%%%%%%%%%%%%%%%%%%%%%%%%%%%%%%%%%%%%%%%%%%%%%%%%%%%%%%%%%%%%%%%%%%%%%%%%%
%%%%%%%%%%%%%%%%%%%%%%%%%%%%%%%%%%%%%%%%%%%%%%%%%%%%%%%%%%%%%%%%%%%%%%%%%%%%%%

\setcounter{section}0
\def\thesection{Appendix~\Alph{section}}
\def\thesubsection{\Alph{section}.\arabic{subsection}}
\def\numparts{\refstepcounter{equation}%
     \setcounter{eqnval}{\value{equation}}%
     \setcounter{equation}{0}%
     \def\theequation{\Alph{section}.\arabic{eqnval}{\it\alph{equation}}}}
\def\endnumparts{\def\theequation{\Alph{section}.\arabic{equation}}%
     \setcounter{equation}{\value{eqnval}}}
\def\theequation{\Alph{section}.\arabic{equation}}

%%%%%%%%%%%%%%%%%%%%%%%%%%%%%%%%%%%%%%%%%%%%%%%%%%%%%%%%%%%%%%%%%%%%%%%%%%%%%%
\section{Notation and frequently used formulae}
\label{s:notations}

We denote the closure, interior and boundary of
a domain $D$ respectively by $\=D$, $D^o$ and $\partial D$,
where as usual $\partial D$ is oriented so as to leave $D$ to its left.
We also occasionally refer to punctured regions of the complex plane,
which we denote as $R\punct=R{-}\{0\}$.
As usual, $[\_A,\_B]=\_A\_B-\_B\_A$ is the commutator of two matrices
$\_A$ and $\_B$.
We use a superscript asterisk to denote
the complex conjugate $z^*$ of a complex number~$z$,
and $|z|^2=z^*z$.
Throughout, $\Real^+=\{x\in\Real:x>0\}$ and $\Real^+_0=\Real^+\cup\{0\}$.
Similarly,
$\Natural=\{1,2,3,\dots\}$ and $\Natural_0=\Natural\cup\{0\}$.
Finally, we denote by $\Complex_\I,\dots,\Complex_\IV$ the first,
second, third and fourth quadrants of the complex plane:
$\Complex_\I=\{k\in\Complex:\Re k>0\,\wedge\,\Im k>0\}$, etc.
Similarly,
we denote by $\Complex_{\I+\II}=\{k\in\Complex:\Im k>0\}$ and
$\Complex_{\III+\IV}=\{k\in\Complex:\Im k<0\}$
the upper-half and lower-half planes, respectively.

The nonlinear Schr\"odinger (NLS) equation~\eref{e:NLS} is a reduction
of the system
\bse
\label{e:NLSsystem}
\bea
iq_t + q_{xx} + 2q^2p=0\,,
\\
-ip_t + p_{xx} + 2p^2q=0\,.
\eea
\ese
%where subscripts $x$ and $t$ denote partial derivatives.
That is, \eref{e:NLS} follows by imposing $p(x,0)=\nu q^*(x,0)$
in~\eref{e:NLSsystem}, which then implies that $p(x,t)=\nu q^*(x,t)$
$\forall t>0$ and $q(x,t)$ is a solution of~\eref{e:NLS}.
A Lax pair for~\eref{e:NLSsystem} is given by:
\bse
\label{e:NLSLP}
\bea
\Phi_x - ik\sigma_3\Phi = \_Q\,\Phi\,,
\label{e:NLSLP1}
\\
\Phi_t + 2ik^2\sigma_3\Phi= \_H\,\Phi\,,
\label{e:NLSLP2}
\eea
\ese
where $\Phi(x,t,k)$ is either a 2-component vector or a $2\times2$ matrix,
and where
\bse
\label{e:NLSLPpotentials}
\bea
\sigma_3= \begin{pmatrix}1 &0\\0&-1\end{pmatrix},\qquad
\_Q(x,t)= \begin{pmatrix}0 &q\\p&0\end{pmatrix},
\label{e:sigma3Q}
\\
\_H(x,t,k)= %-i\_Q\_Q\,\sigma_3-i\_Q_x\sigma_3-2k\_Q=
  i\sigma_3(\_Q_x - \_Q^2)-2k\_Q=
\begin{pmatrix} -iqp& \!\!\!iq_x-2kq\\ -ip_x-2kp &iqp\end{pmatrix}.
\eea
\ese
(The present pair differs from that in Ref.~\cite{CMP230p1} by
the rescaling $k\to-k$,
and from that in Ref.~\cite{APT2003} by $k\to-k$ and $t\to-t$.)
Similarly, the integrable discrete NLS equation~\eref{e:IDNLS} is a
reduction of the system of differential-difference equations
\bse
\label{e:AL}
\bea
i\.q_n+ (q_{n+1}-2q_n+q_{n-1})-q_np_n(q_{n+1}+q_{n-1})=0\,,
\\
i\.p_n+ (p_{n+1}-2p_n+p_{n-1})-p_nq_n(p_{n+1}+p_{n-1})=0\,.
\eea
\ese
That is, imposing $p_n(0)=\nu\,q_n^*(0)$ on~\eref{e:AL}
yields~$p_n(t)=\nu\,q_n^*(t)$ $\forall t>0$,
with $q_n(t)$ satisfying~\eref{e:IDNLS}.
In the literature,
the name Ablowitz-Ladik (AL) is associated to both~\eref{e:IDNLS}
and \eref{e:AL}.
To avoid confusion, here we
will simply refer to~\eref{e:IDNLS} as the IDNLS equation,
reserving the name AL for the more general system~\eref{e:AL}.
A Lax pair for the AL system~\eref{e:AL} is:
\bse
\label{e:ALLP}
\bea
\Phi_{n+1} - \_Z \Phi_n = \_Q_n\,\Phi_n\,,
\label{e:ALLP1}
\\
\.\Phi_n - \txtfrac i2(z-1/z)^2\sigma_3\,\Phi_n =  \_H_n\,\Phi_n\,,
\label{e:ALLP2}
\eea
\ese
where $\Phi_n(z,t)$ is either a two-component column vector
or a $2\times2$ matrix,
and where
\bse
\label{e:QH}
\bea
\_Z= \e^{\sigma_3\,\log z}= \begin{pmatrix}z &0\\ 0 &1/z\end{pmatrix},
%\sigma_3= \begin{pmatrix}1 &0\\0 &\!-1\end{pmatrix},
\qquad
\_Q_n(t)= \begin{pmatrix} 0 & q_n \\ p_n & 0 \end{pmatrix},
\label{e:ZQn}
\\
\_H_n(z,t)= %i\big(\_Z\sigma_3\_Q_n + \_Q_{n-1}\_Z\sigma_3 - \sigma_3 \_Q_n\_Q_{n-1} \big)
  i\sigma_3\big( \_Q_n\_Z^{-1}\! - \_Q_{n-1}\_Z - \_Q_n\_Q_{n-1}\big)
%\nonumber\\[0ex]\kern12em{ }
  = i\begin{pmatrix} -q_np_{n-1} &zq_n-q_{n-1}/z\\
    zp_{n-1}-p_n/z  &p_nq_{n-1} \end{pmatrix}. \label{e:defHn}
\nonumber\\[0ex]\kern12em{ }
\eea
\ese
In sections~\ref{s:IDNLS} and~\ref{s:continuum}
we make frequent use of the integrating factors
\bea
\Zhat(\_A)=\_Z\,\_A\,\_Z^{-1}=
  \begin{pmatrix}a_{11}&z^2a_{12}\\a_{21}/z^2&a_{22}\end{pmatrix},
\qquad
\hsigma\_A=
\begin{pmatrix}a_{11}&-a_{12}\\-a_{21}&a_{22}\end{pmatrix},
\\
\esh{i\theta}(\_A)=\e^{i\theta\sigma_3}\_A\,\e^{-i\theta\sigma_3}=
  \begin{pmatrix}a_{11}&\e^{2i\theta}a_{12}\\\e^{-2i\theta}a_{21}&a_{22}\end{pmatrix}\,.
\label{e:ehs}
\eea

For any matrix $\_A$, we write $\_A=(\_A\o{L},\_A\o{R})$,
where the superscripts $L$~and~$R$ (left and right)
denote respectively the first and second column of $\_A$.
We also write $\_A= \_A_D+\_A_O$, where $\_A_D$ and $\_A_O$
denote respectively the diagonal and off-diagonal part of~$\_A$.
Note that
\bse
\label{e:AOD}
\bea
(\_A\mu)_D=\_A_D\mu_D + \_A_O\mu_O \,,\quad
(\_A\mu)_O=\_A_O\mu_D+\_A_D\mu_O \,,
\\
(\_Q\mu)_D=\_Q\mu_O \,,  \qquad (\_Q\mu)_O=\_Q\mu_D \,,
\eea
\ese
and in particular %when $\_H(n,z,t)=\_H_n(z,t)$,}
\bea
\_H_{n,D}(z,t)= -i\sigma_3\_Q_n\_Q_{n-1}\,,\quad
\_H_{n,O}(z,t)= i\,\big(\_Z\sigma_3\_Q_n+\_Q_{n-1}\_Z\sigma_3\big)\,.
\label{e:HOD}
\eea
Note also that $\_Z\_A_O=\_A_O\_Z^{-1}$ and $\sigma_3\_A_O= -\_A_O\sigma_3$.
%(while $\_Z\_A_D=\_A_D\_Z$ and $\sigma_3\_A_D= \_A_D\sigma_3$, obviously).

The ``involution symmetry'' of the scattering problems of
NLS and IDNLS is expressed through the matrix
\bse
\bea
\sigma_\nu= \begin{pmatrix}0 &1\\\nu&0\end{pmatrix}.
\label{e:sigmanudef}
\eea
That is, when $p(x,t)=\nu q^*(x,t)$ in~\eref{e:sigma3Q},
or $p_n(t)=\nu q_n^*(t)$ in~\eref{e:ZQn}, it is, respectively:
\bea
\sigma_\nu \_Q^*= \_Q\sigma_\nu\,,\qquad
\sigma_\nu \_Q_n^*= \_Q_n\sigma_\nu\,.
\eea
\ese
Note also that
$\sigma_\nu\_Z= \_Z^{-1}\sigma_\nu$,
$\sigma_\nu\sigma_3= -\sigma_3\sigma_\nu$, and
$\sigma_\nu^{-1}=\sigma_\nu^t=\nu\sigma_\nu$.

When discussing the asymptotic behavior of the eigenfunctions,
the behavior of the matrix product $\_A\_Z$
motivates the following definitions:
for any matrix $\_A=(\_A\o L,\_A\o R)$, we write
$\_A=O(\_Z^m)$ as $z\to(0,\infty)$ if $\_A\o L=O(z^m)$ as $z\to 0$
and $\_A\o R=O(1/z^m)$ as $z\to \infty$. Similarly, we write
$\_A=O(\_Z^m)$ as $z\to(\infty,0)$ if $\_A\o L=O(z^m)$ as $z\to
\infty$ and $\_A\o R=O(1/z^m)$ as $z\to 0$.

%%%%%%%%%%%%%%%%%%%%%%%%%%%%%%%%%%%%%%%%%%%%%%%%%%%%%%%%%%%%%%%%%%%%%%%%%%%%%%
\section{Spectral analysis of the $t$-part of the Lax pair of the DLS}
\label{e:ztransfinverse}

The inversion formulae for the spectral functions~\eref{e:DLSztransforms}
in the linear problem can be obtained by performing
spectral anlaysis of the individual parts of the Lax
pair~\eref{e:LaxpairL}.
The first of~\eref{e:LSinvztransf} can be
derived from similar steps as in section~\ref{s:1.3}.
As for the second of~\eref{e:LSinvztransf},
consider the following spectral problem
\be
\mu_t+i\omega(z)\mu = f(t)\,,
\label{e:simpleLPt}
\ee
where $\omega(z)=2-(z+1/z)$.
The Jost solutions are easily obtained,
%using the transformation $\mu(z,t)=\e^{-i\omega(t)}\psi(z,t)$,
and are:
%\bse
\bea
\mu\o1(z,t)=\int_0^t \e^{-i\omega(z)(t-t')}f(t')\,\d t'\,,
\qquad
\mu\o2(z,t)=-\int_t^T \e^{-i\omega(z)(t-t')}f(t')\,\d t'\,.
\nonumber
\eea
%\ese
Note that $\mu\o1$ and $\mu\o2$ are analytic for $z\notin D_+$ and
$z\in D_+$, respectively, where $D_+$ is the same as in
section~\ref{s:1.4}. Also, the jump condition is
%\bse
\bea \mu\o1-\mu\o2=\e^{-i\omega(z)t}\^f(z,T)\,,\quad z\in \partial
D_+\,, \label{e:RHPsimpleLPt}
\\
\noalign{\noindent where}
\^f(z,t)=\int_0^t \e^{i\omega(z)t'}f(t')\,
\d t'\,.
\nonumber
\eea
%\ese
Using integration by parts, one can show that $\mu^\pm =O(1/z)$ as
$z\to\infty$ in their corresponding domains. Hence the solution of
the RHP~\eref{e:RHPsimpleLPt} is given by
\[
\mu(z,t)=\frac1{2\pi i} \int_{\partial D_+}
{\e^{-i\omega(\zeta)t}\^f(\zeta,T) \over \zeta -z}\,\d\zeta\,.
%\label{e:simpleRHPsoln}
\]
Substituting this %~\eref{e:simpleRHPsoln}
into~\eref{e:simpleLPt},
we then find the reconstruction formula
\[
f(t)=-\frac1{2\pi i}\int_{\partial D_+}{\omega(\zeta)-\omega(z)
\over \zeta -z} \e^{-i\omega(\zeta)t}\^f(\zeta,T)\,\d\zeta\,.
\]
Recall that $\partial D_+=\partial D_{+\#in}\cup\partial D_{+\#out}$.
Also note that $\partial D_{+\#in}$ can be deformed to $\partial D_{+\#out}$
by letting $z\to 1/z$,
and $\omega(z)$ and $\^f(z,t)$ are invariant under this transformation.
After some algebra, we then obtain
\[
f(t)=\frac1{2\pi}\int_{\partial D_{+\#out}}\bigg( \frac1{z^2}-1\bigg)\,\e^{-i\omega(z)t}\^f(z,T)\,\d z\,.
\]
Replacing $f(z,t)$ by $q_n(t)$, we finally obtain
the second of~\eref{e:LSinvztransf}.

Both of~\eref{e:LSinvztransf} could also be obtained by more
direct methods.
The first of~\eref{e:LSinvztransf} of course just defines
the coefficients of the principal part in the Laurent expansion
of $\^q(z,t)$.
As for the second of~\eref{e:LSinvztransf}, it can be obtained
as follows.  Define $\~q(t)$ to be the function which equals $q_n(t)$
for $0\le t\le T$ and is 0 otherwise. Also, let\, $\~Q(\omega)=
\int\nolimits_{-\infty}^\infty \e^{i\omega t}\~q_n(t)\,\d t$\, be
its Fourier transform. Then, for all $0<t<T$ it is $q_n(t)=
(1/2\pi)\int\nolimits_{-\infty}^\infty \e^{-i\omega
t}\~Q(\omega)\,\d\omega$. Note however that the transformation
$z\to\omega(z)$ maps $\partial D_{+\#out}$ onto the real $\omega$-axis,
with $\omega(z)$ decreasing monotonically as $\Re\,z$ increases.
Moreover, $\~Q(\omega(z))= \^f_n(z,T)$. Hence we can rewrite the
previous integral as $q_n(t)= (1/2\pi)\int\nolimits_{\partial D_{+\#out}}
\omega'(z)\,\e^{-i\omega(z)t}\^f_n(z,T)\,\d z$.

%%%%%%%%%%%%%%%%%%%%%%%%%%%%%%%%%%%%%%%%%%%%%%%%%%%%%%%%%%%%%%%%%%%%%%%%%%%%%%
\section{IBVPs for DLS with Robin-type boundary conditions}
\label{s:Robin}

Consider the DLS equation~\eref{e:DLS} for $n\in\Natural_0$ and
$t\in\Real^+$ with mixed BCs.
The spectral transform of~\eref{e:DLSRobinBC} yields,
$\forall z\in\Complex\punct$,
\be
\^f_{-1}(z,t) - \alpha\^f_0(z,t)= \^h(z,t)\,,
\label{e:DLSRobinBCtransf}
\ee
where the $\^f_j(z,t)$ are given by~\eref{e:DLSztransforms},
and $\^h(z,t)$ is defined similarly.
Recall that the reconstruction formula~\eref{e:IBVPsoln}
contains the quantity $\^F(z,t)= i(z\^f_0(z,t)-\^f_{-1}(z,t))$.
Use of~\eref{e:DLSRobinBCtransf} and the transformed
global relation~\eref{e:fm1LS} allows one to eliminate $\^f_0(z,t)$ and
$\^f_{-1}(z,t)$ and express $\^F(z,t)$, for all $0<|z|\le1$, as
\be
\^F(z,t)= \frac{\^G(z,t)}{1/z-\alpha}
  - \frac{z-\alpha}{1/z-\alpha}\e^{i\omega(z)t}\^q(1/z,t)\,,
\label{e:DLSRobinF} \ee where $\^G(z,t)$, which contains the known
portion of the RHS, was given in~\eref{e:DLSRobinGdef}. Now recall
that, in~\eref{e:IBVPsoln}, $\^F(z,t)$ is integrated along $\partial
D_{\!+\#in}$. Three possible situations can arise: (i)~$\alpha\in
D_{\!+\#out}$, (ii)~$\alpha\in\partial D_{\!+\#out}$,
(iii)~$\alpha\notin\=D_{\!+\#out}$. We discuss each of these cases
in turn.

If $\alpha\notin\=D_{\!+\#out}$, the denominator
of~\eref{e:DLSRobinF} never vanishes in $\=D_{\!+\#in}$. Thus the
second part of the RHS of~\eref{e:DLSRobinF}, when inserted
in~\eref{e:IBVPsoln}, gives rise to an integrand that is analytic
and bounded in $\=D_{\!+\#in}$.  Hence, that part of the integral is
zero. As a result, the solution of the IBVP is simply \bea \fl
q_n(t)= \frac1{2\pi i}\!\oint_{|z|=1}
    z^{n-1}\e^{-i\omega(z)t}\,\^q(z,0)\,\d z
  - \frac1{2\pi i}\!\int_{\partial D_{\!+\#in}} z^{n-1}\e^{-i\omega(z)t}\,
     \frac{\^G(z,T)}{1/z-\alpha}\,\d z\,,
\label{e:IBVPsolnRobin1} \eea with $\^G(z,t)$ again given
by~\eref{e:DLSRobinGdef}. Now suppose $\alpha\in D_{\!+\#out}$. In
this case $1/z-\alpha$ vanishes at $z=1/\alpha\in D_{\!+\#in}$. Even
though each of the two terms in the RHS of~\eref{e:DLSRobinF} has a
simple pole at this point, however, their sum is finite there, since
$\^F(z,t)$ is analytic in $\Complex\punct$. Thus, \bea \fl
\frac1{2\pi i}\!\int_{\partial D_{\!+\#in}}
\!z^{n-1}\,\frac{z-\alpha}{1/z-\alpha}\^q(1/z,t)\,\d z=
\Res_{z=1/\alpha}\bigg[
z^{n-1}\frac{z-\alpha}{1/z-\alpha}\^q(1/z,t)\bigg]
\nonumber\\[-1ex]\kern7.6em{ }
  = \Res_{z=1/\alpha}\bigg[z^{n-1}\e^{-i\omega(z)t}\frac{\^G(z,t)}{1/z-\alpha}\bigg]
  = -\alpha^{-n-1}\e^{-i\omega(\alpha)t}\^G(1/\alpha,t)\,,
\nonumber
\eea
which implies the solution of the IBVP as
\bea
\fl
q_n(t)= \frac1{2\pi i}\!\oint_{|z|=1}
    z^{n-1}\e^{-i\omega(z)t}\,\^q(z,0)\,\d z
  - \frac1{2\pi i}\!\int_{\partial D_{\!+\#in}} z^{n-1}\e^{-i\omega(z)t}\,
     \frac{\^G(z,T)}{1/z-\alpha}\,\d z
     -\alpha^{-n-1}\e^{-i\omega(\alpha)t}\^G(1/\alpha,t)\,.
\nonumber\\[-2ex]
\label{e:IBVPsolnRobin2}
\eea
Finally, if $\alpha\in\partial D_{\!+\#out}$, the pole is along
the integration contour.
In this case one should go back to the RHP and subtract the pole contribution.
In this way, the solution of the IBVP can be obtained as
\bea
\fl
q_n(t)= \frac1{2\pi i}\!\oint_{|z|=1}
    z^{n-1}\e^{-i\omega(z)t}\,\^q(z,0)\,\d z
  - \frac1{2\pi i}\!\int_{\partial D_{\!+\#in}} z^{n-1}\e^{-i\omega(z)t}\,
     \frac{\^G(z,T)}{1/z-\alpha}\,\d z
     -\frac12\alpha^{-n-1}\e^{-i\omega(\alpha)t}\^G(1/\alpha,t)\,.
\nonumber\\[-2ex]
\label{e:IBVPsolnRobin3}
\eea
Combining \eref{e:IBVPsolnRobin1}, \eref{e:IBVPsolnRobin2}
and \eref{e:IBVPsolnRobin3} one then obtains~\eref{e:DLSRobinsoln}.

%%%%%%%%%%%%%%%%%%%%%%%%%%%%%%%%%%%%%%%%%%%%%%%%%%%%%%%%%%%%%%%%%%%%%%%%%%%%%%
\section{Asymptotic behavior of the eigenfunctions of the IBVP}
\label{s:asymptotics}

%We first discuss the asymptotics in the IBVP for DLS.
%We then derive the corresponding results in the the IBVP for IDNLS.
%Similar techniques are used to obtain the equivalent results in
%the continuum cases.

\paragraph{DLS.}

We first compute the asymptotics for for $n=0$
(where no summation is present),
then consider the case $n\ge1$.
Note that $\omega(z)= -1/z+O(1/z^2)$ as $z\to0$.
Integration by parts yields, as $z\to0$
with $\Im z\le0$,
\bea
\phi_0\o1(z,t)
%= i\int_0^t\e^{-i\omega(z)(t-t')}\big(q_0(t')-q_{-1}(t')/z\big)\,\d t'
= q_{-1}(t)-\e^{-i\omega(z)t}q_{-1}(0)+O(z)\,,
\nonumber
\\
\noalign{\noindent while as $z\to0$ with $\Im z\ge0$ it is}
\phi_0\o3(z,t)=
q_{-1}(t)-\e^{i\omega(z)(t-T)}q_{-1}(T)+O(z)\,.
\nonumber
\eea
Using these in~\eref{e:PhiIBVP} with $n\ge1$ we then have immediately
$\phi_n\o{j}(z,t)= q_{n-1}(t)+O(z)$ as $z\to0$ with $\Im z\le0$ for $j=1$
and $\Im z\ge0$ for $j=3$.
Note also that
$\phi_0\o1(z,t)-\phi_0\o3(z,t)= -\e^{-i\omega(z)t}
 \big(q_{-1}(0)-\e^{i\omega(z)T}q_{-1}(T)\big)+O(z)$ as $z\to0$,
implying that the ratio $\^F(z,T)/z$ in~\eref{e:Phijumps13} remains
bounded as $z\to0$ along the real axis.
As for $\phi_n\o2(z,t)$,
\eref{e:PhiIVPL} implies immediately $\phi_n\o2(z,t)= O(1/z)$
as $z\to\infty$.

\paragraph{IDNLS.}
\label{s:asymptoticsdiscrete}

The determination of the asymptotic behavior in the nonlinear case
is considerably more involved, and requires the use of a Neumann
series approach:
\be
\mu_n\o{j}(z,t)=\sum_{m=0}^{\infty}\mu_n\o{j,m}(z,t)\,.
\label{e:Nseriesmunj}
\ee
We now show that, $\forall n\in\Natural_0$, $m\ge 0$ and $j=1,3$,
as $z\to (\infty, 0)$ it is
\bse
\label{e:NeumannAsympmunj}\bea
\mu_{n,D}\o{j,2m-1}(z,t)=O(\_Z^{-2m})\,,\qquad
\mu_{n,O}\o{j,2m-1}(z,t)=O(\_Z^{-2m+1})\,,
\\
\mu_{n,D}\o{j,2m}(z,t)=O(\_Z^{-2m})\,,\qquad
\mu_{n,O}\o{j,2m}(z,t)=O(\_Z^{-2m-1})\,.
\eea
\ese
The proof
proceeds by induction. Consider $\mu_n\o1(z,t)$ first.
Separating~\eref{e:ALmu1IBVPsolns} into its diagonal and
off-diagonal components then yields $\mu_{n,D}\o{1,0}(z,t)=\_I$ and
$\mu_{n,O}\o{1,0}(z,t)=\_O$, as well as
\bse
\label{e:ALIBVPmu1}
\bea \fl
\mu_{n,D}\o{1,m+1}(z,t)=
  \sum_{n'=0}^{n-1}\_Q_{n'}(t)\mu_{n',O}\o{1,m+1}(z,t)\_Z^{-1}
%\nonumber\\[-1ex]\kern10em{ }
  + \int_0^t(\_H_{0,D}\mu_{0,D}\o{1,m}+\_H_{0,O}\mu_{0,O}\o{1,m+1})(z,t') \,\d t' \,,
\label{e:ALIBVPmu1D}
\\
\fl
\mu_{n,O}\o{1,m+1}(z,t)=
  \sum_{n'=0}^{n-1}\_Q_{n'}(t)\mu_{n',D}\o{1,m}(z,t)\_Z^{-2(n-n')+1}
\nonumber\\[-1ex]\kern4em{ }
  + \int_0^t\e^{-i\omega(z)(t-t')\^\sigma_3}\big(\_H_{0,O}\mu_{0,D}\o{1,m}+\_H_{0,D}\mu_{0,O}\o{1,m}\big)(z,t')\,\_Z^{-2n}\,\d t' \,.
%\nonumber\\[-2ex]
\label{e:ALIBVPmu1O}
\eea
\ese
Note that
\[
\frac1{2\omega(z)}\,\_I=-\_Z^{-2}+O(\_Z^{-4})\,,\quad{\rm as}~z\to (\infty,0).
%\label{e:asymptoticomega}
\]
First consider the case $n=0$. Using integration by parts
in~\eref{e:ALIBVPmu1O}, we obtain, as $z\to(\infty,0)$,
\bse
\label{e:estimatemun1}
\bea
\fl
\mu_{0,O}\o{1,m+1}(z,t)=\big\{\_Q_{-1}(t)\mu_{0,D}\o{1,m}(z,t)
  -\e^{-i\omega(z)t\^\sigma_3}\big[\_Q_{-1}(0)\mu_{0,D}\o{1,m}(z,0)\big]\big\}\,\_Z^{-1}
\nonumber\\{ }
  + \big\{(\_Q_0\_Q_{-1})(t)\mu_{0,O}\o{1,m}(z,t)-\e^{-i\omega(z)t\^\sigma_3}
  \big[(\_Q_0\_Q_{-1})(0)\mu_{0,O}\o{1,m}(z,0)\big]
  \big\}\,\_Z^{-2}\,,
\label{e:estimatemun1O}
\eea
plus higher order terms.
Substituting~\eref{e:estimatemun1O} into~\eref{e:ALIBVPmu1D} with $n=0$,
one finds
\bea
\fl
\mu_{0,D}\o{1,m+1}(z,t)=-i\int_0^t
  \sigma_3(\_Q_0\_Q_{-1})(t')\mu_{0,D}\o{1,m}(z,t') \,\d t' +i\int_0^t
  \sigma_3\_Q_0(t')\mu_{0,O}\o{1,m+1}(z,t')\_Z \,\d t'
\nonumber\\{ }
  - i\int_0^t \sigma_3\_Q_{-1}(t')\mu_{0,O}\o{1,m+1}(z,t')\_Z^{-1} \,\d t' \,.
\label{e:estimatemun1D}
\eea
\ese
Using~\eref{e:estimatemun1}
one can then obtain~\eref{e:NeumannAsympmunj} for $n=0$
and all $m\in\Natural_0$ inductively.
Note also that, for $m=0$, \eref{e:estimatemun1O} yields
\eref{e:mu1Oasymp}. Similarly, repeating the same arguments, one
obtains~\eref{e:mu3Oasymp}.

Next consider the case $n\ge 1$. The integrals in~\eref{e:ALIBVPmu1}
are exactly the same as when $n=0$ except for the fact that the one
in~\eref{e:ALIBVPmu1O} is followed by $\_Z^{-2n}$. Using the same
arguments as before, we obtain  \bse \label{e:estmun1sum}\bea
\mu_{n,O}\o{1,m+1}(z,t)=\_Q_{n-1}(t)\mu_{n-1,D}\o{1,m}(z,t)\_Z^{-1}
+ \mu_{0,O}\o{1,m+1}(z,t)\_Z^{-2n} \,+\,\cdots
\,,\label{e:estmun1Osum}
\\
\mu_{n,D}\o{1,m+1}(z,t)=\sum_{l=0}^{n-1}\_Q_l(t)\mu_{l,O}\o{1,m+1}(z,t)\_Z^{-1}
+\mu_{0,D}\o{1,m+1}(z,t)\label{e:estmun1Dsum}\,. \eea \ese The
induction with~\eref{e:estmun1sum}, one can
derive~\eref{e:NeumannAsympmunj} for $n\ge 1$.  Similarly, one
obtains~\eref{e:NeumannAsympmunj} for $\mu_n\o3$. This completes the
proof of~\eref{e:NeumannAsympmunj}.

The above results imply that $\mu_n\o1(z,t)=\_I+O(\_Z^{-1})$ as
$z\to (\infty,0)$.
%from~\eref{e:Nseriesmun1}.
In particular,
%following the calculation to the next order and
computing the $O(\_Z^{-1})$ terms explicitly one obtains the first
of~\eref{e:ALIBVPmu13asymp}. Similarly, using the same arguments,
one can show that $\mu_n\o3(z,t)=\_I+O(\_Z^{-1})$ as $z\to
(\infty,0)$ and verify the second of~\eref{e:ALIBVPmu13asymp}. In
the IVP, the integrals in the RHS of~\eref{e:ALIBVPmu1D}
and~\eref{e:ALIBVPmu1O} are absent, and the summation starts from
$n'=-\infty$.  Hence in this case one simply obtains
\eref{e:ALasymp}.

The determination of the asymptotic behavior of $\mu_n\o2(z,t)$
requires a slightly different approach,
since following the above steps for $\mu_n\o2(z,t)$,
yields a $O(1)$ term involving the summation of $\_Q_n$ in the RHS.
To circumvent this difficulty, note that~\eref{e:ALLP1} implies
$\mu_n\o2=\big(\_Z+\_Q_n(t)\big)^{-1}\mu_{n+1}\o2\,$.
For $\~\mu_n(z,t)=C_n\mu_n\o2(z,t)$ we have
\be
\~\mu_n-\Zhat^{-1}\~\mu_{n+1}=-\_Q_n\~\mu_n\_Z\,,
\label{e:modifiedmu2}
\ee
with $\~\mu_n(z,t)\to\_I$ as $n\to\infty$
thanks to \eref{e:ALmu1IBVPsolns} and \eref{e:ALphidet}. Introducing
the auxiliary function $\Psi_n(z,t)=\Zhat^{-n}\~\mu_n(z,t)$, it is
easy to check that $\Psi_n(z,t)$ satisfies the equation
$\Psi_{n+1}-\Psi_n= \_Z\,\Zhat^{-(n+1)}(\_Q_n)\Psi_{n+1}\,$, which
can be integrated to obtain the modified Jost solution as
\be
\~\mu_n(z,t)=
\_I-\_Z\sum_{n'=n+1}^\infty\Zhat^{n-n'}\big(\_Q_{n'-1}(t)\~\mu_{n'}(z,t)\big)\,.
\label{e:mutilde}
\ee
Then, applying the same Neumann series
approach as described above to~\eref{e:mutilde}, one finds the
asymptotic expansion for $\mu_n\o2$ as \eref{e:ALasympmu2}.
%Note that~\eref{e:NeumannAsympmunj} is also valid with $\~\mu_n(z,t)$.
Since $\mu_n\o2(z,t)$ is the same in the IVP and in the IBVP;
this asymptotic behavior applies to both problems.

Note that the above results also imply that $a(z)$ and $d(z)$ are
even functions in $D_{\pm\#in}$ and the following symmetries of
$\_M_n^{\pm}$: \bse \label{e:symmMn} \bea
\_M_{n,11}^+(-z,t)=\_M_{n,11}^+(z,t)\,,\quad
\_M_{n,12}^+(-z,t)=-\_M_{n,12}^+(z,t)\,, \\
\_M_{n,11}^-(-z,t)=\_M_{n,11}^-(z,t)\,,\quad
\_M_{n,12}^-(-z,t)=-\_M_{n,12}^-(-z,t)\,.
\eea
\ese

%%%%%%%%%%%%%%%%%%%%%%%%%%%%%%%%%%%%%%%%%%%%%%%%%%%%%%%%%%%%%%%%%%%%%%%%%%%%%%
\section{Independence of the solution on $T$}
\label{s:IndependenceT}
\let\phi=\varphi

The solution of a DDE does not depend on future values of the BCs.
Hence, for any $T_0<T$ the solution of the IBVP resulting from the RHP
obtained by replacing $T$ with~$T_0$ must be equivalent for all $0<t<T_0$
to the solution of the IBVP obtained from the original RHP.
We show next that is indeed the case because the RHP obtained from $T_0$ and
$T$ are related.
%As usual, we first discuss the situation in the continuum case,
%then proceed to the discrete case.

Let $\_M_n(z,t)$ satisfy the RHP \eref{e:IBVPALsystemRHP}, and let
$\_M_n^{\pm\#in}(z,t)$ and $\_M_n^{\pm\#out}(z,t)$ denote the
restrictions of $\_M_n(z,t)$ to the domains $D_{\pm\#in}$ and
$D_{\pm\#out}$, respectively. Moreover, let $A(z,T_0)$ and
$B(z,T_0)$ be the spectral coefficients obtained by replacing $T$
with $T_0$ in~\eref{e:ALdefinitionAB}, and let
$\~{\_J\!}_n\o1(z,t),\dots,\~{\_J\!}_n\o4(z,t)$ denote the jump
matrices obtained by replacing $A(z,T)$ and $B(z,T)$ with $A(z,T_0)$
and $B(z,T_0)$. Finally, let $\~{\_M}_n(z,t)$ satisfy the RHP with
the jump matrices $J_n\o1,\dots,J_n\o4$ replaced by
$\~{\_J}_n\o1,\dots,\~{\_J}_n\o4$. It is straightforward to see the
relations \bea \_M_n^{+\#in}=\~{\_M}_n^{+\#in}\,(
  \_I-\~{\_J}_n\o1)\,\big(\_I-\_J_n\o1\big)^{-1}\,,\qquad
\_M_n^{-\#in}=\~{\_M}_n^{-\#in}\,,
\nonumber
\\
\_M_n^{+\#out}=\~{\_M}_n^{+\#out}\,,\qquad
\_M_n^{-\#out}=\~{\_M}_n^{-\#out}\,
  \big(\_I-\~{\_J}_n\o3\big)^{-1}\,(\_I-\_J_n\o3)\,.
\nonumber \eea Now recall that $q_n(t)$ can be obtained from the
eigenfunctions via~\eref{e:ALreconstruction} or
\eref{e:ALIBVPmu13asymp} with $j=1$.
Note also that
$\mu_n\o{1,R}(z,t)$ enters $\_M_n^{-\#in}$ via~\eref{e:IBVPALdefM2}.
Below, we show that the matrices
$(\_I-\~{\_J}_n\o1)\,\big(\_I-\_J_n\o1\big)^{-1}$ and
$\big(\_I-\~{\_J}_n\o3\big)^{-1}(\_I-\_J_n\o3)$ are analytic and
bounded for $z\in D_{+\#in}$ and $z\in D_{+\#out}$, respectively.
Since $\_M_n(z,t)=\~{\_M}_n(z,t)$ for $z\in D_{-\#in}$, it then
follows that the solutions $q_n(t)$ obtained from $\_M_n$ and
$\~{\_M}_n$ coincide.

To show that $(\_I-\~{\_J}_n\o1)\,\big(\_I-\_J_n\o1\big)^{-1}$
is analytic and bounded for $z\in D_{+\#in}$,
note first that
\be
(\_I-\~{\_J_n\o1\!\!})\, \big(\_I-\_J_n\o1\big)^{-1}=
  \begin{pmatrix} 1 &
    \nu z^{2n}\e^{-2i\omega(z)t}\big(\Gamma^*(1/z^*,T)-\Gamma^*(1/z^*,T_0)\big)
  \\ 0 & 1 \end{pmatrix} \,,
\label{e:jumpmatrixY1}
\ee
and the $(1,2)$ component of \eref{e:jumpmatrixY1} can be written as
\be
X_n(z)=
  \nu z^{2n}\e^{-2i\omega(z)t}{A^*(1/z^*,T_0)B^*(1/z^*,T)-A^*(1/z^*,T)B^*(1/z^*,T_0)
    \over d(z,T)d(z,T_0)}\,.
%\e^{2ikx+4ik^2(T_0-t)}\,{A(k,T_0)B(k,T)-A(k,T)B(k,T_0) \over d^*(k^*,T)d^*(k^*,T_0)}\e^{-4ik^2T_0} \,,
\label{e:ALjumpdiff12comp}
\ee
Now note that~\eref{e:ALscattdef} and~\eref{e:ALIBVPscattdef}
define the scattering data $A(z,T)$ and $B(z,T)$ as
\be
\mu_0\o{1,R}(z,T)=\begin{pmatrix} -\nu\e^{-2i\omega(z)T}B^*(1/z^*,T) \\
A(z,T) \end{pmatrix} =:\begin{pmatrix} \mu_1(z,T) \\
\mu_2(z,T) \end{pmatrix}\,,
\label{e:ALdefinitionAB}
\ee
Hence
\bea
%\fl
X_n(z)=  z^{2n}\e^{2i\omega(z)(T_0-t)}
  {\mu_2^*(1/z^*,T)\mu_1(z,T_0) -\mu_2^*(1/z^*,T_0)\mu_1(z,T)\e^{2i\omega(z)(T-T_0)}
    \over d(z,T)d(z,T_0)}\,.
%\label{e:equationAB1}
\nonumber
\eea
Also, $\mu_0\o{1,R}(z,t)$
satisfies the second column of the $t$-part of the Lax pair~\eref{e:ALLP}
at $n=0$:
%denoting by $\mu_1(z,t)$ and $\mu_2(z,t)$ the first and second component of
\bse \label{e:equationsmu} \bea \.\mu_1(z,t)+2i\omega(z)\mu_1(z,t) =
  H_{0,11}(z,t)\mu_1(z,t)+H_{0,12}(z,t)\mu_2(z,t)\,,\\
\.\mu_2(z,t)= H_{0,21}(z,t)\mu_1(t,k)+H_{0,22}(z,t)\mu_2(z,t)\,.
\eea
\ese
Then, introducing
\bse
\bea
\phi_1(z,t)=\mu_2^*(1/z^*,T)\mu_1(z,t)
-\mu_1(z,T)\mu_2^*(1/z^*,t)\,\e^{2i\omega(z)(T-t)}\,,\\
\phi_2(z,t)=\mu_2^*(1/z^*,T)\mu_2(z,t)
-\nu\mu_1(z,T)\mu_1^*(1/z^*,t)\,\e^{2i\omega(z)(T-t)}\,,
\eea
\ese
we can rewrite the $(1,2)$ component of
$(\_I-\~{\_J}_n\o{1})\,\big(\_I-\_J_n\o1\big)^{-1}$ as
\be
X_n(z)= {z^{2n}\e^{2i\omega(z)(T_0-t)}
  \over d(z,T)d(z,T_0)} \,\phi_1(z,T_0)\,.
\label{e:ALequation12comp} \ee It is therefore enough to show that
$\phi_1(z,t)$ is analytic and bounded for $z\in D_+$. The symmetries
of $\_H_0(z,t)$ [namely, $\_H_{0,12}(z,t)=\nu\_H_{0,21}^*(1/z^*,t)$
and $\_H_{0,11}(z,t)=\_H_{0,22}(1/z^*,t)$] imply that
$(\phi_1,\phi_2)^t$ satisfies the $t$-part of the Lax
pair~\eref{e:ALLP2} with $n=0$. Since $\phi_1(z,T)=0$ and
$\phi_2(z,T)=1$, we then have the following linear integral
equations \bse \bea \phi_1(z,t)=-\int_t^T \e^{2i\omega(z)(t'-t)}
\big(\_H_{0,11}\phi_1+\_H_{0,12}\phi_2\big)(z,t') \,\d t' \,,\\
\phi_2(z,t)=1-\int_t^T
\big(\_H_{0,21}\phi_1+\_H_{0,22}\phi_2\big)(z,t') \,\d t'\,, \eea
\ese From here one can show that $\phi_1$ and $\phi_2$ are analytic
and bounded for $z\in D_+$. As a result, the RHS
of~\eref{e:ALjumpdiff12comp} is analytic and bounded for $z\in D_+$.
Thus $(\_I-\~{\_J}_n\o1)\,\,\big(\_I-\_J_n\o1\big)^{-1}$ is analytic
and bounded for $z\in D_{+\#in}$. The result for
$\big(\_I-\~{\_J}_n\o3\,\,\big)^{-1}\,(\_I-\_J_n\o3)$ follows from
symmetry considerations.

%%%%%%%%%%%%%%%%%%%%%%%%%%%%%%%%%%%%%%%%%%%%%%%%%%%%%%%%%%%%%%%%%%%%%%%%%%%%%%
\section{Linearizable BCs for $T<\infty$}

Here we verify that~\eref{e:ALratioAB} can be used to express
$\Gamma^*(1/z^*)$ also when $T<\infty$.
To do so, we use the same approach that we used to show that
the solution of the IDNLS equation does not depend on $T$.
Denote by $X_n(z)$
the difference between the contributions to the RHP obtained
from $T=\infty$ and $T<\infty$, namely:
\be
X_n(z)=
\nu z^{2n}\e^{-2i\omega(z)t}\big(\Gamma^*(1/z^*)-\Gamma_o^*(1/z^*)\big)\,,
\label{e:ALjump12diff}
\ee
where $\Gamma_o^*(1/z^*)$ is obtained by neglecting the second term
in the RHS of~\eref{e:ALratioAB1}. We can write~\eref{e:ALjump12diff} as
\[
X_n(z)=
  \nu z^{2n} \e^{-2i\omega(z)t}\,{R(z,T)-R_o(z,T)\over d(z)d_o(z)/A^*(1/z^*,T)A_o^*(1/z^*,T)} \,,
\]
with $R(z,T)= B^*(1/z^*,T)/A^*(1/z^*,T)$ as before, and where
$R_o(z)= B_o^*(1/z^*)/A_o^*(1/z^*)$ is computed using only the first
term in the RHS of~\eref{e:ALratioAB1} and
$d_o(z)=a(z)A_o^*(1/z^*,T)-\nu b(z)B_o^*(1/z^*,T)$. Also,
$A_o^*(1/z^*,T)$ and $B_o^*(1/z^*,T)$ are defined by~\eref{e:ALABLin}.
Now, using~\eref{e:ALratioAB1}, we find
\be
X_n(z)=
%\nu z^{2n}\e^{-2i\omega(z)t}
%  \big(\Gamma^*(1/z^*)-\Gamma_o^*(1/z^*)\big)=
  z^{2n}\e^{2i\omega(z)(T-t)}\,f(1/z)\,{G(1/z,T) \over d(z)\Delta(1/z)}\,.
\label{e:differenceGammas}
\ee
In the solitonless case, however, we
can assume that $d(z)$ and $\Delta(1/z)$ never vanish in
$\=D_{+\#in}$.
Then the RHS of~\eref{e:differenceGammas} is analytic and bounded in
$D_{+\#in}$ due to the exponential term and now we know that the
additional term in~\eref{e:ALratioAB1} does not affect the solution
of the RHP. Note that $f(1/z)$ has a pole at $z=\pm 1/\chi^{1/2}$.
When $1<\chi$, or $\chi <-1$, these points belong to $D_{+\#in}$.
Note, however, that since $a(z)$ and $b(z)$ are bounded in
$\=D_{\pm\#in}$, if $f(1/z)$ has a pole, $\Delta(1/z)$ does too, and
hence the terms causing the poles in~\eref{e:differenceGammas} to
cancel out.

%%%%%%%%%%%%%%%%%%%%%%%%%%%%%%%%%%%%%%%%%%%%%%%%%%%%%%%%%%%%%%%%%%%%%%%%%%%%%%
%%%%%%%%%%%%%%%%%%%%%%%%%%%%%%%%%%%%%%%%%%%%%%%%%%%%%%%%%%%%%%%%%%%%%%%%%%%%%%
\catcode`\@ 11
\def\journal#1&#2,#3 (#4){\begingroup \let\journal=\d@mmyjournal {\frenchspacing\sl #1\/\unskip\,} {\bf\ignorespaces #2}\rm, #3 (#4)\endgroup}
\def\d@mmyjournal{\errmessage{Reference foul up: nested \journal macros}}
\def\title#1{{``#1''}}
\def\@biblabel#1{#1.}

\end{document}